\documentstyle[12pt,epsfig]{article}
\newcommand{\beq}{\begin{equation}}
\newcommand{\eeq}{\end{equation}}
\newcommand{\be}{\begin{eqnarray}}
\newcommand{\ee}{\end{eqnarray}}
\newcommand{\bea}{\begin{eqnarray}}
\newcommand{\eea}{\end{eqnarray}}

\def\la{\mathrel{\mathpalette\fun <}}

\def\fun#1#2{\lower3.6pt\vbox{\baselineskip0pt\lineskip.9pt
\ialign{$\mathsurround=0pt#1\hfil##\hfil$\crcr#2\crcr\sim\crcr}}}

\begin{document}

\title{$K$-matrix
analysis of the
$(IJ^{PC}=00^{++})$-wave
in the mass region below $1900$ MeV}

\author{V.V. Anisovich and A.V. Sarantsev}

\date{\today}
\maketitle

\begin{abstract}

We present the results of the current analysis of the
partial wave $IJ^{PC} =00^{++}$ based on the
available data for  meson spectra
($\pi\pi , K\bar K , \eta\eta , \eta\eta', \pi\pi \pi\pi $).
In the framework of the $K$-matrix approach, the analytical
amplitude has been  restored in the mass region $280$ MeV$< \sqrt s
<1900$ MeV. The following scalar-isoscalar states are seen:
comparatively narrow resonances
$f_0(980)$, $f_0(1300)$, $f_0(1500)$, $f_0(1750)$
and the broad state
$f_0(1200-1600)$. The positions of the amplitude poles (masses and
total widths of the resonances) are determined as well as  pole
residues (partial widths to meson channels $\pi\pi , K\bar K , \eta\eta
, \eta\eta', \pi\pi \pi\pi $).  The fitted amplitude gives us the
positions of the $K$-matrix poles (bare states) and the values of
bare-state couplings to meson channels thus allowing  the
quark-antiquark nonet classification of  bare states.  On the basis of
the obtained partial widths to the channels $\pi\pi , K\bar K ,
\eta\eta , \eta\eta' $, we estimate the quark/gluonium content of
$f_0(980)$, $f_0(1300)$, $f_0(1500)$, $f_0(1750)$, $f_0(1200-1600)$.
For $f_0(980)$, $f_0(1300)$, $f_0(1500)$ and $f_0(1750)$, their partial
widths testify the  $q\bar q$ origin of these mesons though being
unable to provide precise evaluation of the possible admixture of the
gluonium component in these resonances.
The ratios of the decay coupling constants for the $f_0(1200-1600)$
support the idea about gluonium nature of this broad state.

\end{abstract}

\section{Introduction}

The classification of meson states in the scalar-isoscalar sector
is a key problem for understanding  the
strong QCD being a subject of intensive discussions in recent
years,
see, for example,
\cite{van,klempt,montanet,ochs,ufn,petry,lesniak} and
references therein.

In this paper we present the results of the $K$-matrix analysis
of the $00^{++}$ wave in the invariant mass range 280--1900 MeV.
This analysis is a continuation of earlier work
\cite{ufn,km,km1900,YF99,YF}.
In the latter paper \cite{YF}, the $00^{++}$ wave
had been reconstructed on the basis of the following data set:\\
(1) GAMS data on the $S$-wave two-meson  production in the reactions
 $\pi p\to \pi^0\pi^0 n$,
$\eta\eta n$ and $\eta\eta' n$
at small nucleon momenta transferred, $|t|<0.2$ (GeV/$c$)$^2$
\cite{gams1,gams2};\\
(2) GAMS data on the $\pi\pi$ $S$-wave production in the reaction
 $\pi p\to \pi^0\pi^0 n$
at large momenta transferred,  $0.30<|t|<1.0$ (GeV/$c$)$^2$
\cite{gams1};\\
(3) BNL data on $\pi p^-\to K\bar K n$ \cite{bnl};\\
(4) CERN-M\"unich data on $\pi^+\pi^- \to\pi^+\pi^-$ \cite{C-M};\\
(5)  Crystal Barrel data on
$p\bar p$(at rest, from liquid $H_2$)$\to \pi^0\pi^0\pi^0$,
$\pi^0\pi^0\eta$, $\pi^0\eta\eta$ \cite{cbc}.

Now the experimental basis has been  much broadened, and
additional samples of data are included into present analysis of the
$00^{++}$ wave, as follows:\\
(6) Crystal Barrel data on proton-antiproton annihilation in gas:
$p\bar p$(at rest, from gaseous $H_2$)$\to
\pi^0\pi^0\pi^0$, $\pi^0\pi^0\eta$ \cite{cb,cbc_new},\\
(7) Crystal Barrel data on proton-antiproton annihilation in liquid:
$p\bar p$(at rest, from liquid $H_2$)$ \to
\pi^+\pi^-\pi^0$,
$K^+K^-\pi^0$, $K_SK_S\pi^0$, $K^+K_S\pi^-$ \cite{cb,cbc_new};\\
(8) Crystal Barrel data on neutron-antiproton annihilation in the
liquid deuterium: $n\bar p$(at rest, from liquid
$D_2$)$\to \pi^0\pi^0\pi^-$, $\pi^-\pi^-\pi^+$,
$K_SK^-\pi^0$, $K_SK_S\pi^-$ \cite{cb,cbc_new};\\
(9) E852 Collaboration data on the $\pi\pi$ $S$-wave production in the
reaction $\pi^-p\to \pi^0\pi^0n$ at the nucleon momentum transfers
squared $0<|t|<1.5 \; ({\rm GeV/c})^2$ \cite{E852}.

The production of resonances in the $00^{++}$ wave is accompanied by a
considerable background. This circumstance forces us to carry out a
simultaneous analysis of as much as possible number of spectra, for
only in such an analysis one can single out resonance
parameters reliably.

As compared to the paper \cite{YF}, the reactions of the $p\bar p$
annihilation in gas have been  included into
the present analysis. One should
keep in mind that, while in the liquid hydrogene the $p\bar p$
annihilation is going dominantly from the $S$-state, in the gas there
is a large admixture of the $P$-wave, thus providing  us  an
opportunity to analyse  three-meson
Dalitz plots in more detail.

New Crystal Barrel data allow us to study the two-kaon
channel more reliably as compared to what had been done before. This is
undoubtedly important for the conclusion about the quark-gluonium
content of the $f_0$-mesons under investigation.

The data of the E852 Collaboration on the reaction
$\pi^- p \to \pi^0\pi^0n$  at $p_{lab}=18$ GeV/c \cite{E852} together
with GAMS data on the rection $\pi^- p \to \pi^0\pi^0n$
at $p_{lab}=38$ GeV/c \cite{gams1} give us a good opportunity for the
study the resonances $f_0(980)$ and $f_0(1300)$, for at large
momenta transferred to the nucleon, $|t| \sim 0.5\sim 1.5$ (GeV/c)$^2$,
the production of resonances goes with a small background thus
allowing us to fix convincingly their masses and widths. This is
especially important for $f_0(1300)$: in the compilation \cite{PDG} this
resonance is referred as $f_0(1370)$, with the mass in the
interval $1200-1500$ MeV, though
experimental data favour the mass near 1300 MeV.

In Section 2 we present a set of
formulae for the analysed $K$-matrix amplitudes as follows:
\\ (i) the $K$-matrix amplitudes of
the mass-on-shell transitions into channels $\pi\pi, K\bar K, \eta\eta,
\eta\eta', \pi\pi\pi\pi$; \\
(ii) the $K$-matrix amplitudes for the transitions $\pi\pi^{(R)} (t)\to
\pi\pi$ and $\pi a^{(R)}_1(t)\to \pi\pi$,
where $\pi^{(R)}(t)$ and $a^{(R)}_1(t)$ are
reggeized pion and $a_1$-meson in the
high energy reaction $\pi^- p \to (\pi\pi)_S\; n$; \\
(iii) the $K$-matrix three-meson production amplitudes  in the $p\bar
p$ and $n\bar p$ annihilations.

The $K$-matrix amplitude determines both the amplitude poles
(masses and widths of resonances)
and the $K$-matrix poles (masses of bare states). The
$K$-matrix poles differ from the amplitude poles in two points: \\
(i) The states corresponding to the $K$-matrix poles do not contain any
component
associated with real mesons which are inherent in resonances. The
absence of a cloud of  real mesons allows us to refer conventionally
to these states  as  bare ones \cite{ufn,km1900,YF}. \\
(ii) Due to the transitions {\it bare  state(1) $\to$ real mesons $\to$
bare state(2)} the observed resonances are the mixtures of bare states.
So, for the quark systematics, bare states are the primary objects
rather than  resonances.

In Section 3, we present the main results of the present analysis. As
in previous papers, we have had three solutions which are denoted,
following  \cite{YF}, Solutions I, II-1 and II-2.
In all Solutions, the
characteristics of the resonances
$f_0(980)$, $f_0(1300)$, $f_0(1500)$ and broad state
$f_0(1200-1600)$  coincide with each other and coincide, with a
reasonable
accuracy, with what had been obtained in \cite{YF}. From this point of
view, the performed analysis provided us with unambigously determined
properties of physical states, with an exception for the resonance
$f_0(1750)$, for which the channel  $f_0(1750)\to \pi\pi\pi\pi$ is not
well defined, that resulted in different full widths given by
Solutions I and II. Solutions I, II-1 and II-2 differ by the
characteristics of bare states and background terms in the $K$-matrix,
just as it was the case in \cite{YF},  though it must be particularly
emphasized that the present analysis reveals a noticeable tendency for
the parametrs from different Solutions to become closer to each other.
In the present analysis the distinction in characteristics of bare
states is not dramatic with an exception, again, for  $f_0(1750)$, for
which the coupling constant of the $K$-matrix pole to the channel
$\pi\pi\pi\pi$ given by  Solutions I and II have opposite sign.

The relations between coupling constants of the $q\bar q$ bare state,
$f_0^{bare}$, to channels $\pi\pi$, $K\bar K$, $\eta\eta$, $\eta\eta'$
provide us with an information on the relative weight of the components
$n\bar n=(u\bar u+d\bar d)/\sqrt{2}$ and $s\bar s$. For the gluoniun,
the relations between coupling constants are nearly
the same as for the flavour $q\bar q$ singlet, thus leading to the
ambiguity in the definition of the glueball: Solutions I,
II-1 and II-2 provided us with different possibilities for the
glueball among bare states.

In Solution I, there are two bare states, every of them pretending to be
the glueball: $f^{bare}_0(1220\pm 30)$ and $f^{bare}_0(1635\pm 25)$. In
Solution II-1 only one state, $f^{bare}_0(1225\pm 25)$, satisfies the
constraint inherent in the glueball. In Solution II-2, again there are
two states whose coupling constants are appropriate to the glueball:
 $f^{bare}_0(1230\pm 30)$ and
$f^{bare}_0(1560\pm 25)$ . The uncertainty in the classification of
bare states, which is based on the value of a coupling to channels
$f_0^{bare}\to \pi\pi, K\bar K,\eta\eta,\eta\eta'$, makes us to apply
the other methods to single out the states which are extra ones for the
$q\bar q$-systematics, namely, exotic states. The method based on the
consideration of trajectories in the $(n, M^2)$-plane ($n$ is radial
quantum number of the $q\bar q$ state and $M$ its mass) is discussed
in Section 5.

On the basis of the extracted partial widths for
the channels $\pi\pi$, $K\bar K$, $\eta\eta$, $\eta\eta'$, the
following properties of resonances
$f_0(980)$, $f_0(1300)$, $f_0(1500)$, $f_0(1750)$ and broad state
$f_0(1200-1600)$ are to claim:

{\bf 1. $f_0(980)$}: This resonance is dominantly the $q\bar q$ state,
$q\bar q =n\bar n\sin \varphi+s\bar s \cos \varphi$,
with a large $s\bar s$ component.
Under the assumption that the
admixture of the glueball component is not greater than 15\%,
$W_{gluonium} \la 15\%$,
the hadronic decays give
us the following constraints:  $-95^\circ \le \varphi \la -40^\circ$,
 that is in agreement with
data on radiative decays $f_0(980) \to \gamma\gamma$ and $\phi(1020)
\to\gamma f_0(980)$ \cite{f0gg,f2gg}. Rather large uncertainties in the
determination of mixing angle  are due to the sensitivity of coupling
constants to plausible small admixtures of the gluonium. If
the gluonium component is absent,
hadronic decays provide $\varphi = -67^\circ\pm 10^\circ$.

As concern the characteristics of the $f_0(980)$, we cannot observe any
special property which would single it out from a $q\bar q$ series:
it belongs to linear $q\bar q$ trajectory on the $(n,M^2)$ and $(J,M^2)$
planes and its decay couplings to hadronic channels are of the same
order as for the other $f_0$ resonances,
see Table 5 in Section 4. Besides, this resonance is
produced, without any suppression, in hard processes as well as
in heavy-quark decays, e.g. in the reactions $\pi^-p\to (\pi\pi)_S\; n$
at large $|t|$ \cite{gams1,E852} and $D^+_s\to \pi^+\pi^+\pi^-$
\cite{D+}.

{\bf 2. $f_0(1300)$}: This resonance is the
descendant of the bare $q\bar q $ state
which is close to the  flavour singlet. The resonance $f_0(1300)$
is formed due to a strong mixing with the primary gluonium
and neighbouring $q\bar q$ states.  The $q\bar q$
content, $q\bar q=n\bar n \cos \varphi+s\bar s\sin \varphi$, in the
$f_0(1300)$ strongly depends on the admixture of the gluonium component.
At $W_{gluonium} \la 30\%$ the mixing angle changes, depending on
$W_{gluonium}$, in the interval
$-45^\circ \la \varphi[f_0(1300)] \la 25^\circ$;
at $W_{gluonium} = 0$ the hadronic decays provide
$\varphi[f_0(1300)] = - 7^\circ \pm 10^\circ$.

{\bf 3. $f_0(1500)$}: The resonance is the
descendant of the bare state with a large $n\bar n$ component.
Being similar to $f_0(1300)$, the resonance
$f_0(1500)$ is formed by mixing with the gluonium
as well as with neighbouring $q\bar q$ states. The $q\bar q$
content, $q\bar q=n\bar n \cos \varphi+s\bar s\sin \varphi$,
depends on the admixture of the gluonium:
at $W_{gluonium} \la 30\%$ the mixing angle changes, depending on
$W_{gluonium}$, in the interval
$-20^\circ \la \varphi[f_0(1500)] \la 25^\circ$;
at $W_{gluonium} = 0$ one has
$\varphi[f_0(1500)] = 10^\circ \pm 10^\circ$.

{\bf 4. $f_0(1750)$}: This resonance is the descendant of the bare state
of the radial excitation nonet $2^3P_1 q\bar q$,
which flavour wave function has large $s\bar s$ component.
In Solutions I and II, the resonance $f_0(1750)$ has different
values of  the $s\bar s$ component. In Solution I, the $s\bar s$
component dominates: in the absence of  gluonium component,
$\varphi[f_0(1750)] = - 72^\circ \pm 5^\circ$, and
if the gluonium admixture is $W_{gluonium}\la 30\%$, then
$-110^\circ \la \varphi[f_0(1750)] \la - 35^\circ$.
 In Solution II, in the absence of  gluonium component,
$\varphi[f_0(1750)] = - 27^\circ \pm 10^\circ$, and
with gluonium admixture  $W_{gluonium}\la 30\%$,
$-50^\circ \la \varphi[f_0(1750)] \la 10^\circ$.

{\bf 5. $f_0(1200-1600)$}: The broad state is the descendant of
a primary glueball. The
analysis of hadronic decays  of this resonance confirms its glueball
nature, for the $q\bar q$ component is allowed to be in a state
produced by the glueball:
$(q\bar q)_{glueball}=(\sqrt{2}n\bar n +
\sqrt\lambda \; s\bar s)/\sqrt{2+ \lambda}$ \cite{alexei},
where $\lambda$ defines relative probability to produce the $s\bar s$
pair by gluon field, $ u\bar u: d\bar d: s\bar s=1:1:\lambda$. Different
estimations give $\lambda\simeq 0.5-0.8$ \cite{ufn,peters} thus leading
to $\varphi[f_0(1200-1600)] \simeq 27^\circ -32^\circ$.  Just such a
magnitude  appears in hadronic decays of the broad state, though the
value of a possible admixture of the $(q\bar q)_{glueball}$ cannot be
fixed by hadronic decays.  The impossibility to determine the
quark-antiquark component is due to the fact that the relations between
decay coupling constants  are exactly the same for the gluonium and
$(q\bar q)_{glueball}$.

Determination of the $f_0(1200-1600)$ and its characteristics needs
additional comments. The spectra in the
channels $\pi\pi$, $K\bar K$, $\eta\eta$, $\eta\eta'$ require
the introduction of a broad bump, and this bump occures to be universal
in different reactions thus making it possible to  describe it
as a broad resonance. One of a distinct characteristics of
resonances is a factorization of the resonance amplitudes:
the amplitude, $g_{in}(s-M^2)^{-1}g_{out}$, may be
factorized with universal couplings ($g_{in}$,
 $g_{out}$) for a variety of reactions. The fitting to a large number
of reactions studied here gets along with the factorization
of the broad state. Large width of the braod state does not
allow us to fix reliably the mass of $f_0(1200-1600)$, but its strong
production in a variety of  reactions under investigation
makes it possible to assuredly determine the ratios of couplings
to the channels $\pi\pi$, $K\bar K$, $\eta\eta$, $\eta\eta'$ thus
determining reliably the $q\bar q$/gluonium content of
$f_0(1200-1600)$.

The present $K$-matrix analysis does
not point to the existence of a light
$\sigma$-meson which is actively discussed now
(see e.g. \cite{van} and references therein),
 in particular in
connection with the recently reported signals in  $D^+ \to
\pi^+\pi^+\pi^-$ with the amplitude pole
at $M=(480\pm 40)\;  -i(160\pm 30)$ MeV \cite{D+}, in
$J/\Psi \to\pi\pi\omega$ with  the pole at
$M=(390\; ^{+60}_{-36}) \;  -i(141\; ^{+38}_{-25})$ MeV \cite{BES},
and in
$\tau \to\pi\pi\pi\nu$ with the pole at
$M\simeq 555  -i270$ MeV \cite{CLEO}.
Possible explanation of this contradiction may lay in a strong
suppression of the light $\sigma$-meson production in the annihilation
process $p\bar p\to \pi\pi\pi$, though there is no visible reason for
this suppression (recall that the statistics for the Crystal Barrel
reactions is by two orders of magnitude larger than in
\cite{D+,BES,CLEO}).
Alternative explanation can be related to a restricted validity of
the $K$-matrix approach at small invariant energy squared $s$: due to
the left-hand cuts in the partial amplitude, which were not
properly taken into
account in the $K$-matrix approach, one may believe that
the analysis does not reconstruct anlytical amplitude in due course at
${\rm Re \,}s \la 4m^2_\pi$, i.e. at $({\rm Re \,}M)-({\rm Im \,}M)^2
\la 4m^2_\pi$.  In a number of analyses, including those performed in
the dispersion relation technique where the left-hand cut can be
accounted for in one way or another, the pole ascribed to the
$\sigma$-meson had been obtained just at ${\rm Re \,}s \la 4m^2_\pi$,
see e.g.  \cite{sigma_dis,sigma80,sigma,sigma_N/D}.

In Section 3, to illustrate the level of accuracy for the Dalitz-plot
description in the reactions
$p\bar p \to three\, mesons$ and
$n\bar p \to three\, mesons$ we
demonstrate experimental spectra in various reactions versus fitted
curves of Solution II-2. We also show the $\pi\pi$ spectra in the
reactions $\pi^- p \to (\pi\pi)_S \, n$ measured by GAMS \cite{gams1}
and E852 \cite{E852} and their description in one of the fits (a
detailed discussion of these spectra is presented in a separate article
\cite{bnl-gams}). One important statement of this analysis is worth
being stressed here:  for a combined description of the GAMS and
E852 data, one needs to introduce the $t$-channel exchanges by the
leading and daughter Regge trajectories: $\pi_{leading}$,
$a_{1(leading)}$ and $\pi_{daughter}$, $a_{1(daughter)}$.

The parameters of the $K$-matrix amplitude obtained directly
from the fit (they are given in Tables 2, 3, 4 for Solutions I, II-1
and II-2) serve us as characteristics of bare states.
To define coupling constants of the $f_0$-resonances to
channels $\pi\pi$, $K\bar K$, $\eta\eta$, $\eta\eta'$, $\pi\pi\pi\pi$
one needs to calculate the amplitude residues, see Section 4.
In this Section, on the basis of calculated couplings, we give partial
widths of resonsnces $f_0(980)$, $f_0(1300)$, $f_0(1500)$  and
$f_0(1750)$.

In Section 5, the calculated couplings are analysed in terms of quark
combinatorics rules to define the quark-antiquark and gluonium content
of resonances.

The analysis performed here does not give us a precise
correlation between the $n\bar n$, $s\bar s$ and gluonium
components.
The reason of this is not a possible  insufficient accuracy
of data but the structure of $f_0$-mesons themselves in
the presence of gluonium admixtures --- this problem is emphasized in
Conclusion.

In Appendices A, B, C we present, in more detail, the formulae used in
this analysis.

\section{$(IJ^{PC}=00^{++})$-wave: the $K$-matrix amplitude and
quark-combinatorics relations for the decay couplings}

Here we set out the
$K$-matrix formulae for the analysis of the
$(IJ^{PC}=00^{++})$-wave and present
the quark-combinatorics relations for the $K$-matrix couplings
which are used for the study the $q\bar q$/gluonium content of the
states under consideration.

\subsection{$K$-matrix amplitude }

The $K$-matrix technique is used for the description
of the two-meson coupled channels:
\beq
\hat A=\hat  K(\hat I-i\hat{\rho} \hat K)^{-1},
\label{1}
\eeq
where $\hat K$ is $n\times n$ matrix (here $n$ is the number of channels
under consideration)
and $\hat I$ is unit matrix.
The phase
space matrix is diagonal:  $\hat{\rho}_{ab}=\delta_{ab}\rho_a$. The
phase
space factor $\rho_a$ is responsible for the threshold singularities
of the amplitude: to keep the amplitude analytical in the
physical region under consideration we use analytical continuation for
$\rho_a$ below threshold.  For example, the $\eta\eta$ phase space
factor $\rho_{\eta\eta}=(1-4m_\eta^2/s)^{1/2}$ is equal to
$i(4m_\eta^2/s-1)^{1/2}$  below the $\eta\eta$ threshold
($s$ is the two-meson invariant energy squared).
To avoid false singularity in the physical region,
we use for the
$\eta\eta'$ channel the following phase space factor
$\rho_{\eta\eta'}=(1-(m_\eta+m_{\eta'})^2/s)^{1/2}$.

For the multi-meson phase volume, we
use the four-pion phase space defined as phase space of either $\rho
\rho$ or  $\sigma \sigma$, where $\sigma$ denotes the
$S$-wave $\pi\pi$ amplitude below  1.2 GeV.  The result does not depend
practicaly on whether we use $\rho \rho$ or $\sigma \sigma$ state for
the description of multi-meson channel: below we write formulae
and the values of obtained parameters for the $\rho \rho$ case, for
which the fitted expressions  are less cumbersome.

\subsubsection{Scattering amplitude}

For the $S$-wave scattering
amplitude in the scalar-isoscalar sector,
we use the parametrization similar
to that of \cite{ufn,YF99,YF}:
\be
K_{ab}^{00}(s)
=\left ( \sum_\alpha \frac{g^{(\alpha)}_a
g^{(\alpha)}_b}
{M^2_\alpha-s}+f_{ab}
\frac{1\;\mbox{GeV}^2+s_0}{s+s_0} \right )\;
\frac{s-s_A}{s+s_{A0}}\;\;,
\label{2}
\ee
where $K_{ab}^{IJ}$ is a 5$\times $5 matrix ($a,b$ = 1,2,3,4,5), with
the following notations for meson states: 1 = $\pi\pi$, 2 = $K\bar K$,
3 = $\eta\eta$, 4 = $\eta\eta'$ and 5 =  multimeson states
(four-pion state mainly at $\sqrt{s}<1.6\; \mbox{GeV}$).
The $g^{(\alpha)}_a$ is the coupling constant of the bare state
$\alpha$ to meson channel; the parameters $f_{ab}$ and $s_0$
describe a smooth part of the $K$-matrix elements ($1 \le s_0 \le 5$
GeV$^2$).  We use  the factor $(s-s_A)/(s+s_{A0})$ to suppress false
kinematical singularity at $s=0$ in the physical region near the
$\pi\pi$ threshold. The parameters $s_A$ and $s_{A0}$ are kept to be of
the order of $s_A\sim (0.1-0.5)m^2_\pi$ and $s_{A0}\sim (0.1-0.5)$
GeV$^2$; for these intervals, the results do not depend practically on
precise values of $s_A$ and $s_{A0}$.

For the two pseudoscalar-particle states,
$\pi\pi$, $K\bar K$, $\eta\eta$, $\eta\eta'$,
the  phase space matrix elements  are equal to:
\beq
\rho_a(s)=\sqrt{\frac{s-(m_{1a}+m_{2a})^2}{s}}\qquad , \qquad a=1,2,3,4
\label{3}
\eeq
where $m_{1a}$ and $m_{2a}$ are the masses of pseudoscalars.
The multi-meson phase-space factor is determined as follows:
\beq
\rho_5(s)= \left \{ \begin{array}{cl}
\rho_{51}\;\;\mbox{at} \;\;s<1\;\mbox{GeV}^2,\\
\rho_{52}\;\;\mbox{at} \;\;s>1\;\mbox{GeV}^2,
 \end{array} \right.
\eeq
$$
\rho_{51}=\rho_0\int\frac{ds_{1}}{\pi}\int\frac{ds_{2}}{\pi}
 M^2\Gamma(s_{1})\Gamma(s_{2})
\sqrt{(s+s_{1}-s_{2})^2-4ss_{1}}
$$
$$
\times s^{-1} [(M^2-s_{1})^2+M^2\Gamma^2(s_{1})
]^{-1}
[(M^2-s_{2})^2+M^2\Gamma^2(s_{2}) ]^{-1} ,
$$
$$
\rho_{52}= \left (\frac{s-16m_\pi^2}{s}\right )^n \ .
$$
Here $s_1$ and $s_2$ are the two-pion energies
squared, $M$ is the $\rho$-meson mass and
$\Gamma(s)$ its energy-dependent width,
$\Gamma(s)=\gamma\rho_1^3(s)$. The factor $\rho_0$ provides the
continuity of $\rho_5(s)$  at $s=1$ GeV$^2$.
The power parameter $n$ is taken to be 1, 3, 5
for different variants of the fitting;  the results are weakly
dependent on these values (in our
previous analysis \cite{YF} the value $n=5$ was used).

\subsubsection{High-energy production of the $S$-wave mesons
$\pi\pi$, $K\bar K$, $\eta\eta$ and
$\eta\eta'$ in the $\pi p$ collisions}

Here we present the formulae for the high-energy $S$-wave
production of $\pi\pi$, $K\bar K$, $\eta\eta$, $\eta\eta'$
at small and moderate momenta transferred to the nucleon.
In \cite{gams1,gams2,bnl,E852},
the $\pi p$ collisions were studied at $p_{beam}\sim (15-40)$ GeV/c
(or $s_{\pi N}\simeq 2m_N p_{beam} \sim 30-80 $ GeV$^2$). At such
energies, two pseudoscalar mesons are produced
due to the $t$-channel exchange by  reggeized mesons
belonging to the $\pi $ and $a_1$ trajectories, leading and daughter
ones.

The $\pi$ and $a_{1}$ reggeons have different signatures,
$\xi_\pi =+1$ and $\xi_{a1} =-1$. Accordingly, we
write  the $\pi$ and $a_{1}$ reggeon propagators as
\be
e^{i\frac{\pi}{2}\alpha_\pi (t)}
\frac{s^{\alpha_\pi (t)}_{\pi N}}{\sin (\frac{\pi}{2}\alpha_\pi (t))}
\quad ,
\qquad
ie^{-i\frac{\pi}{2}\alpha_{a1} (t)}
\frac {s^{\alpha_{a1}  (t)}_{\pi N}}{\cos (\frac{\pi}{2}\alpha_{a1}
(t))}\quad \ .
\label{7}
\ee
Following \cite{syst}, we use for
leading trajectories:
\be
\alpha_{\pi (leading)} (t)\simeq -0.015 +0.72 t , \quad
\alpha_{a1 (leading) } (t) \simeq   -0.10 +0.72 t,
\label{8}
\ee
and for  daughter ones:
\be
\alpha_{\pi ( daughter)} (t)\simeq -1.10 +0.72 t , \quad
\alpha_{a1( daughter)} (t) \simeq   -1.10 +0.72 t\ .
\label{9}
\ee
Here the slope parameters are in (GeV/c)$^{-2}$.
In the centre-of-mass frame, which is the most convenient for the
consideration of  reggeon exchanges, the
incoming particles move along the $z$-axis with a large momentum $p$.
In the leading order of the $1/p$ expansion,
the spin factors for $\pi$ and $a_1$ trajectories read:
\be
\pi-{\rm trajectory}: \qquad
 (\vec \sigma \vec q_\perp)\ ,
\label{10}
\ee
$$
a_1-{\rm trajectory}:\qquad
i (\vec \sigma \vec n_z)\ ,
$$
where $\vec n_z=\vec p/p$ and $\vec q_\perp$ is
the momentum transferred to the nucleon ($t\simeq -q^2_\perp$).  The
Pauli matrices $\vec \sigma$ work in the two-component spinor space for
the incoming and outgoing nucleons:  $(\varphi _{out}^* \vec \sigma
\varphi_{in})$ (for  more detail see, e.g.,
\cite{kaidalov,alkhazov}). Consistent removal from the vertices
(\ref{10}) of terms decreasing with $p\to \infty$ is necessary for
the correct account for daughter trajectories which should obey,
similarly to leading ones, the constraints imposed by the
$t$-channel unitarity.

In our calculations, we modify
reggeon propagators in (\ref{7})  conventionally by replacing
\be
s_{\pi N}\to \frac{ s_{\pi N}}{ s_{\pi N0}}\ ,
\label{11}
\ee
where the normalization parameter $ s_{\pi N0}$ is of the order
of 2--20 GeV$^2$. To eliminate the poles at $t<0$ we introduce
additional factors (Gamma-functions) into  reggeon
propagators by substituting in (\ref{7}) as follows:
\be
\sin \left (\frac{\pi}{2}\alpha_\pi (t)\right )
\to
\sin \left (\frac{\pi}{2}\alpha_\pi (t)\right ) \;
\Gamma \left (\frac{ \alpha_\pi (t)}{2} +1\right )\ ,
\label{12}
\ee
$$
\cos \left (\frac{\pi}{2}\alpha_{a1} (t)\right )
\to
\cos \left (\frac{\pi}{2}\alpha_{a1} (t)\right )  \;
\Gamma \left (\frac {\alpha_{a1} (t)}{2} +\frac 12\right )\, .
$$
The $K$-matrix amplitude for         the
transitions $\pi R(t)\to \pi\pi$,
$K\bar K$, $\eta\eta$, $\eta\eta'$, $\pi\pi\pi\pi$,
where $R(t)$ refers to reggeon, reads:
\be
\hat A_{\pi R}=\hat K_{\pi R}(\hat I-i\hat{\rho}
\hat K)^{-1},
\label{13}
\ee
where $\hat K_{\pi R}$ is the following vector:
\be
K_{\pi R,b}^{00}
=\left ( \sum_\alpha \frac{G^{(\alpha)}_{\pi R}(t)
g^{(\alpha)}_b}
{M^2_\alpha-s}+F_{\pi R,b}(t)
\frac{1\;\mbox{GeV}^2+s_{R0}}{s+s_{R0}} \right )\;
\frac{s-s_A}{s+s_{A0}}\;\; .
\label{14}
\ee
Here $G^{(\alpha)}_{\pi R}(t)$ and $F_{\pi R,b}(t)$
are the reggeon $t$-dependent form factors.
The following limits are imposed on the form factors:
\be
G_{\pi\pi}^{(\alpha)}(t\to m^2_\pi)=
g^{(\alpha)}_{\pi\pi}\, ,
\qquad F_{\pi
\pi,b}(t\to m^2_\pi)=f_{\pi\pi,b}\ ,
\label{t-limit}
\ee
where
$g^{(\alpha)}_{\pi\pi}$ and $f_{\pi\pi,b}$ enter
the matrix element (\ref{2}).

\subsection{Three-meson production amplitudes}

In this Section
we present the formulae for the reactions $p\bar p\to
\pi^0\pi^0\pi^0$, $\pi^0\pi^0\eta$, $\pi^0\eta\eta$
from the liquid $H_2$, when annihilation occurs
from the $^1S_0p\bar p$ state and scalar
resonances, $f_0$ and $a_0$,  are formed
in the final state only. Of course, this is only one
sub-process from a number of reactions considered here; still, it is
rather representative with respect to the applied technique of the
three-meson production reaction. A full set of amplitude terms taken
into account in the present analysis (production of vector and tensor
resonances, $p\bar p$ annihilation from the $P$-wave states $^3P_1$,
$^3P_2$,  $^1P_1$) is given in Appendices A and B.

For the transition $p\bar p\; (^1S_0)\to \pi^0\pi^0\pi^0$,
the amplitude
 has the following structure:
\be
A_{p\bar p\; (^1S_0)\to \pi^0\pi^0\pi^0}=
\left (\bar\psi(-q_2)\frac{i\gamma_5}{2\sqrt 2 m_N} \psi(q_1)\right )
\cdot
\label{15pi}
\ee
$$
\left [A_{p\bar p\; (^1S_0)\pi^0,\pi^0\pi^0}(s_{23})
+A_{p\bar p\; (^1S_0)\pi^0,\pi^0\pi^0}(s_{13})
+A_{p\bar p\; (^1S_0)\pi^0,\pi^0\pi^0}(s_{12})\right ] \; .
$$
The amplitude $A_{p\bar p\; (^1S_0)\pi^0,\pi^0\pi^0} (s_{ij})$
describes the diagrams with
interacting mesons, with the last
interaction for the particles $i$ and $j$.
In the initial state, the four-spinors
$\bar\psi(-q_2)$ and $ \psi(q_1)$ refer to antiproton
and proton, respectively.

The amplitudes for the transitions $p\bar p\; (^1S_0)\to
\eta\pi^0\pi^0,\pi^0\eta\eta$ have similar form:
\be
A_{p\bar p\; (^1S_0)\to \eta\pi^0\pi^0}=
\left (\bar\psi(-q_2)\frac{i\gamma_5}{2\sqrt 2 m_N} \psi(q_1)\right )
\label{15eta}
\ee
$$
\times\left [A_{p\bar p\; (^1S_0)\eta,\pi^0\pi^0}(s_{23})
+A_{p\bar p\; (^1S_0)\pi^0,\eta\pi^0}(s_{13})
+A_{p\bar p\; (^1S_0)\pi^0,\eta\pi^0}(s_{12})\right ] \; ,
$$
and
\be
A_{p\bar p\; (^1S_0)\to \pi^0\eta\eta}=
\left (\bar\psi(-q_2)\frac{i\gamma_5}{2\sqrt 2 m_N} \psi(q_1)\right )
\label{15etaeta}
\ee
$$
\times\left [A_{p\bar p\; (^1S_0)\pi^0,\eta\eta}(s_{23})
+A_{p\bar p\; (^1S_0)\eta,\eta\pi^0}(s_{13})
+A_{p\bar p\; (^1S_0)\eta,\eta\pi^0}(s_{12})\right ] \; .
$$
The following amplitude is used for the two-meson interaction block
in the scalar-isoscalar state $(IJ=00)$
in the reactions  $p\bar p\to \pi^0\pi^0\pi^0$ and
 $p\bar p\to \pi^0\eta\eta$:
\be
A_{p\bar p\; (^1S_0)\pi^0,b}(s_{23})
=\sum\limits_{a}\widetilde
K_{p\bar p(^1S_0)\pi^0,a}^{00} (s_{23})
\left[\hat I-i \hat\rho
\hat K^{00}(s_{23})\right]^{-1}_{ab}\; .
\label{16}
\ee
Here
$b=\pi^0\pi^0$ stands for
the $\pi^0\pi^0\pi^0$ production, and $b=\eta\eta$
for $\pi^0\eta\eta$.
The $\widetilde K$-matrix terms which describe
the prompt $f_0$-production
in the $p\bar p$ annihilation have  the following
form:
\be
\widetilde K_{p\bar p(^1S_0)\pi^0,a}^{00}(s_{23})=\left (
\sum_\alpha \frac{\Lambda^{(\alpha)}_{p\bar p(^1S_0)\pi^0}[00]
g^{(\alpha)}_a}
{M^2_\alpha-s_{23}} +\phi_{p\bar p(^1S_0)\pi^0,a}[00]\;
\frac{1\; \mbox{GeV}^2+s_0}{s_{23}+s_0} \right)
\left (\frac{s_{23}-s_A}{s_{23}+s_{A0}}\right )\;,
\label{18}
\ee
The parameters $\Lambda^\alpha_{p\bar p(^1S_0)\pi^0}[00]$ and
$\phi_{p\bar p(^1S_0)\pi^0}[00]$
(or $\Lambda^\alpha_{p\bar p(^1S_0)\eta}[00]$
and $\phi_{p\bar p(^1S_0)\eta}[00]$)
can be complex-valued, with different phases due to
 three-particle interactions.
The matter is that in the final state interaction term (\ref{18}) we
take into account the leading (pole) singularities only.  The
next-to-leading singularities are accounted for effectively, by
considering the vertices $p\bar p\to mesons$ as complex factors, for
more detail see \cite{ABSZ}.

Here the formulae are written for the production of two mesons in the
$(IJ^{PC}=00^{++})$-state: this very wave has been fitted in the present
analysis. In the final state of reactions under consideration, vector
and tensor mesons are produced. As was said above, the parameters for
vector and tensor resonances have not been fitted in this analysis;
we have used those found in \cite{YF}. The formulae for
 vector and tensor resonances used in fitting procedure are given
in Appendices A and B together with formulae for the annihilation from
higher states.

\subsection{Quark-combinatorics rules for the decay couplings}

The decay couplings of the
$q\bar q$-meson and glueball to a pair of mesons are determined
by the planar diagrams with $q\bar q$-pairs
produced by  gluons: these diagrams provide
the leading terms in the 1/N expansion \cite{t'h}, while
non-planar diagrams give the
next-to-leading contribution.

The production of soft $q\bar q$ pairs by
gluons violates flavour symmetry, with the following ratios of the
production probabilities:
\be
u\bar u:d\bar d:s\bar s=1:1:\lambda \; .
\label{lambda}
\ee
Suppression parameter $\lambda$ for the production of strange
quarks varies in the interval $0\leq \lambda\leq 1$. An estimate
performed for the
high energy collisions gives $\lambda=0.4-0.6$ \cite{lambda-hec}, in
hadron decays it was evaluated as $\lambda\simeq 0.8$ \cite{peters}.

We impose for the decay couplings of bare states, $g^{(\alpha)}_a$,
the quark-combinatorics relations. The rules
of quark combinatorics were
previously suggested for the high energy hadron production \cite{as-bf}
and then extended for hadronic $J/\Psi$ decays \cite{wol}.
The quark combinatoric relations were used
for the decay couplings of the scalar-isoscalar states
in the analysis of the quark-gluonium content of resonances in
\cite{glp} and later on in a set of papers
\cite{km,km1900,YF99,YF,aas-z,AlAn,dmatr1,dmatr2}.  We present the
decay couplings for scalar mesons, which are a subject of our analysis,
in Appendix C.

The wave function of the $f_0$-state is supposed to be a mixture of
the quark-antiquark and gluonium componets as follows:
\be
q\bar q\cos\alpha+ gg \sin\alpha \ ,
\label{3.1}
\ee
where the $q\bar q$-state is a mixture of
non-strange and strange quarks, $n\bar n=(u\bar
u+d\bar d)/\sqrt{2}$ and $s\bar s$:
\be
q\bar q=
n\bar n \cos\varphi+s\bar s \sin\varphi    \; .
\label{3.2}
\ee
Using formulae given in Appendix C for the vertices $q\bar q\to
\pi\pi$, $K\bar K$, $\eta\eta$, $\eta\eta'$ together with analogous
couplings for the transition {\it gluonium$\to$two-meson state}, we
obtain the following coupling constants squared for the decays
$f_0\to \pi\pi$, $K\bar K$, $\eta\eta$, $\eta\eta'$:
\be
g^2_{ \pi\pi}=\frac 32\left
(\frac{ g}{\sqrt 2}\cos\varphi +\frac{ G}{\sqrt{2+\lambda}} \right )^2,
\label{gpipi}
\ee
$$
g^2_{K\bar K} =2
\left (\frac{g}{2} (\sin\varphi+\sqrt{\frac{ \lambda}{2}}\cos\varphi)
+G \sqrt{\frac{\lambda}{2+\lambda}}\; \right )^2\ ,
$$
$$
g^2_{\eta\eta}=\frac 12 \left (g\; (\frac{\cos^2\Theta}{\sqrt
2}\cos\varphi +\sqrt{\lambda}\;\sin\varphi\;\sin^2\Theta\; )+
\frac{G}{ \sqrt{2+\lambda}}\;
 (\cos^2\Theta+ \lambda\sin^2\Theta\ )\right )^2 \ ,
$$
$$
g^2_{\eta\eta'} =\sin^2\Theta\;\cos^2\Theta\; \left (
g\;  (\frac{1}{\sqrt 2}\cos\varphi-
\sqrt{\lambda}\;\sin\varphi  ) +G\;
\frac{1-\lambda}{\sqrt{2+\lambda}}
 \right )^2\ .
$$
Here $g=g_0 \cos\alpha $ and $G=G_0 \sin\alpha $, where $g_0$ is
universal constant for all nonet members and $G_0$ is universal
decay constant for the gluonium state. The value
$g^2_{\pi\pi}$ is determined as
a sum of couplings squared for the transitions to
$\pi^+\pi^-$ and $\pi^0\pi^0$, when  the identity
factor for $\pi^0\pi^0$ is taken into account.
Likewise, $g^2_{K\bar K}$ is the sum of coupling constants squared for
the transitions to $K\bar K$ and $K^0\bar K^0$. The angle $\Theta$
stands for the mixing of $n\bar n$ and $s\bar s $ components in the
$\eta$ and $\eta'$ mesons (see Appendix C).

The quark combinatorics make it  possible
to perform the nonet classification of
bare states. In doing that we
refer to $f^{(bare)}_0$'s as pure states, either
$q\bar q$ or glueball.  For the $q\bar q$ states that means:

\begin{description}
\item[(1)] The angle difference between
isoscalar nonet partners should be
$90^\circ$:
\be
\varphi [f^{(bare)}_0(1)]-\varphi [f^{(bare)}_0(2)]=
90^\circ\pm 5^\circ \; .
\label{25}
\ee
\item[(2)] Coupling constants $g_0$ presented in Appendix C
should be roughly equal to each other for all nonet
partners:
\be
g_0[f^{(bare)}_0(1)] \simeq g_0[f^{(bare)}_0(2)]
\simeq g_0[a^{(bare)}_0]
\simeq g_0[K^{(bare)}_0] \; .
\label{26}
\ee
\item[(3)]
Decay couplings for  bare gluonium should obey relations
for glueball (see Appendix C).
\end{description}
 Conventional quark model requires exact coincidence of the
couplings $g_0$
but the energy dependence of the decay loop diagram,
$B(s)$, may violate the coupling-constant balance because of the mass
 splitting inside a nonet. The $K$-matrix coupling constant  contains
additional $s$-dependent factor as compared to the coupling of the
N/D-amplitude
\cite{aas-z}: $g^2(K) =  g^2(N/D)/[1+B'(s)]$. The factor
$[1+B'(s)]^{-1}$ mostly affects the low-$s$ region due to the threshold
and left-hand side singularities of the partial amplitude.
Therefore, the coupling constant equality is mostly violated for the
lightest $00^{++}$ nonet, $1^3P_0$ $q\bar q$. We allow for the members
of this nonet $1\leq g[f_0(1)]/g[f_0(2)]\leq 1.3$.
For the $2^3P_0$ $q\bar q$ nonet members, we set the
two-meson couplings  to be equal both for isoscalar and isovector
mesons.

\section{Description of data and the results for the $00^{++}$-wave}

 Table \ref{table1} demonstrates the data sets which have been fitted in
the present analysis; here the number of fitted points for each
reaction is also shown.

For the description of the $00^{++}$ wave in the mass region below
1900 MeV, five $K$-matrix poles are needed (a four-pole amplitude fails
to describe well the data set under consideration). Accordingly,
five bare states are
introduced. As in the previous analysis we have found
three solutions.  All Solutions, which we denoted, following
\cite{YF}, Solutions I,
 II-1 and II-2, are similar in the pricipal points
to that found in \cite{YF}.

The $\chi^2$ values for each
reaction in every solution are given in Table 1.

\begin{table}
\begin{center}
\caption\protect{List of reactions and $\chi^2$ values for the
$K$-matrix solutions.}
\label{table1}
\vskip 0.5cm
\begin{tabular}{|l|c|c|c|c|}
\hline
~ & ~ &  ~ & ~ & \\
~ &  Solution I &  Solution II-1 &  Solution II-2&  Number of\\
~ & ~ &  ~ & ~& points\\
\hline
 ~& \multicolumn{4}{|c|}{The Crystal Barrel data} \\
\hline
from liquid $H_2$:              & ~     &~      &~     & ~   \\
$\bar pp\to \pi^0\pi^0\pi^0$    & 1.30  &1.39   &1.32  & 7110\\
$\bar pp\to \pi^0\eta\eta$      & 1.32  &1.30   &1.32  & 3595\\
$\bar pp\to \pi^0\pi^0\eta$     & 1.22  &1.31   &1.34  & 3475\\
\hline
from gaseous $H_2$:             &~      &      &     &  ~ \\
$\bar pp\to \pi^0\pi^0\pi^0$    & 1.38  &1.35  &1.39 &  4891\\
$\bar pp\to \pi^0\eta\eta$      & 1.25  &1.24  &1.26 &  1182\\
$\bar pp\to \pi^0\pi^0\eta$     & 1.17  &1.18  &1.20 &  3631\\
\hline
with charge pions,              &~      &      &     &  ~ \\
from liquid $H_2$:              & ~     &~     &~    &  ~ \\
$\bar pp\to \pi^+\pi^0\pi^-$    & 1.38  &1.42  & 1.46&    1334\\
from liquid $D_2$:              & ~     &~     & ~   &   ~ \\
$\bar pn\to \pi^0\pi^0\pi^-$    & 1.46  &1.43  & 1.48&    825 \\
$\bar pn\to \pi^-\pi^-\pi^+$    & 1.51  &1.57  & 1.58&    823 \\
\hline
with kaons,                     & ~     &~     &~    &  ~   \\
from liquid $H_2$:              & ~     &~     &~    &  ~   \\
$\bar pp\to K_SK_S\pi^0$        & 1.08  &1.08  &1.10 &   394\\
$\bar pp\to K^+K^-\pi^0$        & 0.94  &0.91  &0.99 &   521 \\
$\bar pp\to K_LK^\pm\pi^\mp$    & 0.79  &0.81  &0.80 &   737 \\
from liquid $D_2$:              & ~     &~     &~    &  ~ \\
$\bar pn\to K_S K_S\pi^-$       & 1.73  &1.61  &1.69 &   396\\
$\bar pn\to K_S K^-\pi^0$       & 1.33  &1.40  &1.22 &   378 \\
\hline
\hline
 CERN-M\"unich data          & ~     &~     &~    &   ~ \\
$\pi^+\pi^-\to\pi^+\pi^-$ (all waves)
                                & 1.82  &1.70  &1.68 &  705 \\
\hline
 GAMS data                   &~      &      &     &   ~ \\
$\pi\pi\to \pi^0\pi^0$~(S-wave) & 1.15  &1.13  &1.28 &  16 \\
$\pi\pi\to \eta\eta$~~~~(S-wave)& 1.16  &1.35  &0.92 &  16 \\
$\pi\pi\to \eta\eta'$~~~(S-wave)& 0.54  &1.76  &0.67 &  8  \\
\hline
 BNL data                    &  ~    & ~    & ~   &  ~   \\
$\pi\pi\to K\bar K$~(S-wave)    & 1.15  &0.75  &1.10 &   35 \\
\hline
 SLAC-NAGO-CINC-INUS    &  ~    & ~    & ~   &  ~   \\
 data                   &  ~    & ~    & ~   &  ~   \\
$K^-\pi^+\to K^-\pi^+$~(S-wave) & 2.80  &2.80  &2.80 &   46 \\
\hline
\end{tabular}
\end{center}
\end{table}

Figures 1-11 demonstrate the two-meson spectra and angle
distributions for the reactions
$p\bar p\to three\; mesons$ and $n\bar p\to three\; mesons$.
The scattering amplitudes $\pi\pi\to \pi\pi$, $\pi\pi\to \eta\eta$,
and so on, are restored on the same level
of accuracy as in \cite{YF}, and we
do not show them.

In Figs. 1, 2, 3 one can see the
$M^2_{\pi^0\pi^0}$, $M^2_{\eta\pi^0}$, $M^2_{\eta\eta}$
distributions in the reactions
$p\bar p(liquid\, H_2) \to \pi^0\pi^0\pi^0
\; \pi^0\eta\eta,\;\pi^0\pi^0\eta$. The Dalitz-plots for these
reactions were the corner-stones for the
$K$-matrix fits to the $00^{++}$ wave
at the early stage of our study \cite{km,km1900}: Figs 1, 2, 3
demonstrate the level of accuracy of the present fit.
Figure 4
shows similar mass distributions in the reaction
$p\bar p(liquid\, H_2) \to \pi^+\pi^0\pi^-$.

Figures 5 and 6 demonstrate the distributions for the reactions
$n\bar p(liquid\, D_2) \to \pi^0\pi^0\pi^-$, $\pi^-\pi^-\pi^+$. The
angle distributions stand for  bands with the production of
$f_2(1285)+f_0(1300)$ (Fig. 5c) and $\rho(1450)$ (Figs. 5d, 6d). In
connection with the discussion of a possible resonance structure in the
$I=2$ channel, we show in Fig. 6c the angle distribution along the band
$M_{\pi^-\pi^-}\sim 1300$ MeV (recall that resonance states with $I=2$
are not included into  our fit).

In Figs. 7 and 8,
one can see the mass and angle distributions in the reactions
$p\bar p(liquid\, H_2) \to K_S K_S\pi^0, K^+ K^-\pi^0$
(parameters for $a_0^{(bare)}$ and $K_0^{(bare)}$ have been fixed,
correspondingly, in \cite{YF} and \cite{AlAn}
($K$-matrix re-analysys of the SLAC-NAGO-CINC-INUS data \cite{aston}
for $K\pi\to K\pi$).
Angle distributions in Figs. 7d and 8d are selected
for the demonstration, in connection with the
discussed possibility for the
$a_0$-resonance to exist in the region $\sim 1300$ MeV (in the fit performed
in \cite{YF}, the $a_0$-state had not been found
in the mass region $\sim 1300$ MeV).

In Fig. 9, the distributions in
the reactions
$p\bar p(liquid\, H_2) \to K_L K^+\pi^- (K_L K^-\pi^+), K_S K^-\pi^+$
are shown. The distributions in the channel $K_S K^-\pi^+$ have not
been fitted because of the acceptance-correction problem: the curve is
the result of the fit carried out without taking account of this
channel data; nevertheless,
as is seen from the figure, the description is rather good.

In Figs. 10
and 11, we show distributions in the reactions $n\bar p(liquid\, D_2)
\to K_S K_S\pi^- , K_S K^-\pi^0$. The angle distributions (Figs. 10d,
11d) are presented for the band where the production of $a_0(1300)$ can
be seen; yet, we do not see any signal from $a_0(1300)$.

\subsection{Bare $f_0$-states and resonances}

In the $K$-matrix analysis of the $00^{++}$-wave,
five bare states have been found, see Tables 2, 3 and 4.
The bare states can be classified as
nonet partners of the $q\bar q$ multiplets $1^3P_0$ and $2^3P_0$
or a scalar glueball. The
$K$-matrix solutions give us two variants for the glueball
definition: either it
is a bare state with the mass near 1250 MeV, or it is located near 1600
MeV.

After having imposed the constrains (\ref{25}) and
(\ref{26}), we  found the following variants for the nonet
classification.

Solution {\bf{I}}:

$f_0^{\rm bare}(700\pm 100 )$ and $f_0^{\rm bare}(1245\pm 40)$ are
$1^3P_0$ nonet partners with
$\varphi [f_0^{\rm bare}(700)] = -70^\circ\pm 10^\circ$ and
$\varphi [f_0^{\rm bare}(1245)] = 20^\circ\pm 10^\circ$.\\
For members of the $2^3P_0$ nonet, there are two variants:\\
1) either $f_0^{\rm bare}(1220\pm 40)$ and $f_0^{\rm bare}(1750\pm 30)$
are  $2^3P_0$ nonet partners,
with $\varphi [f_0^{\rm bare}(1220)] = 33^\circ\pm 8^\circ$ and
$\varphi [f_0^{\rm bare}(1750)] = 60^\circ\pm 10^\circ$, while
$f_0^{\rm bare}(1630\pm 40)$ is the glueball, with
$\varphi [f_0^{\rm bare}(1630)] = 27^\circ\pm 10^\circ$; or\\
2) $f_0^{\rm bare}(1630\pm 40)$ and $f_0^{\rm bare}(1750\pm 30)$
are  $2^3P_0$ nonet partners, and
$f_0^{\rm bare}(1220\pm 40)$ is the glueball.

Solution {\bf{II-1}}:

$f_0^{\rm bare}(670\pm 100 )$ and $f_0^{\rm bare}(1215\pm 40)$ are
$1^3P_0$ nonet partners with
$\varphi [f_0^{\rm bare}(670)] = -65^\circ\pm 10^\circ$
and
$\varphi [f_0^{\rm bare}(1215)] = 15^\circ\pm 10^\circ$;
\\ $f_0^{\rm bare}(1560\pm 40)$ and $f_0^{\rm
bare}(1830\pm 40)$ are $2^3P_0$ nonet partners
with $\varphi [f_0^{\rm bare}(1560)] = 15^\circ\pm 10^\circ$
and $\varphi
[f_0^{\rm bare}(1830)] =-80^\circ\pm 10^\circ$,
\\ $f_0^{\rm bare}(1220)$ is the glueball,
$\varphi [f_0^{\rm bare}(1220)] = 40^\circ\pm 10^\circ$.

Solution {\bf{II-2}}:

$f_0^{\rm bare}(700\pm 100 )$ and $f_0^{\rm bare}(1220\pm 40)$ are
$1^3P_0$ nonet partners with
$\varphi [f_0^{\rm bare}(700)] = -70^\circ\pm 10^\circ$
and
$\varphi [f_0^{\rm bare}(1220)] = 15^\circ\pm 10^\circ$.\\
In this Solution there are two variants  for the $2^3P_0$ nonet:\\
1) either $f_0^{\rm bare}(1230\pm 30)$  and $f_0^{\rm
bare}(1830\pm 40)$ are $2^3P_0$ nonet partners
with
$\varphi [f_0^{\rm bare}(1230)] = 43^\circ\pm 8^\circ$
and $\varphi
[f_0^{\rm bare}(1830)] =-60^\circ\pm 10^\circ$,
\\ $f_0^{\rm bare}(1560\pm 30)$ is the glueball, with
$\varphi [f_0^{\rm bare}(1560)] = 15^\circ\pm 10^\circ$, or
\\
2) $f_0^{\rm bare}(1560\pm 30)$  and $f_0^{\rm
bare}(1830\pm 40)$ are nonet partners
and $f_0^{\rm bare}(1230)$ is the glueball.

Tables \ref{table2}, \ref{table3} and \ref{table4} present parameters
which correspond to these three Solutions.

\begin{table}
\caption\protect{Masses, coupling constants (in GeV) and mixing angles
(in degrees)
for the $f_0^{\rm bare}$-resonances for Solution {\bf I}.
The errors reflect the boundaries for a satisfactory
description of the data.
The sheet II  is under the $\pi\pi$ and $4\pi$
cuts; the sheet IV is under the $\pi\pi$, $4\pi$, $K\bar K$ and $\eta
\eta$ cuts;
the sheet V is under the $\pi\pi$, $4\pi$, $K\bar K$, $\eta
\eta$ and $\eta \eta'$ cuts.}\\ ~\\ \label{table2}
\begin{tabular}{|l|ccccc|}
\hline
~& ~ &\multicolumn{3}{c}{Solution {\bf I}} &~  \\
\hline
~ & $\alpha=1$ &$\alpha=2$ & $\alpha=3$ & $\alpha=4$ & $\alpha=5$ \\
\hline
~         &~ & ~ & ~ & ~ & ~ \\
M              & $0.650^{+.120}_{-.050}$ &$1.245^{+.040}_{-.030}$ &
$1.220^{+.030}_{-.030}$ & $1.630^{+.030}_{-.020}$ &
$1.750^{+.040}_{-.040}$ \\
~         &~ & ~ & ~ & ~ & ~ \\
$g^{(\alpha)}$ &$0.940^{+.80}_{-.100}$ &$1.050^{+.080}_{-.080}$
&$0.680^{+.060}_{-.060}$
&$0.680^{+.060}_{-.060}$ &$0.790^{+.080}_{-.080}$\\
~         &~ & ~ & ~ & ~ & ~ \\
$g_{5}^{(\alpha)}$& 0  & 0 &$0.960^{+.100}_{-.150}$&
$0.900^{+.070}_{-.150}$ & $0.280^{+.100}_{-.100}$  \\
~         &~ & ~ & ~ & ~ & ~ \\
$\varphi_\alpha (deg) $  & -(72$^{+5}_{-10}$) & 18.0$^{+8}_{-8}$
&33$^{+8}_{-8}$  & 27$^{+10}_{-10}$ & -59$^{+10}_{-10}$\\
~         &~ & ~ & ~ & ~ & ~ \\
\hline
~ & $a=\pi\pi$ &$a=K\bar K$ & $a=\eta\eta$ & $a=\eta\eta'$ & $a=4\pi$ \\
\hline
~         &~ & ~ & ~ & ~ & ~ \\
$f_{1a} $ &$-0.050^{+.100}_{-.100}$ &$ 0.250^{+.100}_{-.100}$ &
$0.440^{+.100}_{-.100}$ &$0.320^{+.100}_{-.100}$ &
$-0.540^{+.100}_{-.100}$ \\
~ &  ~ & $f_{ba}=0$ &$b=2,3,4,5$ & ~ & ~ \\
~         &~ & ~ & ~ & ~ & ~ \\
\hline
~ & ~&\multicolumn{3}{c}{Pole position}& ~  \\
sheet II     & $1.031^{+.008}_{-.008}$& ~ & ~ & ~ & ~ \\
 ~ & $-i(0.032^{+.008}_{-.008})$ & ~ & ~ & ~ & ~\\
\hline
sheet IV       & ~ &$1.306^{+.020}_{-.020}$&$1.489^{+.008}_{-.004}$
              &$1.480^{+.100}_{-.150}$ & ~ \\
~     & ~ &$-i(0.147^{+.015}_{-.025})$&$-i(0.051^{+.005}_{-.005})$
              &$-i(1.030^{+.080}_{-.170})$ & ~ \\
\hline
sheet V    & ~ & ~ & ~ & ~ & $1.732^{+.015}_{-.015}$ \\
~       & ~ & ~ & ~ & ~  & $-i(0.072^{+.015}_{-.015})$   \\
\hline
\end{tabular}
\clearpage
\end{table}

\begin{table}
\caption\protect{Masses, coupling constants (in GeV) and mixing angles
(in degrees)
for the $f_0^{\rm bare}$-resonances for Solution {\bf II-1}.
The errors reflect the boundaries for a satisfactory
description of the data. The sheet
II  is under the $\pi\pi$ and $4\pi$
cuts;
the sheet IV is under the $\pi\pi$, $4\pi$, $K\bar K$ and $\eta \eta$
cuts; the sheet V is under the $\pi\pi$, $4\pi$, $K\bar K$, $\eta \eta$
and $\eta \eta'$ cuts.}\\ ~\\ \label{table3}
\begin{tabular}{|l|ccccc|}
\hline
~ & ~ &\multicolumn{3}{c}{Solution {\bf II-1}}& ~  \\
\hline
~ & $\alpha=1$ &$\alpha=2$ & $\alpha=3$ & $\alpha=4$ & $\alpha=5$ \\
\hline
~         &~ & ~ & ~ & ~ & ~ \\
M              & $0.670^{+.100}_{-.100}$ &$1.215^{+.40}_{-.040}$ &
$1.220^{+.015}_{-.030}$ & $1.560^{+.030}_{-.040}$ &
$1.830^{+.030}_{-.050}$ \\
~         &~ & ~ & ~ & ~ & ~ \\
$g^{(\alpha)}$ &$0.990^{+.080}_{-.120}$
&$1.100^{+.080}_{-.100}$  &$0.670^{+.100}_{-.120}$
&$0.500^{+.060}_{-.060}$ &$0.410^{+.060}_{-.060}$\\
~         &~ & ~ & ~ & ~ & ~ \\
$g_{5}^{(\alpha)}$& 0 & 0 &$0.870^{+.100}_{-.100}$ &
$0.600^{+.100}_{-.100}$ & $-0.850^{+.080}_{-.080}$  \\
~         &~ & ~ & ~ & ~ & ~ \\
$\varphi_\alpha (deg) $  & -(66$^{+8}_{-10}$)
& 13$^{+8}_{-5}$ & 40$^{+12}_{-12}$
& 15$^{+08}_{-15}$ &-80$^{+10}_{-10}$\\
~         &~ & ~ & ~ & ~ & ~ \\
\hline
~ & $a=\pi\pi$ &$a=K\bar K$ & $a=\eta\eta$ & $a=\eta\eta'$ & $a=4\pi$ \\
\hline
~         &~ & ~ & ~ & ~ & ~ \\
$f_{1a} $ &$0.050^{+.100}_{-.100}$ & $0.100^{+.080}_{-.080}$ &
$0.360^{+.100}_{-.100}$ &$0.320^{+.100}_{-.100}$ &
$-0.350^{+.060}_{-.060}$ \\
~ &  ~ & $f_{ba}=0$ & $b=2,3,4,5$ & ~& ~\\
~         &~ & ~ & ~ & ~ & ~ \\
\hline
~ & ~ & \multicolumn{3}{c}{Pole position}& ~  \\
 sheet II     &$1.020^{+.008}_{-.008}$ & ~ & ~ & ~ & ~\\
  ~           &$-i(0.035^{+.008}_{-.008})$ & ~ & ~ & ~ & ~\\
\hline
sheet IV  & ~ &$1.320^{+.020}_{-.020}$ & $1.485^{+.005}_{-.006}$
              &$1.530^{+.150}_{-.100}$& ~ \\
 ~        & ~ &$-i(0.130^{+.015}_{-.025})$ & $-i(0.055^{+.008}_{-.008})$
              &$-i(0.900^{+.100}_{-.200})$&~ \\
\hline
sheet V   & ~ & ~ & ~ & ~ &$1.785^{+.015}_{-.015}$ \\
 ~        & ~ & ~ & ~ & ~  &$-i(0.135^{+.025}_{-.010})$  \\
\hline
\end{tabular}
\clearpage
\end{table}

\begin{table}
\caption\protect{Masses, coupling constants (in GeV) and mixing angles
(in degrees)
for the $f_0^{\rm bare}$-resonances for Solution {\bf II-2}.
The errors reflect the boundaries for a satisfactory
description of the data. The sheet II is under the $\pi\pi$ and $4\pi$
cuts; the
 sheet IV is under the $\pi\pi$, $4\pi$, $K\bar K$ and $\eta \eta$
cuts; the sheet V is under the $\pi\pi$, $4\pi$, $K\bar K$, $\eta \eta$
and $\eta \eta'$ cuts.}\\ ~\\ \label{table4}
\begin{tabular}{|l|ccccc|}
\hline
~ & ~ & \multicolumn{3}{c}{Solution {\bf II-2}} & ~ \\
\hline
~ & $\alpha=1$ &$\alpha=2$ & $\alpha=3$ & $\alpha=4$ & $\alpha=5$ \\
\hline
~         &~ & ~ & ~ & ~ & ~ \\
M              & $0.650^{+.120}_{-.050}$ &$1.220^{+.040}_{-.030}$ &
$1.230^{+.030}_{-.030}$ & $1.560^{+.030}_{-.020}$ &
$1.830^{+.040}_{-.040}$ \\
~         &~ & ~ & ~ & ~ & ~ \\
$g^{(\alpha)}$ &$1.050^{+.80}_{-.100}$ &$0.980^{+.080}_{-.080}$
&$0.470^{+.050}_{-.050}$
&$0.420^{+.040}_{-.040}$ &$0.420^{+.050}_{-.050}$\\
~         &~ & ~ & ~ & ~ & ~ \\
$g_{5}^{(\alpha)}$& 0 & 0 & $0.870^{+.100}_{-.100}$ &
$0.560^{+.070}_{-.070}$ & $-0.780^{+.070}_{-.070}$  \\
~         &~ & ~ & ~ & ~ & ~ \\
$\varphi_\alpha (deg)$  & -(68$^{+3}_{-15}$)
& 14$^{+8}_{-8}$ & 43$^{+8}_{-8}$
& 15$^{+10}_{-10}$ &-55$^{+10}_{-10}$\\
~         &~ & ~ & ~ & ~ & ~ \\
\hline
~ & $a=\pi\pi$ &$a=K\bar K$ & $a=\eta\eta$ & $a=\eta\eta'$ & $a=4\pi$ \\
\hline
~         &~ & ~ & ~ & ~ & ~ \\
$f_{1a} $ &$0.260^{+.100}_{-.100}$ &$ 0.100^{+.100}_{-.100}$ &
$0.260^{+.100}_{-.100}$ &$0.260^{+.100}_{-.100}$ &
$-0.140^{+.060}_{-.060}$ \\
~ &  ~ & $f_{ba}=0$ & $b=2,3,4,5$&~ & ~\\
~         &~ & ~ & ~ & ~ & ~ \\
\hline
~ & ~ &\multicolumn{3}{c}{Pole position}& ~  \\
 sheet II     &$1.020^{+.008}_{-.008}$& ~&~&~&~ \\
  ~           &$-i(0.035^{+.008}_{-.008})$& ~&~&~&~\\
\hline
sheet IV   &~ &$1.325^{+.020}_{-.030}$ & $1.490^{+.010}_{-.010}$
              &$1.450^{+.150}_{-.100}$ & ~ \\
  ~        & ~&$-i(0.170^{+.020}_{-.040})$ & $-i(0.060^{+.005}_{-.005})$
              &$-i(0.800^{+.100}_{-.150})$ & ~ \\
\hline
sheet V    & ~ & ~ & ~ & ~     & $1.740^{+.020}_{-.020}$ \\
  ~       & ~ & ~ & ~ & ~     & $-i(0.160^{+.025}_{-.010})$   \\
\hline
\end{tabular}
\end{table}

\subsection{GAMS and E852 data on the $\pi\pi$ $S$-wave
production  $\pi^- p\to (\pi\pi)_S \; n$ at  moderate momenta
transferred to nucleon, $0<|t|<1.5$ (GeV/$c$)$^2$}

The formulae of Section 2.2 allow one to describe simultaneously
the $\pi\pi$ spectra in the
reaction $\pi^- p\to (\pi\pi)_S \, n$ at $p_{lab}=18$ GeV/c \cite{E852}
and $p_{lab}=38$ GeV/c \cite{gams1}.

The leading $\pi$ and $a_1$ trajectories have rather close
intercepts and slopes \cite{syst}, so corresponding reggeon exchanges
do not cause  a change of $t$-distributions with
 energy. However, the experimental data definitely point to a
break of similarity of the spectra: in Fig. 12 the difference of the
$M_{\pi\pi}$-distributions is shown for $|t|\sim (0.3-0.4)$ (GeV/c)$^2$
where one can trace differences in spectra related to E852 and GAMS
experiments. This fact definetely points to a significant contribution
of daughter trajectories.

We have  fitted to data \cite{gams1} and \cite{E852}
under a variety of assumptions on the $|t|$-exchange structure;
these versions of fitting are discussed in a separate publication
\cite{bnl-gams}. Here, in Figs. 13 and 14,
we demonstrate  typical description of the data.
Figure 13 represents the $M_{\pi\pi}$ spectra of
E852 Collaboration for different $|t|$-intervals, and figure 14 shows
the same for GAMS group (calculated spectra refer to Solution
II-2).

In Fig. 15, we show  form factors for $\pi$ and $a_1$ exchanges,
which at large $|t|$ describe effectively the multi-reggeon
interactions. We show the form factors which correspond
to the so-called Orear behaviour of the scattering amplitude
\cite{Orear} at large $|t|$
(see also \cite{X} and references therein):
\be
G^{(\alpha)}_{\pi\pi}(t)
=g_{\pi\pi}
\left [ \exp \left ( \beta_1^{(\alpha)} (t-m^2_\pi) \right ) +
\Lambda (t-m^2_\pi) \exp \left (- \beta_2^{(\alpha)}
\sqrt{|t-m^2_\pi|} \right ) \right ]\ .
\label{Orear}
\ee
Another types of the behaviour of
$G^{(\alpha)}_{\pi\pi}(t)$ were analysed in
 \cite{bnl-gams}.

\section{$f_0$-resonances: masses, decay couplings and\\ partial
widths}

The resonance masses and decay couplings are not
determined directly in the fitting
procedure. To calculate these quantities one needs to perform
 analytical continuation of the $K$-matrix amplitude into lower
half-plane of the complex plane $s$. One is allowed to do it, for the
$K$-matrix amplitude correctly takes into account the threshold
singularities related to the
$\pi\pi$, $\pi\pi\pi\pi$, $K\bar K$, $\eta\eta$, $\eta\eta'$ channels
which are important in the $00^{++}$-wave.

\subsection{Masses of resonances}

The complex masses
of the resonances
$f_0(980)$, $f_0(1300)$, $f_0(1500)$, $f_0(1200-1600)$
obtained in Solutions I, II-1 and II-2 do not differ
strongly. Solutions I and II are essentially different in the
characteristics of the $f_0(1750)$

For the pole positions
the following values, in MeV, have been found:
\bea
{\rm Solution\; I:}
&f_0(980)&\to   1031-i\; 32
\\ \nonumber
&f_0(1300)&\to  1306-i\; 147
\\ \nonumber
&f_0(1500)&\to  1489-i\; 51
\\ \nonumber
&f_0(1750)&\to  1732-i\; 72
\\ \nonumber
&f_0(1200-1600)&\to  1480-i\; 1030
\eea

\bea
{\rm Solution\; II-1:}
&f_0(980)&\to   1020-i\; 34
\\ \nonumber
&f_0(1300)&\to  1320-i\; 130
\\ \nonumber
&f_0(1500)&\to  1485-i\; 55
\\ \nonumber
&f_0(1750)&\to  1785-i\; 135
\\ \nonumber
&f_0(1200-1600)&\to   1530-i\; 930
\eea

\bea
{\rm Solution\; II-2:}
&f_0(980)&\to   1020-i\; 35
\\ \nonumber
&f_0(1300)&\to  1325-i\; 170
\\ \nonumber
&f_0(1500)&\to  1490-i\; 60
\\ \nonumber
&f_0(1750)&\to  1740-i\; 160
\\ \nonumber
&f_0(1200-1600)&\to   1450-i\; 800
\eea
We see that Solutions I and II
 give different magnitudes for total width of the
$f_0(1750)$.

\subsection{Decay couplings and partial widths}

We determine the coupling constants and partial decay widths using
the following procedure.
The $00^{++}$-amplitude for the transition $a\to b$,
\be
A_{a\to b}(s)\; , \qquad a,b=
\pi\pi,K\bar K, \eta\eta, \eta\eta', \pi\pi\pi\pi ,
\label{4.1}
\ee
is considered as a function of the invariant energy squared $s$
in the complex-$s$ plane near the pole related to the
resonance $n$. In the vicinity of the pole the amplitude reads:
\be
A_{a\to b}(s)= \frac{g^{(n)}_a g^{(n)}_b}{\mu^2_n -s}
\; e^{i(\theta ^{(n)}_{a}+\theta ^{(n)}_{b})} +
{\rm non-pole \; terms}\ .
\label{4.2}
\ee
Here $\mu_n$ is the resonance complex mass $\mu_n =M_n -i\Gamma_n/2\;$;
$g^{(n)}_a$ and $ g^{(n)}_b$ are the couplings for
the transitions $f_0 \to a$ and $f_0 \to b$.
In (\ref{4.2}), the non-pole background terms  are omitted.

The decay
coupling constants squared, $g^{(n)\; 2}_a$, are shown in Table 5
for $a=\pi\pi$, $K\bar K$, $\eta\eta$, $\eta\eta'$, $\pi\pi\pi\pi$.
The couplings are determined with
the normalization of the amplitude used in Section 2:
for example, we write
the $\pi\pi$ scattering amplitude
(\ref{4.2}) as $$A_{\pi\pi\to \pi\pi}(s)= (
\eta^0_0 \exp{(2i\delta^0_0)}-1)/2i\rho_{\pi\pi}(s)\ ,$$ where
$\eta^0_0$
and $\delta^0_0$ are
the inelasticity parameter and phase shift for the
$00^{++}$ $\pi\pi$-wave, reprectively.
The coupling constants $g^{(n)\; }_a$ are
found by calculating
the residues of the amplitudes $\pi\pi\to \pi\pi$,
$K\bar K$, $\eta\eta$, $\eta\eta'$, $\pi\pi\pi\pi$.
Also we check
the factorization property for the pole terms by
calculating residues for other reactions,
such as $K\bar K \to K\bar K $, and so on.

The resonance $f_0(980)$ is located
near the strong $K\bar K$ threshold, therefore two poles
are related to $f_0(980)$:
for example, in  Solution II-2 the nearest
one is on the 3rd sheet
(under $\pi\pi$ and $\pi\pi\pi\pi$ cuts, at $M\simeq 1020-i35$ MeV) and
a remote pole on the 4th sheet (under $\pi\pi$, $\pi\pi\pi\pi$ and
$K\bar K$ cuts, at $M\simeq 940-i240$ MeV). Coupling constants for
$f_0(980)$ are determined as residues of the nearest pole which is
located on the 3rd sheet.

\begin{table}
\caption{Coupling constants squared (in GeV$^2$) of
scalar-isoscalar resonances to hadronic channels $\pi\pi$, $K\bar
K$, $\eta\eta$, $\eta\eta'$ and $\pi\pi\pi\pi$ for different $K$-matrix
solutions.}
\label{table5}
\begin{tabular}{l|ccccc|c}
\hline
Pole position&$\pi\pi$&$K\bar K$&$\eta\eta$&$\eta\eta'$&$\pi\pi\pi\pi$
&Solution\\
\hline
$f_0(980)$ &      &       &        &     &           &    \\
$ 1031-i32 $  &0.056 & 0.130 &  0.067 &  -- &   0.004   &  I \\
$ 1020-i35 $  &0.051 & 0.115 &  0.051 &  -- &   0.003   &II-1\\
$ 1020-i35 $  &0.054 & 0.117 &  0.139 &  -- &   0.004   &II-2\\
\hline
$f_0(1300)$&      &        &        &       &       &    \\
$ 1306-i147$   &0.036 &  0.009 &  0.006 & 0.004 & 0.093 &I   \\
$ 1320-i130$   &0.038 &  0.003 &  0.005 & 0.012 & 0.109 &II-1\\
$ 1325-i170$   &0.053 &  0.003 &  0.007 & 0.013 & 0.226 &II-2\\
\hline
$f_0(1500)$&      &        &        &          &        &\\
$ 1489-i51 $   & 0.014 &  0.006 &   0.003 &  0.001 &  0.038 &I\\
$ 1485-i55 $   & 0.020 &  0.007 &   0.004 &  0.003 &  0.049 &II-1\\
$ 1490-i60 $   & 0.018 &  0.007 &   0.003 &  0.003 &  0.076 &II-2\\
\hline
$f_0(1750)$&      &        &        &          &        &\\
$ 1732-i72 $   & 0.013 &  0.062 &   0.002 &  0.032 & 0.002 &I\\
$ 1785-i135$   & 0.066 &  0.003 &   0.007 &  0.033 & 0.080 &II-1\\
$ 1740-i160$   & 0.089 &  0.002 &   0.009 &  0.035 & 0.168 &II-2\\
\hline
$f_0(1200-1600)$&      &        &        &          &        &\\
$ \;1480-i1000$   & 0.364 &  0.265 &   0.150  &0.052&  0.524 &I\\
$ 1530-i900 $  & 0.325 &  0.235 &   0.086  &  0.015 &  0.474 &II-1\\
$ 1450-i800 $  & 0.179 &  0.204 &   0.046  &  0.005 &  0.686 &II-2\\
\hline
\end{tabular}
\end{table}

The partial width for the decay  $f_0 \to a$ is determined as a
product of the coupling  constant
squared, $g^{(n)\; 2}_a$, and  phase space,
$\rho_a (s)$, averaged over the resonance density:
\be
\Gamma_a (n)= C_n \int \limits_{s>s_{th}} \frac{ds}{\pi} \,
\frac{g^{(n)\; 2}_a \;\rho_a (s)}{(Re \;\mu^2_n -s)^2+
({\rm Im} \,\mu^2_n)^2}
\ .
\label{4.3}
\ee
The resonance density factor,
$\left [(Re \;\mu^2_n -s)^2+({\rm Im} \,\mu^2_n)^2\right ]^{-1}$,
guarantees rapid convergence of the integral (\ref{4.3}).
The normalization constant $C_n$ is determined by the requirement that
the sum
of all hadronic partial widths is equal to the total width of the
resonance:
\be
\Gamma (n) =\sum \limits_{a} \Gamma_a (n)\ .
\label{4.5}
\ee
In Table 6 we show the values of partial widths
for the resonances
$f_0(980)$, $f_0(1300)$, $f_0(1500)$, $f_0(1750)$.
Partial widths for
$f_0(1300)$, $f_0(1500)$, $f_0(1750)$
are calculated within standard formulae for the
Breit-Wigner resonances (\ref{4.3}).

\begin{table}
\caption{Partial widths of scalar-isoscalar resonances (in MeV)
in hadronic channels $\pi\pi$, $K\bar K$, $\eta\eta$, $\eta\eta'$ and
$\pi\pi\pi\pi$ for different $K$-matrix solutions.}
\label{table6}
\begin{tabular}{l|ccccc|l|c}
\hline
 &$\pi\pi$&$K\bar K$&$\eta\eta$&$\eta\eta'$&$\pi\pi\pi\pi$&
$\Gamma_{tot}/2$&Solution\\
\hline
$f_0(980)$&52&10&--&--&2& 32  &I\\
          &57& 9&--&--&2& 34  &II-1\\
          &58&10&--&--&2& 35  &II-2\\
\hline
$f_0(1300)$&77&14&7&--&196& 147 &I\\
           &66&4 &4&--&186& 130 &II-1\\
           &63&3 &5&--&269& 170 &II-2\\
\hline
$f_0(1500)$&23&6&3&0.0&70& 51 &I\\
           &25&7&3&0.1&75& 55 &II-1\\
           &22&6&2&0.1&88& 59 &II-2\\
\hline
$f_0(1750)$&31 &96&5&7&5  &72 &I\\
           &114&3 &8&7&138&135&II-1\\
           &103&1 &8&4&204&160&II-2\\
\hline
\end{tabular}
\end{table}

Because of a strong $K\bar
K$ threshold near the pole, the resonance term
for $f_0(980)$
should be used not as the Breit-Wigner amplitude but in a
more complicated form.
Instead of the Breit-Wigner pole term
$R^{(ab)}_n =g^{(n)}_a g^{(n)}_b/(\mu^2_n -s)$ entering
equation (\ref{4.2}),
the following resonance amplitudes can be
written for the
$\pi\pi\to \pi\pi$
and $K\bar K\to K\bar K$
transitions near $f_0(980)$ \cite{width}:
\be
R^{(\pi\pi, \pi\pi)}_{f_0(980)}
=\left [G^2+i\frac{\sqrt{s-4m^2_K}}{m_0}
(2GG_{K\bar K}f+f^2(m^2_0-s))
\right ]\,\frac 1D\ ,
\label{star}
\ee
$$
R^{(K\bar K, K\bar K)}_{f_0(980)}
=\left [G^2_{K\bar K}+i(
2GG_{K\bar K}f+f^2(m^2_0-s))
\right ]\, \frac 1D\ ,
$$
where
\be
D=m^2_0-s-iG^2-i
\frac{\sqrt{s-4m^2_K}}{m_0}
\left [G^2_{K\bar K}+i(2GG_{K\bar K}f+f^2(m^2_0-s))
\right ]\ .
\label{star2}
\ee
Here  $m_0$ is the input mass of $f_0(980)$; $G$ and $G_{K\bar K}$
are coupling constants to pion and kaon channels.
The dimensionless constant
$f$ stands for the prompt transition $ K\bar K\to\pi\pi$:
the value $f/m_0$ is the "transition length" which is analogous
to the scattering length of the low-energy hadronic interaction.
The constants
$m_0$, $G$, $G_{K\bar K}$, $f$
are parameters which are to be chosen to reproduce
the $f_0(980)$ charactristics. Equations (\ref{star}) and (\ref{star2})
are written for $s>4m^2_K$, at $s<4m^2_K$ one needs to
replace $\sqrt{s-4m^2_K}\to i\sqrt{4m^2_K-s}$.

In \cite{width}, two sets of parameters
were obtained, with sufficiently correct
values for the $f_0(980)$ pole position and couplings. They are
equal (in GeV) to:
\bea
&&Solution \,A: \quad m_0=1.000\ ,\,f=0.516\ ,\,G=0.386\ ,\,
G_{K\bar K}=0.447\ ,\nonumber \\
&&Solution \,B: \quad m_0=0.952\ ,\,f=-0.478\ ,\,G=0.257\ ,\,
G_{K\bar K}=0.388\ .
\label{AB}
\eea

Partial widths of $f_0(980)$ are calculated with
the expression similar to (\ref{4.3}), with the replacement of the
integrand denominator  as follows:
\be
({\rm Re}\,\mu_n^2-s)^2+ ({\rm Im}\,\mu_n^2)^2 \to |D|^2.
\ee
For both sets of parameters
(\ref{AB}) the calculated partial widths are close to
each other. The values of partial widths for $f_0(980)$ averaged
over Solutions $A$ and $B$ are presented  in Table 6.

The total hadron width of $f_0(980)$ is defined in the
same way as for
the other $f_0$-mesons, namely, by using the position of pole in the
complex-$M$ plane: the imaginary part of the mass is equal to a
half-width of the resonance. For the Breit--Wigner resonance this
definition is in accordance with what is observed from resonance
spectrum (provided there is no interference with the background). If
the resonance is located in the vicinity of a strong threshold, the
visible resonance width differs significantly from that given by
 pole position.

The results of our calculations of partial decay widths are
presented below, the magnitudes are in MeV. In the
brackets we demonstrate the values obtained in \cite{width} on the
basis of previous analysis \cite{YF}:

\noindent
\hspace{-0.25cm}
\begin{tabular}{lccccccc}
Resonance &$\Gamma_{\pi\pi}$&$\Gamma_{K\bar K}$&
$\Gamma_{\eta\eta}$&$\Gamma_{\eta\eta'}$&$\Gamma_{\pi\pi\pi\pi}$&
$\Gamma_{tot}/2$&Solution \\

$f_0(980):$ &$55\pm 5$ & $10\pm 1$ & -- & -- & $2\pm 1$ & $34\pm 5$
&I,II\\
          &$(64\pm 8)$ &$(12\pm 1)$& -- &--& $(4\pm 2)$&$(40\pm 5)$
&\cite{YF,width}\\
$f_0(1300):$&$66\pm 10$ & $6\pm 4$ & $5\pm 2$  & -- & $230\pm 50$ &
$150\pm 20$&I,II\\
      &$(46\pm 12)$ & $(5\pm 3)$ & $(4\pm 2)$  & -- & $(171\pm 10)$ &
$(113\pm 10)$&\cite{YF,width}\\
$f_0(1500):$&$23\pm 5$ & $6\pm 2$ & $3\pm 1$ & $0.1\pm 0.1$ &
$80\pm 10$ & $55\pm 5$&I,II\\
          &$(37\pm 2)$ & $(7\pm 2)$ & $(4\pm 1)$ & $(0.2\pm 0.1)$ &
$(79\pm 6)$ & $(62\pm 3)$&\cite{YF,width}\\
$f_0(1750):$&$30\pm 5$ &  $100\pm 10$ & $5\pm 3$ & $7\pm 3$ &
$5\pm 5$ &$70\pm 20$&I\\
   &$105\pm 10$ &  $2\pm 1$ & $8\pm 1$ & $5\pm 3$ &
$170\pm 40$ &$150\pm 25$&II\\
       &$(74^{+15}_{-30})$ &$(11^{+17}_{-9})$ & $(7\pm 1)$&$(3\pm 1)$ &
$(91^{+30}_{-60})$ &$(93^{+20}_{-40})$&\cite{YF,width}\\
\end{tabular}
\be
\;
\label{table}
\ee
In Solutions I and II,
 partial widths for $f_0(980)$, $f_0(1300)$, $f_0(1500)$
are agreeably coincide but for $f_0(1750)$ the decay
characteristics are different. So we present separately
the $f_0(1750)$ partial widths for both Solutions, I and II.

\section{ Classification of scalar resonances}

The $K$-matrix approach works ingenuously with the bare states
thus directly providing
the $q\bar q$-classification
for these states.
Concerning the resonances, the classifications
of the $a_0$ and $K_0$ mesons in terms of bare states
and real resonances are similar, while $f_0$ mesons give us an
opposite case: the matter is that there is an exotic state, gluonium,
in the region under consideration, whose mixing with $q\bar q$ state is
not forbidden by the $1/N$ expansion rules (for more detail see
\cite{ufn,aas-z}).

In the present Section, on the basis of the values for
couplings to channels
$\pi\pi$, $K\bar K$, $\eta\eta$, $\eta\eta'$
calculated in Section 4, we analyse the quark-gluonium content of
resonances $f_0(980)$, $f_0(1300)$, $f_0(1500)$, $f_0(1750)$ and the
broad state $f_0(1200-1600)$.

\subsection{ Overlapping of the  $f_0$-resonances in the mass region
1200--1700 MeV: accumulation of widths of the $q\bar q$ states by the
glueball }

The existence of the broad resonance is not an eventual
phenomenon: it originated due to the mixing of states in the
 decay processes, namely, transitions
$f_0(m_1)\to real\; mesons\;
\to f_0(m_2)$. These transitions
result in  a specific phenomenon, that is, when several resonances
overlap, one of them accumulates the widths of neighbouring resonances
and transforms into the broad state.

This phenomenon had been observed in
\cite{km,km1900} for scalar-isoscalar states,
and  the following scheme has been suggested in
\cite{aas-z,glueball}: the broad state $f_0(1200-1600)$
is the descendant of the pure glueball which being in the neighbourhood
of $q\bar q$ states accumulated their widths and transformed into the
mixture of gluonium and $q\bar q$ states. In \cite{aas-z},
this idea had been modelled for four resonances
$f_0(1300)$, $f_0(1500)$, $f_0(1200-1600)$ and $f_0(1750)$,
by using the language of the quark-antiquark and two-gluon states,
$q\bar q$ and $gg$: the decay processes were considered as the
transitions $f_0 \to q\bar q, gg$, correspondingly; the
same processes are responsible for the mixing of resonances.
In this model, the gluonium component is mainly   dispersed
over three resonances,  $f_0(1300)$, $f_0(1500)$,
$f_0(1200-1600)$, so every  state is a mixture
of $q\bar q$ and $gg$ components, with roughly equal percentage
of gluonium (about 30-40\%).

 Accumulation of widths of overlapping resonances by one of them is a
well-known effect
in nuclear physics \cite{Shapiro,Okun,Stodolsky}.
In meson physics this phenomenon can play rather important role, in
particular for exotic states which are beyond the $q\bar q$
systematics. Indeed, being among  $q\bar q$ resonances, the exotic
state creates a group of overlapping resonances.
The exotic state, which is not orthogonal
to its neighbours, after accumulating the "excess" of widths
turns into the broad one. This broad resonance should be accompanied
by narrow states which are the descendants of states from which the
widths have been taken off. In this way, the existence of a broad
resonance accompanied by narrow ones may be a signature of
exotics.  This possibility, in context of searching for exotic states,
was discussed in \cite{YF99,PR-exot}.

The broad state may be one of the components which form the
confinement barrier:  the broad states, after accumulating the widths of
neighbouring resonances, play for these latter the role of locking
states. Evaluation of the mean radii squared of the broad state
$f_0(1200-1600)$ and its neighbouring resonances argues in
favour of this idea, for the radius of $f_0(1200-1600)$
is significantly larger than that of $f_0(980)$ and $f_0(1300)$
\cite{YF99,rad_pl}, thus making $f_0(1200-1600)$
 to be the locking state. The data on the $t$-dependence
in the reaction
 $\pi^- p\to (\pi\pi)_S\, n$ at $p_{beam}=
18$ (GeV/c)$^2$ \cite{E852} also support this idea, for the signal from
the broad state (background in the spectra) falls with the increase of
$|t|$ much quicker than that of the $f_0(980)$ and $f_0(1300)$.

The $K$-matrix solutions obtained here (Section 3.1)
give us two variants                for the
transformation of bare states into real resonances coming after
the onset of the decay channels. They are as follows:

\be
\begin{array}{cl}
f_0^{bare}(700)\pm 100) &\to \; f_0(980)\ ,\\
f_0^{bare}(1220\pm 40)  &\to \;f_0(1300)\ , \\
f_0^{bare}(1230\pm 40) &\to \;{\rm broad \,state}\,f_0(1200-1600)\ ,
\\ f_0^{bare}(1580\pm 40) &\to \;f_0(1500)\ , \\
f_0^{bare}(1800\pm 40) &\to \;f_0(1750)\ .
\end{array}
\ee
and
\be
\begin{array}{cl}
f_0^{bare}(700)\pm 100) &\to \; f_0(980)\ ,\\
f_0^{bare}(1220\pm 40) &\to \;f_0(1300)\ , \\
f_0^{bare}(1230\pm 40) &\to \;f_0(1500)\ , \\
f_0^{bare}(1580\pm 40) &\to \;{\rm broad \,state}\,f_0(1200-1600)\ ,\\ 
f_0^{bare}(1800\pm 40) &\to \;f_0(1750)\ .  
\end{array} 
\ee 
The evolution of bare states into real resonances is illustrated by Fig.
16: the shifts of  amplitude poles in the complex-$M$ plane correspond
to a  gradual onset of the decay channels. Technically it is done by
replacing  the phase spaces $\rho_a$ for $a=\pi\pi$, $\pi\pi\pi\pi$,
$K\bar K$, $\eta\eta$, $\eta\eta'$ in the $K$-matrix amplitude as
follows: $\rho_a\to \xi \rho_a $, where the parameter $\xi$ runs in the
interval $0\le \xi \le 1$. At $\xi \to 0$ one has bare states, while the
limit $\xi \to 1$ gives us the positions of real resonances.

\subsection{Hadronic decays
and evaluation of the quark-gluonium content of scalar-isoscalar
resonances}

Here, on the basis of the quark combinatorics for the decay
coupling constants, we analyse
 quark-gluonium content of resonances $f_0(980),
f_0(1300), f_0(1500), f_0(1750)$ and the
broad state $f_0(1200-1600)$.

The comparison of
Tables 2, 3, 4 with Table 5 demonstrates a strong change of couplings
during the  evolution from bare states to real resonances;
the same type of strong deviation of couplings has been observed
before, for previous $K$-matrix solutions, see \cite{content}.
Note that the
change occurs not only in  absolute values of couplings but
in relative magnitudes as well
that means the change of the quark-antiquark content of states in the
evolution caused by the decay onset.

Let us look at the proportion of $s\bar s$, $n\bar n$
and gluonium components given by quark-combinatorics relations
(\ref{gpipi})
for the studied resonances. The coupling
constants squared for $f_0 \to \pi\pi$, $K\bar K$, $\eta\eta$,
$\eta\eta'$ in (\ref{gpipi}) are expressed as a sum of two terms
which correspond to transitions of quarkonium and gluonium components
into two pseudoscalar mesons.

First, let us find the mean value of mixing angle, $\langle\varphi\rangle$,
for the $n\bar n$/$s\bar s$ components in the intermediate state.
 Because of the two-stage mechanism of the gluonium decay
$gluonium\to quark-antiquark \;pair\to two\; mesons$
(see \cite{ufn} and Appendix C), we determine $\langle\varphi\rangle$
in the intermediate state as follows:
\be
f_0\to {\rm gluonium}+q\bar q \to n\bar n \cos \langle\varphi\rangle +s\bar s \sin
\langle\varphi\rangle \to {\rm two\; mesons}.
\ee
So, $\langle\varphi \rangle$ is the angle corresponding to the
coupling constants squared (\ref{gpipi}) at $G=0$.

Based on the values of couplings given in Table 5, we have found the
following values of $\langle \varphi \rangle$ and $\lambda$.\\
Solution I:
\be
f_0(980):&\quad \langle\varphi \rangle\simeq -68^\circ\ ,
&\lambda\simeq 0.5-1.0\ ,\\
\nonumber
f_0(1300):&\quad \langle\varphi \rangle\simeq (-3^\circ)-4^\circ\ ,
&\lambda\simeq 0.5-0.9\ ,\\
\nonumber
{\rm Broad\;\;  state\;\;}
f_0(1200-1600):&\quad \langle\varphi \rangle\simeq 27^\circ\ ,
&\lambda\simeq 0.54\ , \\
\nonumber
f_0(1500):&\quad \langle\varphi \rangle\simeq 12^\circ-19^\circ\ ,
&\lambda\simeq 0.5-1.0\ ,\\
\nonumber
f_0(1750):&\quad \langle\varphi \rangle\simeq -72^\circ\ ,
&\lambda\simeq 0.5-0.7\ ,\\
\nonumber
\label{SolutionI-mean-phi}
\ee
Solution II-1:
\be
f_0(980):&\quad \langle\varphi \rangle\simeq -67^\circ\ ,
&\lambda\simeq 0.5-1.0\ ,\\
\nonumber
f_0(1300):&\quad \langle\varphi \rangle\simeq (-21^\circ)-
(-10^\circ)\ ,
&\lambda\simeq 0.5-1.0\ ,\\
\nonumber
{\rm Broad\;\;  state\;\;}
f_0(1200-1600):&\quad \langle\varphi \rangle\simeq 28^\circ\ ,
&\lambda\simeq 0.55\ ,\\
\nonumber
f_0(1500):&\quad \langle\varphi \rangle\simeq 0^\circ-11^\circ\ ,
&\lambda\simeq 0.5-1.0\ ,\\
\nonumber
f_0(1750):&\quad \langle\varphi \rangle\simeq -37^\circ\ ,
&\lambda\simeq 0.45-0.55\ ,\\
\nonumber
\ee
Solution II-2:
\be
f_0(980):&\quad \langle\varphi \rangle\simeq -67^\circ\ ,
&\lambda\simeq 0.6-1.0\ ,\\
\nonumber
f_0(1300):&\quad \langle\varphi \rangle\simeq (-16^\circ)-
(-13^\circ)\ ,
&\lambda\simeq 0.5-0.6\ ,\\
\nonumber
{\rm Broad\;\;  state\;\;}
f_0(1200-1600):&\quad \langle\varphi \rangle\simeq 33^\circ\ ,
&\lambda\simeq 0.85\ ,\\
\nonumber
f_0(1500):&\quad \langle\varphi \rangle\simeq 2^\circ-11^\circ\ ,
&\lambda\simeq 0.6-1.0\ ,\\
\nonumber
f_0(1750):&\quad \langle\varphi \rangle\simeq -18^\circ\ ,
&\lambda\simeq 0.5\ ,\\
\nonumber
\ee
Note that the relations (\ref{gpipi}) provide one more solution for
$f_0(980)$, with $\langle\varphi \rangle\simeq 40^\circ$.
However this solution contradicts to $\varphi[f_0^{bare}(700)]\simeq
-67^\circ$ obtained within the $K$-matrix fit (Tables 3, 4, 5): for
details see \cite{content}, where the transformation of bare poles into
resonances after the onset of the decay channels,
was traced.

When the resonance $f_0$ is considered as a system $gluonium+q\bar
q$, where $q\bar q=u\bar u\cos \varphi+s\bar s\sin \varphi$, the
coupling constants given in Table 5 determine $\varphi$ as a function
of the ratio $G/g$. The results of fitting to coupling constants for
comparatively narrow resonances $f_0(980)$, $f_0(1300)$, $f_0(1500)$,
$f_0(1750)$ and the broad state $f_0(1200-1600)$ are presented in Fig.
17. In Fig. 17a,c,e, the curves present
 $\varphi$ as a function of $G/g$, with  different $\lambda$'s from the
intervals given by Eqs. (41)--(43)
for  the resonances
$f_0(980)$, $f_0(1300)$, $f_0(1500)$, $f_0(1750)$: a bunch of curves
provides us  the values $(\varphi,G/g)$ which describe well
the  couplings squared of Table 5. One can see that
the correlation curves for all resonances
$f_0(980)$, $f_0(1300)$, $f_0(1500)$, $f_0(1750)$ are, with a good
accuracy, the straight band.
As is stressed below, such a behaviour of correlation curves
is a signature of the $q\bar q$ origin of the resonances.

The magnitudes $g^2$ and $G^2$ are proportional  to  $W_{q\bar q}$ and
$W_{gluonium}$ which are the
probabilities for quark and gluonium components
to be present  in the considered resonance:
\be
g^2=g^2_{q\bar q} W_{q\bar q}\; , \qquad
G^2=G^2_{gluonium}W_{gluonium}\ .
\ee
According to the
rules of $1/N$ expansion \cite{t'h}, the coupling constants,
$g^2_{q\bar q}$  and    $G^2_{gluonium}$,  are of the
same order
(for more detail see \cite{ufn,lecture-95}), therefore  we accept as a
rough estimation:
\be G^2/g^2\simeq W_{gluonium}/W_{q\bar q}\ .
\label{G/g}
\ee
Varying
$G/g$ in the interval $-0.8 \le G/g \le0.8$ corresponds to a plausible
    admixture of the gluonium
component up to 40\%, $W_{gluonium}\la 0.40$.

Figures 17a,c,e provide the following intervals
$\varphi$ in the resonances
$f_0(980)$, $f_0(1300)$, $f_0(1500)$, $f_0(1750)$,
after the admixture of the  gluonium component.

Solution I:
\be
&W_{gluonium}[f_0(980)]\la 15\%\; :& \quad
-93^\circ \la \varphi [f_0(980)] \la -42^\circ , \\
\nonumber
&W_{gluonium}[f_0(1300)]\la 30\% \; :& \quad
-25^\circ \la \varphi [f_0(1300)] \la 25^\circ\ , \\
\nonumber
&W_{gluonium}[f_0(1500)]\la 30\% \; :& \quad
-2^\circ \la \varphi [f_0(1500)] \la 25^\circ\ , \\
\nonumber
&W_{gluonium}[f_0(1750)]\la 30\% \; :& \quad
-112^\circ \la \varphi [f_0(1750)] \la -32^\circ\ . \\
\nonumber
\label{SolutionI}
\ee

Solution II-1:
\be
&W_{gluonium}[f_0(980)]\la 15\%\; :& \quad
-92^\circ \la \varphi [f_0(980)] \la -42^\circ , \\
\nonumber
&W_{gluonium}[f_0(1300)]\la 30\% \; :& \quad
-54^\circ \la \varphi [f_0(1300)] \la 13^\circ\ , \\
\nonumber
&W_{gluonium}[f_0(1500)]\la 30\% \; :& \quad
-19^\circ \la \varphi [f_0(1500)] \la 21^\circ\ , \\
\nonumber
&W_{gluonium}[f_0(1750)]\la 30\% \; :& \quad
-73^\circ \la \varphi [f_0(1750)] \la -2^\circ\ . \\
\nonumber
\ee

Solution II-2:
\be
&W_{gluonium}[f_0(980)]\la 15\%\; :& \quad
-90^\circ \la \varphi [f_0(980)] \la -43^\circ , \\
\nonumber
&W_{gluonium}[f_0(1300)]\la 30\% \; :& \quad
-42^\circ \la \varphi [f_0(1300)] \la 10^\circ\ , \\
\nonumber
&W_{gluonium}[f_0(1500)]\la 30\% \; :& \quad
-18^\circ \la \varphi [f_0(1500)] \la 23^\circ\ , \\
\nonumber
&W_{gluonium}[f_0(1750)]\la 30\% \; :& \quad
-46^\circ \la \varphi [f_0(1750)] \la 7^\circ\ . \\
\nonumber
\ee

The broad state $f_0(1200-1600)$ demonstrates another type of the
$(\varphi, G/g)$ correlation, for all Solutions I, II-1, II-2,
when the constraints (\ref{gpipi}) are imposed on the coupling
constants of Table 5 (see Fig. 17b,d,f). The correlation curves form
a typical cross which, as is seen from the explanation given below, is
the glueball signature. Indeed, the
glueball can mix with the $q\bar q$ state because of the transition
$glueball\to (q\bar q)_{glueball}\to glueball$.
The state $(q\bar q)_{glueball} $ is close to
the the SU(3) flavour singlet $(q\bar q)_{singlet} =(u\bar u+d\bar
d+s\bar s)/\sqrt{3}$ but is slightly different. In the transition
$glueball\to (q\bar q)_{glueball}\to glueball$
the production of the $s\bar s$
pair can be suppressed, and it looks reasonable to assume that this
suppression is of the same order as it is in meson $q\bar q$ decays,
where the the new $q\bar q$ pair is created  by the gluon field:
$u\bar u:d\bar d:s\bar s=1:1:\lambda$, with $\lambda\simeq 0.5-0.8$.
Then
$(q\bar q)_{glueball}=(u\bar u +d\bar d +
\sqrt\lambda \; s\bar s)/\sqrt{2+ \lambda}$ \cite{alexei}.
In terms of the $n\bar n$ and $s\bar s$ states,
\be
(q\bar q)_{glueball}  =
n\bar n \cos\varphi_{glueball}+s\bar s \sin\varphi_{glueball}    \; ,
\label{phi-glue}
\ee
where $\varphi_{glueball} ={\rm tg}^{-1}\sqrt{\lambda /2}
\simeq 27^\circ -33^\circ$ for $\lambda \simeq 0.5-0.8$.
The glueball descendant is a mixture of the gluonium $(gg)$ and
quarkonium $(q\bar q)_{glueball}$ components:
\be
gg \cos\gamma + (q\bar q)_{glueball} \sin \gamma
\label{gamma-glue}
\ee
Since the ratios of couplings for the transitions
$(gg)\to \pi\pi$, $K\bar K$, $\eta\eta$, $\eta\eta'$
and
$(q\bar q)_{glueball}\to \pi\pi$, $K\bar K$, $\eta\eta$, $\eta\eta'$
are quite the same, one cannot find out the mixing angle $\gamma$.
But this very property -- the similarity of ratios of coupling
constants for the gluonium and quarkonium component -- gives rise to
special form of correlation curves on the $(\varphi, g/G)$-plot,
the cross in Fig. 17b,d,f. Vertical part of the line means that the
glueball descendant may have any noticeable proportion of the
$(q\bar q)_{glueball}$ state. Horizontal part of the curve corresponds
to the dominantly gluonium component.

The couplings from Table 5 for $f_0(1200-1600)$ allow certain
deviation in the $q\bar q$ component from the state
$(q\bar q)_{glueball}$: the couplings may be described by the values
$(g/G)$ and $\varphi$ belonging to hyperboles of the type shown in Fig.
17b,d,f by dashed and dot-dashed lines. We have a bunch of such
hyperboles which go off the centre of the cross, when $\lambda$
deviates from its central value ($\lambda=0.55$ for Solution I,
$\lambda=0.54$ for Solution II-1 and
$\lambda=0.85$ for Solution II-2).

An appearence of the gluonium cross
on the $(\varphi, g/G)$-plot, when the formula (\ref{gpipi}) is used to
determine coupling constants,  is the glueball (or glueball descendant)
signature in case of a strong mixing of the gluonium and quarkonium
components. On the contrary, the absence of the gluonium cross in the
correlation curves should point to a $q\bar q$ origin of the considered
$f_0$ state.

Therefore, the broad state $f_0(1200-1600)$ in all Solutions is the
glueball descendant, for the transition couplings
$f_0(1200-1600)\to \pi\pi$, $K\bar K$, $\eta\eta$, $\eta\eta'$ point
directly to this fact.

The states
$f_0(980)$, $f_0(1300)$, $f_0(1500)$, $f_0(1750)$ cannot pretend to be
the glueball descendants.

The states $f_0(1300)$ and  $f_0(1500)$ are dominantly the $n\bar n$
states. Though,
in Solution II-1, the resonance $f_0(1300)$ is allowed to have a
large $s\bar s$ component: with a noticeable admixture of the
gluonium component, the angle $\varphi [f_0(1300)]$ may achieve the
value $\simeq -50^\circ$. As to $f_0(1500)$,
$\varphi[f_0(1300)]$ may reach
$25^\circ$ at $G/g\simeq -0.6$ (Solution I). The
description of couplings squared, $g^2_a$, in this case owes
a strong destructive interference of the decay amplitudes
$(q\bar q)\to two\;pseudoscalars$ and
$(gg)\to two\;pseudoscalars$. So in this case we should not be
tempted by the proximity of $\varphi[f_0(1500)]$ and
$\varphi_{glueball}$ to identify $f_0(1500)$ as the glueball or its
descendant.

\subsection{Systematics of scalar states on the $(n,M^2)$-plane}

The systematics of resonances carried out in
\cite{syst} demonstates that all resonances can be plotted on
linear trajectories at the $(n,M^2)$-plane,
$M^2=M^2_0+(n-1)\mu^2$,  with a universal slope $\mu^2\simeq 1.3$
GeV$^2$.

This empirical property of $q\bar q$ states may serve as an additional
signature for the $q\bar q$ origin of resonances
$f_0(980)$, $f_0(1300)$, $f_0(1500)$, $f_0(1750)$.
These resonances
fit well to linear
trajectories, with the slope $\mu^2\sim 1.3$ GeV$^2$. The figure 18a
demonstrates the $(n,M^2)$-trajectories for resonance states with
$00^{++}$, $10^{++}$ and $\frac12 0^{++}$, if the $f_0(1200-1600)$ is
accepted to be of the glueball origin (recall that a doubling of the
$f_0$-trajectories occurs, due to the existence of two components,
$n\bar n$ and $s\bar s$).  Similar trajectories for bare states are
shown in Fig. 18b, if the $f_0^{bare}(1580)$ is the gluonium.  The
trajectory slopes for real and bare states almost coinside.

In Fig. 18, we present the variant where $f_0^{bare}(1580)$ is the gluonium
 and the broad
state $f_0(1200-1600)$ is its descendant; in this way it is natural
that there is no room for these states to be on  linear trajectories on
the  $(n,M^2)$-plane.

For bare states of Solution II-1, the linearity of trajectories
is completely broken, see Section 3.1, because    the linearity
exists for bare states only when $f_0^{bare}(1580)$ is the glueball.
This is an argument against Solution II-1 as physical solution.

\section{Conclusion}

The $K$-matrix analysis of the $00^{++}$-wave based on the use of the
spectra  $\pi\pi, K\bar K,\eta\eta,\eta\eta'$ in a broad variety
of reactions
(Table 1) provided us with three solutions:
Solution I and  Solutions II-1, II-2 (Tables 2, 3, 4).
All these Solutions, despite  a significant increase of the used
experimental information,  occurred to be similar to Solutions obtained
in the previous analysis \cite{YF}. In all Solutions,  five poles have
been found for the $00^{++}$ amplitude at complex masses in
the studied mass region, $280\le M\le 1900$ MeV; they correspond to
five scalar-isoscalar resonances, four of them being comparatively
narrow resonances, $f_0(980)$, $f_0(1300)$, $f_0(1500)$, $f_0(1750)$,
and a broad state           $f_0(1200-1600)$.

All Solution prove that comparatively narrow resonances
$f_0(980)$, $f_0(1300)$, $f_0(1500)$, $f_0(1750)$ are of the $q\bar q$
origin. Both the ratios of the decay couplings for
$f_0\to \pi\pi$, $K\bar K$, $\eta\eta$, $\eta\eta'$, which result in
linear behaviour of the correlation function $(\varphi, G/g)$ (see
Section 5.2), and the creation of linear trajectories in the
$(n,M^2)$-plane by resonances (Section 5.3) point to this fact.

The broad state $f_0(1200-1600)$ has a gluonium origin: this is
testified by \\
(i) a specific behaviour of the correlation curve --- glueball cross
--- in the $(\varphi, g/G)$ plot (Section 5.2), and \\
(ii) the absence of a room for the broad state on linear trajectories
in the $(n,M^2)$-plane (Section 5.3).

 Solutions II-1 and II-2 provide us with  proximate properties for all
resonances
$f_0(980)$, $f_0(1300)$, $f_0(1500)$, $f_0(1750)$,
and a broad state           $f_0(1200-1600)$
but they differ by the classification of
bare states, i.e. by the "formation history" of these resonances. In
Solution II-1, a pure glueball state is located around $1200- 1250$
MeV, that results in the fact that bare $q\bar q$-states do
not form linear $(n,M^2)$-trajectories. As to Solutions I and II-2, the
pure gluonium state can be located around $1500-1600$ MeV, so the bare
$q\bar q$ states are reliably set on linear trajectories. The criterion
of linearity for bare $q\bar q$ states allows us to consider
as physical Solutions I and II-2.

Solution I differs from Solution II by the content of $f_0(1750)$ which
is dominantly $s\bar s$ state for Solution I and
dominantly $n\bar n$ state for Solution II.

The $K$-matrix amplitudes reconstructed in our analysis do not contain
the pole associated with a comparatively narrow sigma-meson. However,
one should keep in mind that
the $K$-matrix technique  does not allow one to restore analytical
amplitude in the region neighbouring left-hand cuts; this region is
evaluated as ${\rm Re}\, s \la 4m^2_\pi$ or, in terms of the invariant
mass $M$, as
$({\rm Re}\, M)^2  - ({\rm Im}\, M )^2\la
4m^2_\pi$. Because of that our $K$-matrix analysis is unable to provide
definite conclusion about the existence of the  $\sigma$-meson
with a large width. Nevertheless, let us stress that in some variants
of the fit the restored $K$-matrix amplitudes have poles which might be
considered as a light and broad $\sigma$-meson. For example, in
Solution I such a pole is located at
$M^2\simeq 0.25-i\,1.00$ GeV$^2$, while we do not see similar pole
around $Im \,M^2 \sim 0.5-1.5$ GeV$^2$ in Solution II.
But, underline once again, this
region cannot be treated as reliable in reconstructing analytical
amplitude.

We have analysed rich experimental information, yet it allowed us
qualitative evaluation only of  the percentage of the $n\bar n$, $s\bar
s$ and gluonium component in studied $f_0$-mesons.
In this sense, the information provided by hadronic decays of
$f_0$-mesons is exausted: the increase of data accuracy  or of a number
of studied reactions cannot help us to attain a considerable progress
in understanding of $f_0$-mesons in the region under investigation. A
qualitative and precise reconstruction of the content of $f_0$-mesons
can be done after analysing a broad variety of
non-hadronic reactions. It looks like the study of reactions
such as $\gamma\gamma \to \pi\pi, K\bar K,\eta\eta,\eta\eta' $ as
well as $J/\Psi\to \gamma\pi\pi, \gamma K\bar K,\gamma \eta\eta,
\gamma \eta\eta' $  (on the ground of relevant statistics) could
provide the reliable and final magnitudes for the $n\bar n$, $s\bar s$
and gluonium component in $f_0$-mesons, for the production of hadrons
in these reactions is initiated by the $q\bar q$ and gluonium
components, correspondingly.

\section*{Acknowledgement}

We are grateful to A.V. Anisovich, D.V. Bugg, L.G. Dakhno, E. Klempt,
V.A.  Nikonov for useful and stimulating discussions
and L. Lesniak for valuable information. The paper is
supported by the RFBR grant N 01-02-17861. One of us (A.V.S.) thanks
Science Support Foundation (grant for talented young researchers).

\section*{Appendix A: Amplitudes for the partial waves
$02^{++}$, $10^{++}$, $12^{++}$}

Here we present the amplitudes for the waves which have not been fitted
in the present analysis. The parameters of the resonances in these
waves were found in \cite{YF}, and they are used as fixed valyes.

\subsection*{Scattering amplitudes}

Scattering amplitudes for the partial waves
$02^{++}$, $10^{++}$, $12^{++}$ are written as:
\be
\hat A^{(IJ)}=\hat K^{(IJ)}
\left (\hat 1-i \hat \rho \hat K^{(IJ)}\right )^{-1}\ .
\label{A.1}
\ee

\subsubsection*{Isoscalar-tensor, $02^{++}$, partial wave}

The $D$-wave interaction in  isoscalar sector
is parametrized by the 4$\times $4 $K$-matrix where
$1 = \pi\pi$, $2 = K\bar K$,
$3 = \eta\eta$ and $4 =  {\rm multimeson\; states}$:
\be
K_{ab}^{02}(s)=D_a(s)\left ( \sum_\alpha \frac{g^{(\alpha)}_a
g^{(\alpha)}_b}{M^2_\alpha-s}
+f^{(02)}_{ab}\frac{1\,\mbox{GeV}^2+s_2}{s+s_2} \right )
D_b(s)\;.
\label{A.2}
\ee
Factor $D_a(s)$ stands for the $D$-wave centrifugal barrier.
We take this factor in the following form:
\be
D_a(s)=\frac{k_a^2}{k_a^2+3/r_a^2},\quad
a=1,2,3\ ,
\label{A.3}
\ee
where $k_a=\sqrt{s/4-m_a^2} $ is the momentum of the decaying
meson in the centre-of-mass frame  of the resonance. For the
multi-meson decay, the factor $D_4(s)$ is taken to be  1. The used
phase space factors  are the same as those for the isoscalar
$S$-wave channel.

\subsubsection*{Isovector-scalar, $10^{++}$, and isovector-tensor,
$12^{++}$, partial waves}

For the amplitude in the isovector-scalar and isovector-tensor
channels, we use the 4$\times $4 $K$-matrix with
1 = $\pi\eta$, 2 = $K\bar K$, 3 = $\pi\eta'$ and
4 =  multi-meson states:
\be
K_{ab}^{1J}(s)=D_a(s)
 \left ( \sum_\alpha \frac{g^{(\alpha)}_a
g^{(\alpha)}_b}{M^2_\alpha-s}
 +f_{ab}\frac{1.5\; \mbox{GeV}^2+s_1}{s+s_1} \right )D_b(s)\;.
\label{A.4}
\ee
Here $J=0,2$; the factors $D_a(s)$ are equal to 1 for the $10^{++}$
amplitude, while for the
$D$-wave partial amplitude the factor $D_a(s)$ is taken in the form:
\be
D_a(s)=&&\frac{k_a^2}{k_a^2+3/r_3^2}, \;\; a=1,2,3,  \nonumber \\
D_4(s)=&&1\; .
\label{A.5}
\ee

\subsection*{Three-meson production amplitudes}

The partial waves $02^{++}$, $10^{++}$, $12^{++}$ are taken into
account in the three-meson production processes.

Invariant production amplitude
$ A_{p\bar p\; (^{2S+1}L_J1,b)} ^{Ij}(23)$ for the transition
$p\bar p\; (^{2S+1}L_J) \to 1+(2+3)_{Ij}$ where the indices $Ij$ refer
the isospin and total angular momentum of the mesons $b=2+3$ ($Ij=02$
$10$, $12$) reads:
\be
A_{p\bar p\; (^{2S+1}L_J)1,b} ^{Ij}(23)=\sum\limits_{a}\widetilde
K_{p\bar p\; (^{2S+1}L_J)1,b}^{Ij} (s_{23})
 \left[\hat 1-i \hat\rho
\hat K^{Ij}(s_{23})\right]^{-1}_{ab}\; .
\label{A.6}
\ee
Here
\be
\widetilde K_{p\bar p\; (^{2S+1}L_J)1,a} ^{Ij}(s_{23})=\left (
\sum_\alpha \frac{\Lambda_{p\bar p\; (^{2S+1}L_J)1}^{(\alpha)} [Ij]
g^{(\alpha)}_a}
{M^2_\alpha-s_{23}}
+\phi_{p\bar p\; (^{2S+1}L_J)1,a}[Ij]
\frac{1\; \mbox{GeV}^2+s_0}{s_{23}+s_0} \right)
D_a(s_{23})\, ,
\label{A.7}
\ee
and parameters $\Lambda_{p\bar p\; (^{2S+1}L_J)1}^{(\alpha)} [Ij]$
$\phi_{p\bar p\; (^{2S+1}L_J)1,a}[Ij]$
may be complex magnitudes with different phases due to
three-particle interactions.

Equations (\ref{A.6}) and (\ref{A.7}) give us invariant parts of the
final-state interaction block. The factors related to the
angular-momentum expansion are presented in Appendix B.

\section*{Appendix B: Moment-operator expansion for the processes
$p\bar p \to three\;  mesons$}

We analyse the processes $p\bar p \to three\; mesons$ by
using the technique of moment-operator expansion; in this way our work
is grounded
upon  the papers \cite{km,km1900,YF99,YF}. Below the
necessary formulae are given for the reaction $p\bar p\to three\; mesons$
in the liquid and  gas, when the $p\bar p$ annihilation is going
from the lowest waves, $S$ and $P$ and the meson pair is in the $S$-,
$P$- and $D$-states. In \cite{operator}, this technique is presented in
its general form, one may address this paper for  more detail.

The amplitude for the cascade transition
$p\bar p\to resonance+meson \to three\; mesons$ has the following
structure:
\be
\bar\psi(-q_2)Q^{(S,L,J)}_{\mu_1\mu_2\ldots\mu_{J-1}\mu_J}
(q)\psi(q_1)
Q^{(j,J)}_{\mu_1\mu_2\ldots\mu_{J-1}\mu_J}(k_{12},k_3)\ .
\label{B.15}
\ee
The operator
$\bar\psi(-q_2)Q^{(S,L,J)}_{\mu_1\mu_2\ldots\mu_{J-1}\mu_J}\psi(q_1)$
refers to the $p\bar p$ state with the total
angular momentum $J$: here
$S$ is the total spin of fermions, $S=0,1$, and $L$ is
the angular momentum; $q_1$ and $q_2$ are the momenta of fermions
and $q=(q_1-q_2)/2$.
The operator
$Q^{(j,J)}_{\mu_1\mu_2\ldots\mu_{J-1}\mu_J}(k_{12},k_3)$ stands for
the three-meson operator with the production of
particles $1$ and $2$ in the resonance $j$-wave; this operator
depends on relative momentum, $k_{12}$, of mesons
$1$ and $2$, $k_{12}=(k_1-k_2)/2$, and
$k_3$.
Partial--wave amplitude
is a function of the invariant
energy squared $s=P^2$, where $P=k_1+k_2+k_3=q_1+q_2$, and
$s_{12}=(k_1+k_2)^2$.

\subsection*{ Angular-momentum operator
for two particles, $ X^{(L)}_{\mu_1\mu_2\ldots\mu_{L-1}\mu_L}(q)$}

Let us introduce the operator of angular momentum
of two particles, $ X^{(L)}_{\mu_1\mu_2\ldots\mu_{L-1}\mu_L}(q)$, which
is constructed by using  relative momentum
of mesons in the space orthogonal to
the total momentum $P$:
\be
q^\perp_\mu = q_\nu
g^\perp_{\nu\mu},  \qquad
g^\perp_{\nu\mu}=g_{\nu\mu}-\frac{P_\nu
P_\mu}{s}.
\ee
In the
centre-of-mass system, where $P=(P_0,\vec P)=(\sqrt s,0)$, the vector
$q^\perp$ is space-like: $q^\perp=(0,\vec q)$.
We determine the operator
$ X^{(L)}_{\mu_1\mu_2\ldots\mu_{L-1}\mu_L}(q)$
to
be  symmetrical and traceless. It is easy to construct it for the lowest
values of $L$, $L=0,1,2$:
\be
X^{(0)}=1\ , \qquad X^{(1)}_\mu=q^\perp_\mu\ , \qquad
X^{(2)}_{\mu_1 \mu_2}=\frac32\left(q^\perp_{\mu_1}
q^\perp_{\mu_2}-\frac13\, q^2_\perp g^\perp_{\mu_1\mu_2}\right),
\label{B.2}
\ee
Correspondingly, the generalization of
$X^{(L)}_{\mu_1\ldots\mu_L}$ for
$L>1$ reads:
\be
\label{B.3}
X^{(L)}_{\mu_1\ldots\mu_L}&=&q^\perp_\alpha
Z^{(L-1)}_{\mu_1\ldots\mu_L, \alpha} \  , \\
Z^{(L-1)}_{\mu_1\ldots\mu_L, \alpha}&=&
\frac{2L-1}{L^2}\left (
\sum^L_{i=1}X^{{(L-1)}}_{\mu_1\ldots\mu_{i-1}\mu_{i+1}\ldots\mu_L}
g^\perp_{\mu_i\alpha} \right .
\nonumber \\
&&\left . -\frac{2}{2L-1} \sum^L_{i,j=1 \atop i<j}
g^\perp_{\mu_i\mu_j}
X^{{(L-1)}}_{\mu_1\ldots\mu_{i-1}\mu_{i+1}\ldots\mu_{j-1}\mu_{j+1}
\ldots\mu_L\alpha} \right )\ .
\nonumber
\ee
It is seen that the operator
$ X^{(L)}_{\mu_1\mu_2\ldots\mu_{L-1}\mu_L}(q)$
constructed in accordance with (\ref{B.3})
is symmetrical,
\be
X^{(L)}_{\mu_1\ldots\mu_i\ldots\mu_j\ldots\mu_L}\; =\;
X^{(L)}_{\mu_1\ldots\mu_j\ldots\mu_i\ldots\mu_L},
\label{B.12}
\ee
and it works in the space orthogonal to $P$:
\be
P_{\mu_i}X^{(L)}_{\mu_1\ldots\mu_i\ldots\mu_L}\ =\ 0 \ .
\label{B.13}
\ee
The angular-momentum operator
$X^{(L)}_{\mu_1\ldots\mu_L}$ is traceless over any two
indices:
\be
g_{\mu_i\mu_j}X^{(L)} _{\mu_1\ldots\mu_i\ldots\mu_j\ldots\mu_L}\
=\ g^\perp _{\mu_i\mu_j}X^{(L)} _{\mu_1\ldots\mu_i\ldots\mu_j\ldots\mu_L}
\ =\ 0.
\label{B.1}
\ee

\subsection*{Three-particle production amplitude
with resonance in the intermediate state}

The moment-operator for the production of three
spinless particles in the cascade process
$resonance(j^{PC}) +meson$, can be written in terms
of operators $ X^{(L)}_{\mu_1\ldots\mu_L}$.
To be definite, we consider the simplest
case of vector resonance $(j=1)$ and give a
generalization for $j=2$.

For the two-stage reaction of the type $p\bar p\to VP \to PPP$,
we denote the
particle momenta of  pseudoscalars $(P)$ in the final states
as $k_1,k_2,k_3$; the vector resonance $(V)$ is produced
in the channel $1+2$ with the total momentum $p=k_1+k_2$
and relative momentum for the decay products $k_{12}=\frac12 (k_1-k_2)$.
Then the final-state
moment operator in (\ref{B.15}) reads:
\be
Q^{(j=1,J)}_{\mu_1\mu_2\ldots\mu_{J-1}\mu_J}(k_{12},k_3)
=A^{(J,l=J+1)}(s,s_{12})
X^{(J+1)}_{\mu_1\ldots\mu_J\alpha} (k^\perp_{3})
X^{(1)}_\alpha(k^\perp_{12})
\label{B.16a}
\ee
$$
+A^{(J,l=J-1)}(s,s_{12})
X^{(J-1)}_{\mu_1\ldots\mu_{J-1}} (k^\perp_{3})
X^{(1)}_{\mu_{J}}(k^\perp_{12})\, ,
\nonumber
$$
where
$k^\perp_{12\mu}$ is orthogonal to $p=k_1+k_2$,
\be
k^\perp_{12\mu}=\left
(g_{\mu\mu'}-\frac{p_{\mu}p_{\mu'}}{s_{12}}\right )k_{12\mu'}\ .
\label{B.16b}
\ee
 Here $p^2=s_{12}$, and the momentum $k^\perp_3$ is orthogonal to the
total momentum $P=k_1+k_2+k_3$,
\be
k^\perp_{3\mu}=\left
(g_{\mu\mu'}-\frac{P_{\mu}P_{\mu'}}{P^2}\right )k_{3\mu'} \, .
\label{B.16c}
\ee
One has for the state with $l=L$:
\be
A^{(J,l=J)}(s,s_{12})
X^{(J)}_{\mu_1\ldots\mu_{J-1}\alpha} (k^\perp_{3})
X^{(1)}_\beta(k^\perp_{12}) P_\gamma \epsilon_{\alpha\beta\gamma\mu_J}\
,
\label{B.16d}
\ee
where $\epsilon_{\alpha\beta\gamma\mu_J}$ is  the totally
antisymmetrical four-tensor.

The generalization for a higher resonance
is obvious: for the tensor resonance, $j=2$, in the reaction
$ p\bar p\to TS \to PPP $, instead of two terms in (\ref{B.16a})
one has three terms  for the partial-wave amplitude:
\be
A^{(J,l=J+2)}(s,s_{12})
X^{(J+2)}_{\mu_1\ldots\mu_J\alpha_1\alpha_2} (k^\perp_{3})
X^{(2)}_{\alpha_1\alpha_2}(k^\perp_{12})
\label{B.16e}
\ee
$$
+A^{(J,l=J)}(s,s_{12})
X^{(J)}_{\mu_1\ldots\mu_{J-1}\alpha} (k^\perp_{3})
X^{(2)}_{\alpha\mu_{J}}(k^\perp_{12})
\nonumber
$$
$$
+A^{(J,l=J-2)}(s,s_{12})
X^{(J-2)}_{\mu_1\ldots\mu_{J-2}} (k^\perp_{3})
X^{(2)}_{\mu_{J-1}\mu_{J}}(k^\perp_{12})\ .
\nonumber
$$

\subsection*{Moment operator for proton-antiproton
system,
$(\bar\psi(-q_2)Q^{(S,L,J)}_{\mu_1\ldots\mu_J}\psi(q_1))$ }

The partial-wave vertex for this system with the total angular momentum
$J$, orbital momentum $L$
and total spin $S$ is determined by bilinear form
$\bar\psi(-q_2)Q^{(S,L,J)}_{\mu_1\mu_2\ldots\mu_{J-1}\mu_J}\psi(q_1)$
where $(\bar\psi(-q_2)$ and $\psi(q_1))$ are bispinors
and $Q^{(S,L,J)}_{\mu_1\mu_2\ldots\mu_{J-1}\mu_J}$
is the fermion partial-wave operator. This latter operator should be
constructed with the use of the  orbital-momentum operator $X^{(L)}$ and
spin operator for fermion-antifermion system.

We have two fermion spin states, $S=0$ and $S=1$.
For the spin-0 state, $J=L$, and for the spin-1 state, one has
$J=L-1,L,L+1$.

For fermion operator
$Q^{(S,L,J)}_{\mu_1\mu_2\ldots\mu_{J-1}\mu_J}$, one implies
the same constraints
as for the boson one:  the fermion operator should be symmetrical,
$P$-orthogonal and traceless:
\be
Q^{(S,L,J)}_{\mu_1\mu_2\ldots\mu_{J-1}\mu_J}=
Q^{(S,L,J)}_{\mu_2\mu_1\ldots\mu_{J-1}\mu_J}\  ,\qquad
P_\mu Q^{(S,L,J)}_{\mu\mu_2\ldots\mu_{J-1}\mu_J}=0 \  , \qquad
g_{\mu_i\mu_k}Q^{(S,L,J)}_{\mu_1\ldots\mu_J}=0\ .
\label{B.26}
\ee
The spin-0 operator for
proton-antiproton system, $\Gamma^{(0)}$, is proportional to the
$\gamma_5$-matrix.  We normalize $\Gamma^{(0)}$
by the condition:
\be
{\rm Sp}\  \left ( \Gamma^{(0)} (m+\hat q_1)
\Gamma^{(0)} (m-\hat q_2)\right )= 1\ ,
\label{B.27}
\ee
that gives
\be
\Gamma^{(0)}=\frac{i\ \gamma_5}
{\sqrt{2s}} \  .
\label{B.28}
\ee
The angular-momentum operator for the spin-0 state is a product of
$\Gamma^{(0)}$ and the angular-momentum operator $X^{(J)}_
{\mu_1\ldots\mu_J}(q)$:
\be
Q^{(0,L,J)}_{\mu_1\mu_2\ldots\mu_{J-1}\mu_J}(q)=\Gamma^{(0)}
X^{(J)}_{\mu_1\ldots\mu_J}(q)  \  .
\label{B.29}
\ee
The spin-1
operator is constructed as follows:
\be
\Gamma_\alpha^{(1)} =
\frac{1}{\sqrt{2s}}
\left(\gamma^\perp_{\alpha} -
\frac{\hat q q^\perp_\alpha }{m(m+\frac{\sqrt{s}}{2})}
\right)=
\frac{1}{\sqrt{2s}}
\left(\gamma^\perp_{\alpha} -
\frac{ q^\perp_\alpha }{m+\frac{\sqrt{s}}{2}}\right)\ .
\label{B.30a}
\ee
Here we have used that
$\hat q=(\hat k_1 -\hat k_2)/2=m$.
The operator $\Gamma^{(1)}_\alpha$ is orthogonal to the total momentum
$P$ ($P_\alpha \Gamma_\alpha^{(1)} =0 $)
and normalized as follows:
\be
{\rm Sp}\;
\left (\Gamma_\alpha^{(1)} (m+\hat q_1) \Gamma_\beta^{(1)} (m-\hat
q_2)\right )= g^\perp_{\alpha\beta} \ .
\label{normg}
\ee
The operators $\Gamma_\alpha^{(1)}$ and $\Gamma^{(0)}$ are orthogonal
to one another in the spin space:
it means that, if initial fermions are not polarized, the spin-0
and spin-1 states do not interfere with each other in the differential
cross section.

The spin-1 state with $J=L-1$ is constructed from
the operator $\Gamma_\alpha^{(1)}$ and angular--momentum operator
$X^{(J+1)}_{\mu_1\ldots\mu_J\mu_{J+1}}$. It carries $J$ indices, so
two indices, one from $\Gamma_\alpha^{(1)}$
and another from $X^{(J+1)}_{\mu_1\ldots\mu_J\mu_{J+1}}$,
are to be absorbed, that can be done with the help of the metric tensor
$g_{\alpha\mu_{J+1}}$:
\be
Q^{(1,L,J=L-1)}_{\mu_1\ldots\mu_J}(q)=\Gamma_\alpha^{(1)}
X^{(L)}_{\mu_1\ldots\mu_{L-1}\alpha}(q) \  .
\label{B.34}
\ee
For $J=L$,
the construction of the operator $Q^{(1,L,J=L)}_{\mu_1\ldots\mu_J}$
is performed by using
antisymmetrical tensor $\varepsilon_{\mu\nu_1\nu_2\nu_3}$:
the operator
$\Gamma_\nu X^{(J)}_{\mu_1\mu_2\ldots\mu_J}$
must have the same number of indices as the angular--momentum operator.
The only possible non-zero combination of antisymmetrical
tensor, the $X^{(J)}$ and $\Gamma^{(1)}$ operators,
is given by the following convolution:
\be
\varepsilon_{\mu_1\nu_1\nu_2\nu_3}P_{\nu_1}
\Gamma_{\nu_2}^{(1)}
X^{(J)}_{\nu_3\mu_2\ldots\mu_J}(q)  \ .
\label{B.14}
\ee
However, the operator entering equation (\ref{B.14}), being
$P$-orthogonal and traceless, is not symmetrical. The symmetrization
can be performed by using the tensor $Z^{(J-1)}_{\nu\mu_2\ldots\mu_J
,\alpha}$ instead of $X^{(J)}_{\nu\mu_2\ldots\mu_J}$. In this way, we
get:
\be
Q^{(1,L,J=L)}_{\mu_1\ldots\mu_J}(q)=
\varepsilon_{\alpha\nu_1\nu_2\nu_3}P_{\nu_1} \Gamma_{\nu_2}^{(1)}
Z^{(J)}_{\nu_3\mu_1\ldots\mu_J,\alpha} (q)\ .
\label{B.36}
\ee
Following the same procedure, we can easily construct the operator for
the total angular momentum $J=L+1$. One has:
\be
Q^{(1,L,J=L+1)}_{\mu_1\ldots\mu_J}(q)=
\Gamma_\alpha^{(1)}
Z^{(J-1)}_{\mu_1\ldots\mu_J,\alpha}(q)\ .
\label{B.37}
\ee
After putting $L=1$ in (\ref{B.34}), (\ref{B.36})
and (\ref{B.37}), we have the $P$-wave operators which are used in the
analysis of the $p\bar p$ and $n\bar p$ annihilations.

The moment-operator expansion presented above was
used in analysis of the meson spectra in a number of papers
\cite{km,km1900,YF}.

Sometimes this technique
is misleadingly referred as the Zemach expansion method.
Comparing the operators of Eqs. (\ref{B.2})
(or (\ref{B.3})) and (\ref{B.16a}),
one can see the common and different features of the
three-dimensional approach of Zemach \cite{Zemach} and  covariant
method applied here and in
\cite{km,km1900,YF}. For the operator (\ref{B.3})
used in the centre-of-mass frame, the expressions used in both
approaches coincide. Indeed,  the four-momentum
$q^\perp_\mu$ has space-like components only,
$q^\perp_\mu=(0,\vec q)$, so the operator
$X^{(L)}_{\mu_1\ldots\mu_L}(q)$
turns into Zemach's operator. However, for the amplitude (\ref{B.16a}) a
simultaneous equality to zero of operators with zero components is
impossible. In \cite{Zemach} a special procedure was suggested  for
such cases, namely, the  operator is treated in its own
centre-of-mass frame, with subsequent Lorentz boost to a needed
frame. But in the procedure developed in
\cite{km,km1900,YF} and summarized
in \cite{operator}, these additional
manipulations are unnecessary.

The Lorentz boost should be also carried out upon the three-particle
production amplitude considered in terms of spherical wave functions
as well as in its version suggested by \cite{Chung}.

\section*{Appendix C: Quark-combinatorics relation for the decay
couplings}

In Table 7, we give the decay constants for the transition
\be
q\bar q=n\bar n \cos \varphi +s\bar s \sin \varphi\to
{\rm two\,pseudoscalar\,mesons}\ ,
\ee
where $n\bar n=(u\bar u+d\bar d)/\sqrt{2}$, while the angle $\Theta$
defines the quark content of $\eta$ and $\eta'$ mesons assuming them
pure $q\bar q$ states:
$\eta=n\bar n \cos \Theta -s\bar s \sin \Theta$ and
$\eta'=n\bar n \sin \Theta +s\bar s \cos \Theta$, with
$\cos \Theta \simeq 0.8$ and $\sin \Theta \simeq 0.6$.

The relations for the decay constants are given for planar diagrams
which are the leading ones, in terms of the $1/N$ expansion \cite{t'h},
see also \cite{ufn, lecture-95} for details.

The gluball decay is the two-stage process: $glueball \to
(q\bar q)_{glueball}\to mesons$. The transition
$glueball\to (q\bar q)_{glueball} $
is not suppressed in the framework of
$1/N$ expansion rule, see \cite{ufn,lecture-95}, hence the transition
constants
\be
glueball \to
{\rm two\, pseudoscalar\, mesons}
\label{const_glu}
\ee
are of the same order as the decay constants of the $q\bar q$-state. The
constants for the transition (\ref{const_glu}) obey the same relations
as Eq. (75), with a special fixation of $\varphi$. By substituting
\be
\varphi \to \varphi _{glueball}\ , \quad g_0\to G_0\ ,
\ee
we have $\tan\varphi _{glueball} =\sqrt {\lambda/2}$.
Such a definition of the $\varphi _{glueball}$ is due to the equality
$(q\bar q)_{glueball}=
(u\bar u+d\bar d+\sqrt{\lambda}s\bar s)/\sqrt{2+\lambda}$: recall that we assume the
new $q\bar q$-pairs to be produced in the proportion
$u\bar u:d\bar d:s\bar s=1:1:\lambda$.

The coupling constants given in Table 7 satisfy the sum rule:
\be
\sum\limits_{c=\pi\pi, K\bar K , \eta\eta , \eta\eta', \eta'\eta'}
 g^2_0\left ( n\bar n \to c
\right ) I_c+ \sum
\limits_{c= K\bar K , \eta\eta , \eta\eta', \eta'\eta'}
g^2_0\left ( s\bar s \to c \right ) I_c=
\frac 34 g^2_0 (2+\lambda).
\ee
The factor $(2+\lambda)$ is due to the production of one additional
$q\bar q$-pair in the decay of the $q\bar q$-meson; $I_c$ is the
identity factor, see Table 7.

For the glueball decay
the sum of couplings squared over all channels is proportional to
the probability $(2+\lambda)^2$ to produce  two $q\bar q$ pairs.
 So
\be
\sum\limits_{c=\pi\pi, K\bar K , \eta\eta , \eta\eta', \eta'\eta'}
G^2_0(c)I(c)=\frac{1}{2}G^2_0(2+\lambda)^2 \, .
\label{A3.2}
\ee

The relations for the decay coupling constants in case of non-planar
diagrams may be found in \cite{ufn,YF}. The analysis \cite{YF} proved
that the non-planar diagram contribution is suppressed --- just as it
should be within the rules of $1/N$ expansion, so we do not use such
type of terms in this analysis.

\begin{table}
\caption\protect{Coupling constants given by quark combinatorics for
a $q\bar q$-meson
decaying into a pair of pseudoscalar mesons in the leading terms
of the $1/N_c$ expansion.
$\varphi$ is the mixing angle for $n\bar n$ and $s\bar s$ states, and
$\Theta$ is the mixing angle for $\eta -\eta'$ mesons:
$\eta=n\bar n \cos\Theta-s\bar s \sin\Theta$ and
$\eta'=n\bar n \sin\Theta+s\bar s \cos\Theta$.
Glueball decay couplings in the leading terms of $1/N$ expansion
are obtained by the replacements $g_0\cos\varphi\to \sqrt 2 G_0$,
$g_0\sin\varphi\to\sqrt\lambda G_0$.}

\begin{center}
\begin{tabular}{|c|c|c|}
\hline
~      &     ~                    &  ~           \\
~      & The $q\bar q$-meson decay&Identity   \\
~      & couplings in the         &   factor in  \\
Channel& leading terms of $1/N$   &phase space \\
~      & expansion                &~   \\
~      &     ~                    &~           \\
\hline
~      & ~                        & ~  \\
$\pi^0\pi^0$ &  $g_0\cos\varphi/\sqrt{2}$                         & 1/2
\\ ~      & ~                                                     &  ~
\\ $\pi^+\pi^-$ & $g_0\cos\varphi/\sqrt{2}$                          &
1 \\ ~      & ~                                                     & ~
\\ $K^+K^-$ & $g_0 (\sqrt 2\sin\varphi+\sqrt \lambda\cos\varphi)/\sqrt 8
$& 1 \\ ~ & ~ & ~  \\ $K^0K^0$ & $g_0 (\sqrt 2\sin\varphi+\sqrt
\lambda\cos\varphi)/\sqrt 8 $ &  1 \\ ~ & ~ & ~  \\ $\eta\eta$ &
$g_0\left (\cos^2\Theta\;\cos\varphi/\sqrt 2 +
\sqrt{\lambda}\;\sin\varphi\;\sin^2\Theta\right )$                 &1/2
\\ ~ & ~ & ~  \\ $\eta\eta'$ &
$g_0\sin\Theta\;\cos\Theta\left(\cos\varphi/\sqrt 2-
\sqrt{\lambda}\;\sin\varphi\right ) $                              & 1\\
~ & ~ & ~  \\
$\eta'\eta'$ &
$g_0\left(\sin^2\Theta\;\cos\varphi/\sqrt 2+
\sqrt{\lambda}\;\sin\varphi\;\cos^2\Theta\right)$                  &1/2
\\ ~ & ~ & ~  \\ \hline
\end{tabular}
\label{table7}
\end{center}
\end{table}

\newpage

\begin{figure}
\centerline{\epsfig{file=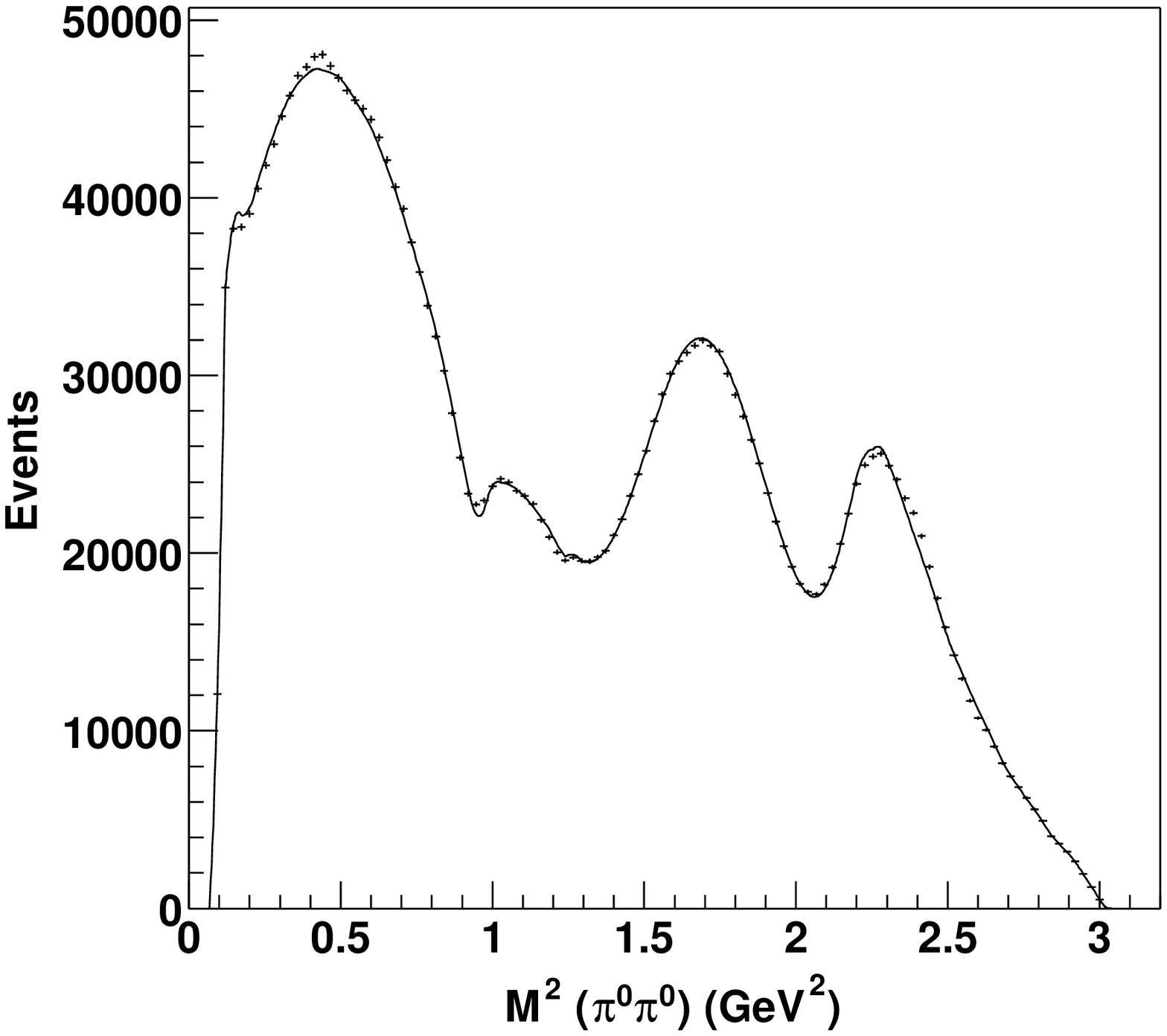,width=13cm}}
\caption{A mass projection of the acceptance-corrected
Dalitz plot for the $p\bar p$ annihilation into
$\pi^0\pi^0\pi^0$ in liquid $H_2$.
The curve corresponds to the fit in
Solution II-2.}
\end{figure}

\newpage
\begin{figure}
\centerline{\epsfig{file=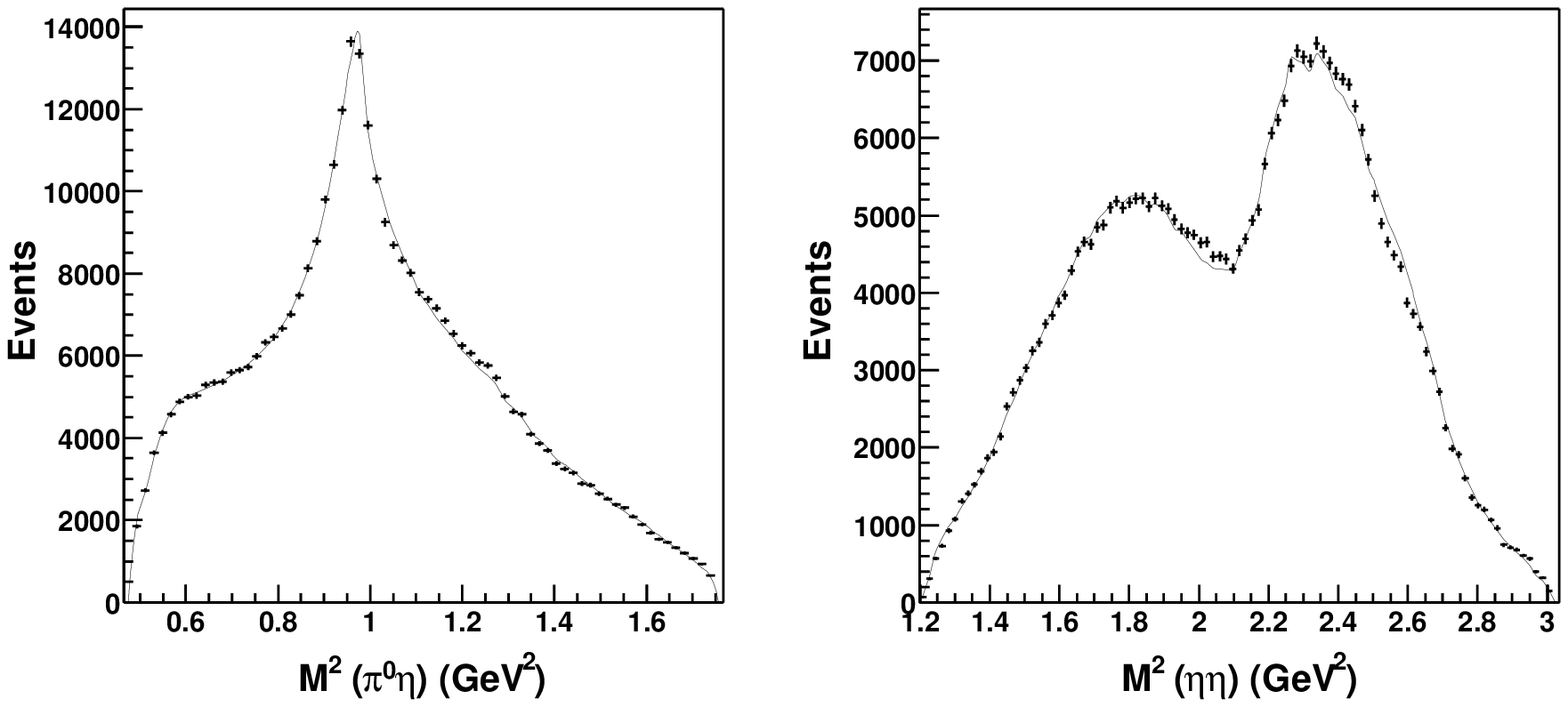,width=16cm}}
\caption{Mass projections of the acceptance-corrected
Dalitz plot for the $p\bar p$ annihilation into
$\pi^0\eta\eta$ in liquid $H_2$.
Curves correspond to the fit in
Solution II-2.}
\end{figure}
\begin{figure}
\centerline{\epsfig{file=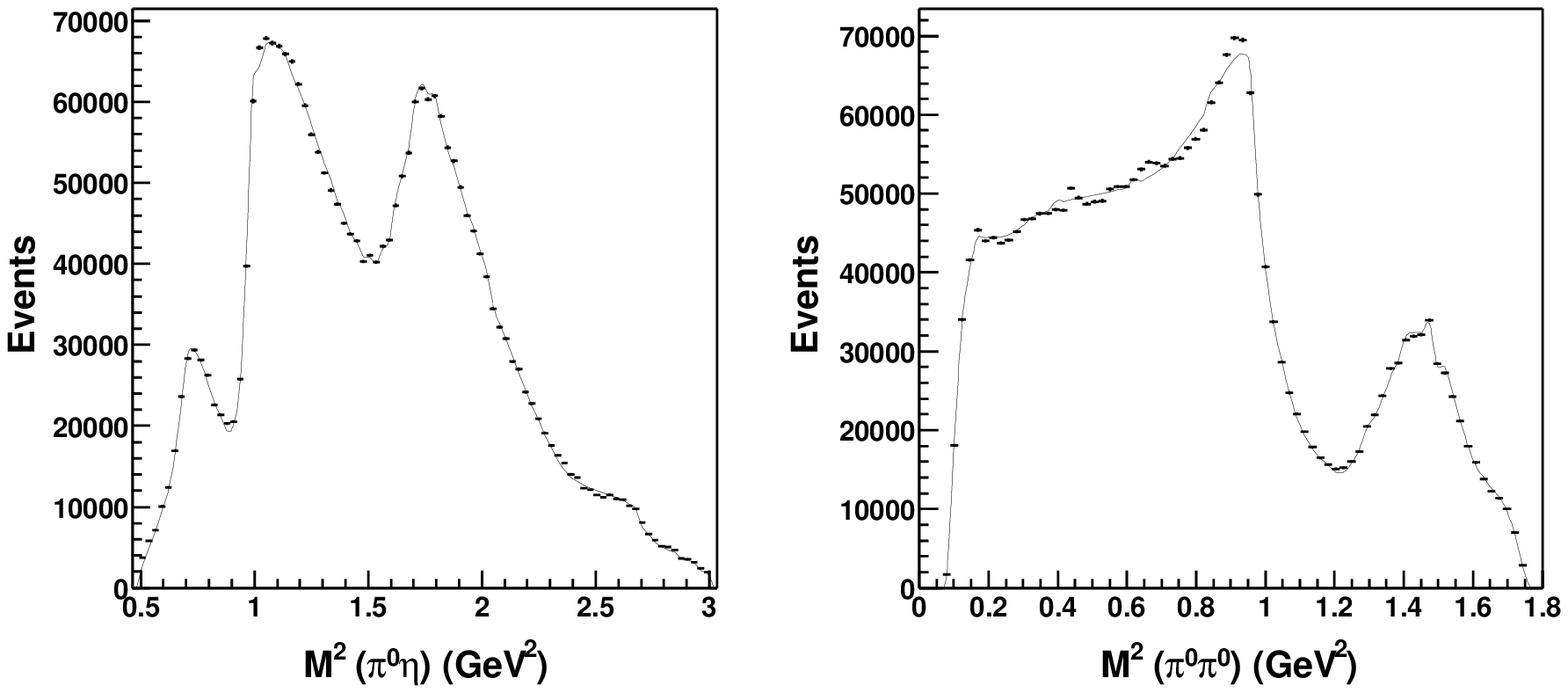,width=16cm}}
\caption{Mass projections of the acceptance-corrected
Dalitz plot for the $p\bar p$ annihilation into
$\pi^0\pi^0\eta$ in liquid $H_2$.
Curves correspond to the fit in
Solution II-2.}
\end{figure}

\newpage
\begin{figure}
\centerline{\epsfig{file=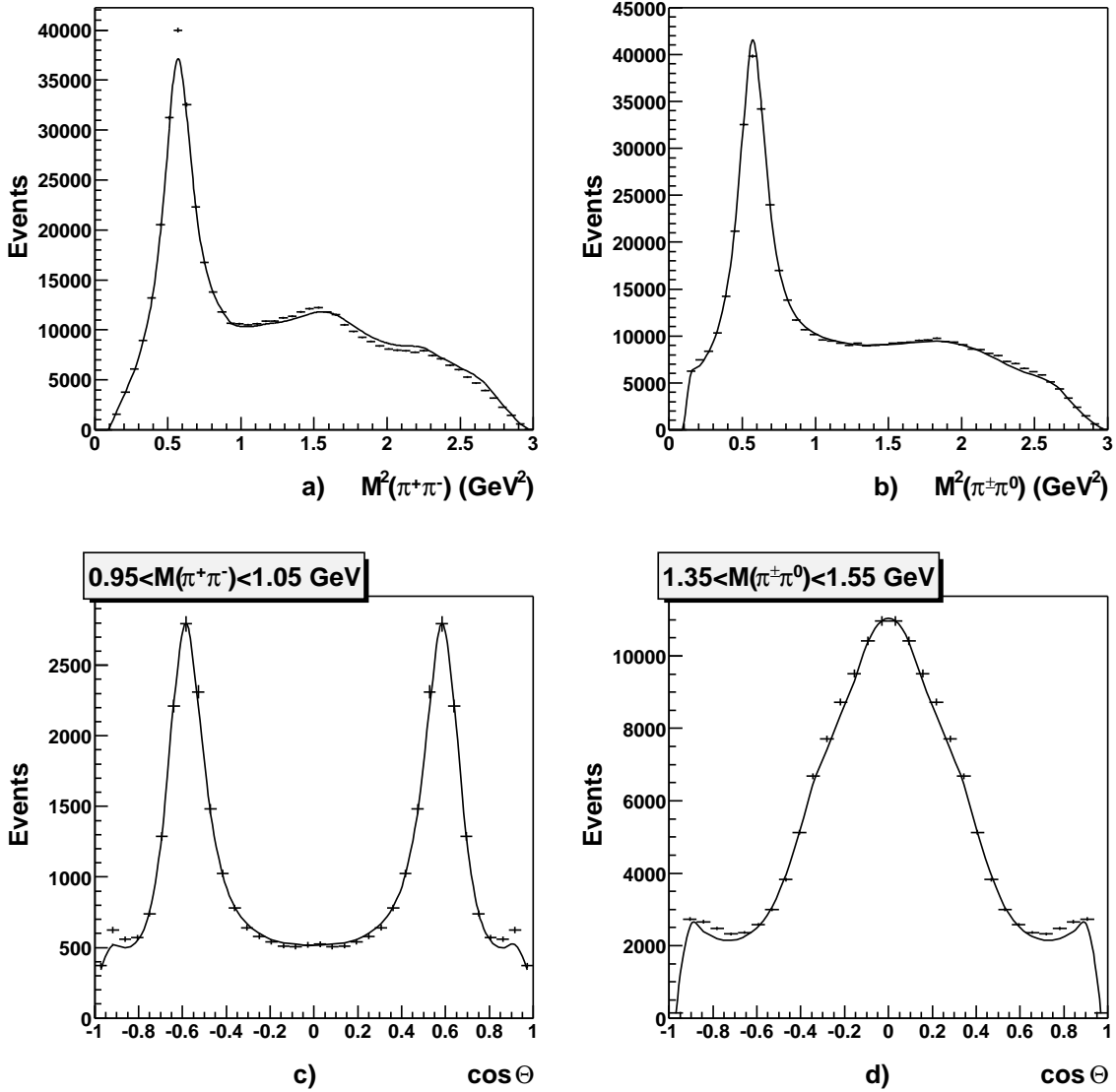,width=16.5cm}}
\caption{a,b) Mass projections of the acceptance-corrected
Dalitz plot for the $p\bar p$ annihilation into
$\pi^+\pi^0\pi^-$ in liquid $H_2$,
c) the angle distribution
between charged and neutral pions in c.m.s. of $\pi^+\pi^-$ system
taken at masses between 0.95 and  1.05 GeV,
d) the angle distribution
between charged pions in c.m.s. of $\pi^\pm\pi^0$ system
taken at masses between 1.35 and  1.55 GeV. Figure 4d shows the event
distribution along the band with the production of $\rho(1450)$.}

 \end{figure}

\begin{figure}
\centerline{\epsfig{file=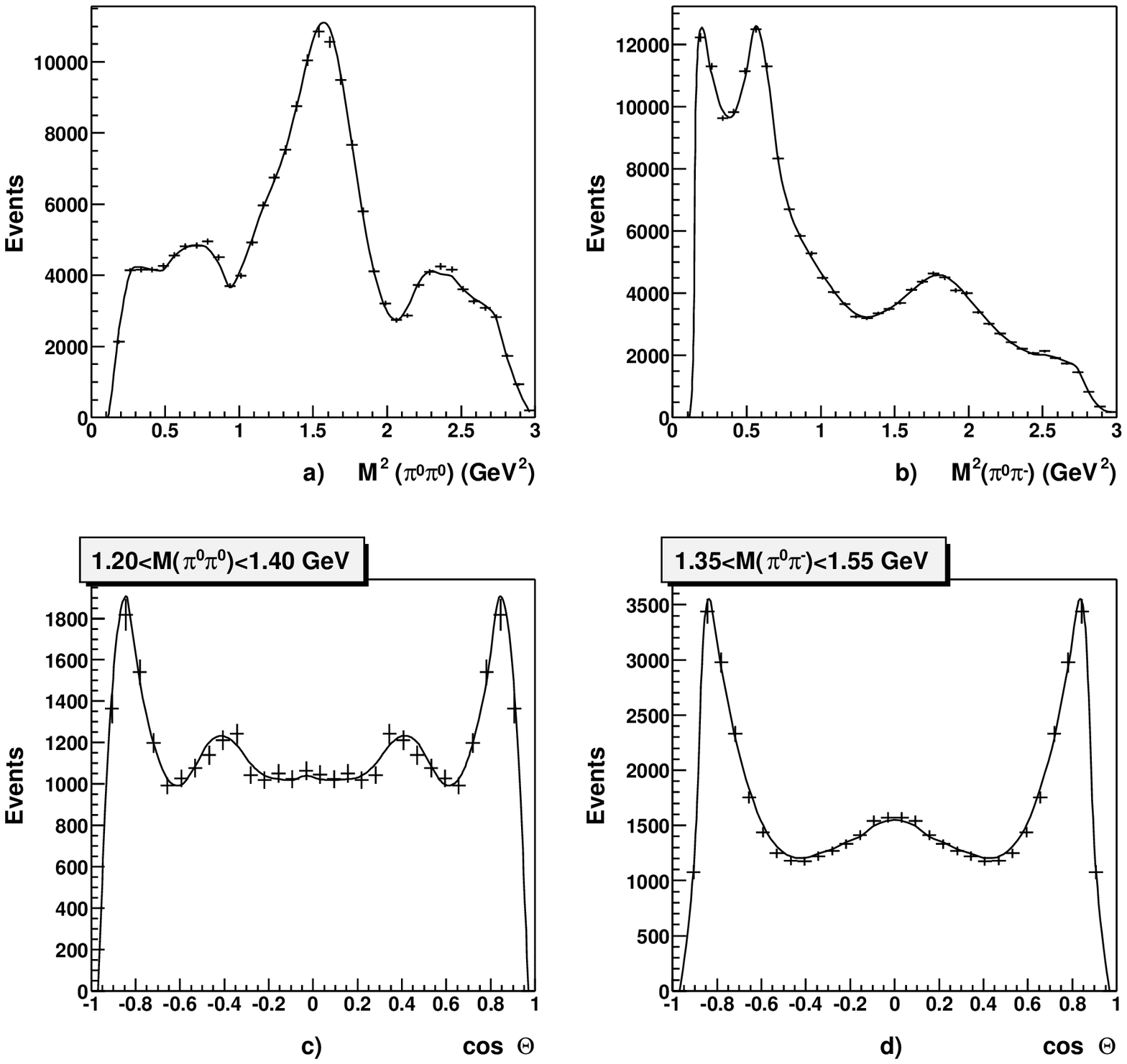,width=16.5cm}}
\caption{a,b) Mass projections of the acceptance-corrected
Dalitz plot for the $p\bar p$ annihilation into
$\pi^0\pi^0\pi^-$ in liquid $D_2$,
c) the angle distribution
between charged and neutral pions in c.m.s. of $\pi^0\pi^0$ system
taken at masses between 1.20 and  1.40 GeV,
d) the angle distribution
between neutral pions in c.m.s. of $\pi^0\pi^-$ system
taken at masses between 1.35 and  1.55 GeV}
\end{figure}

\begin{figure}
\centerline{\epsfig{file=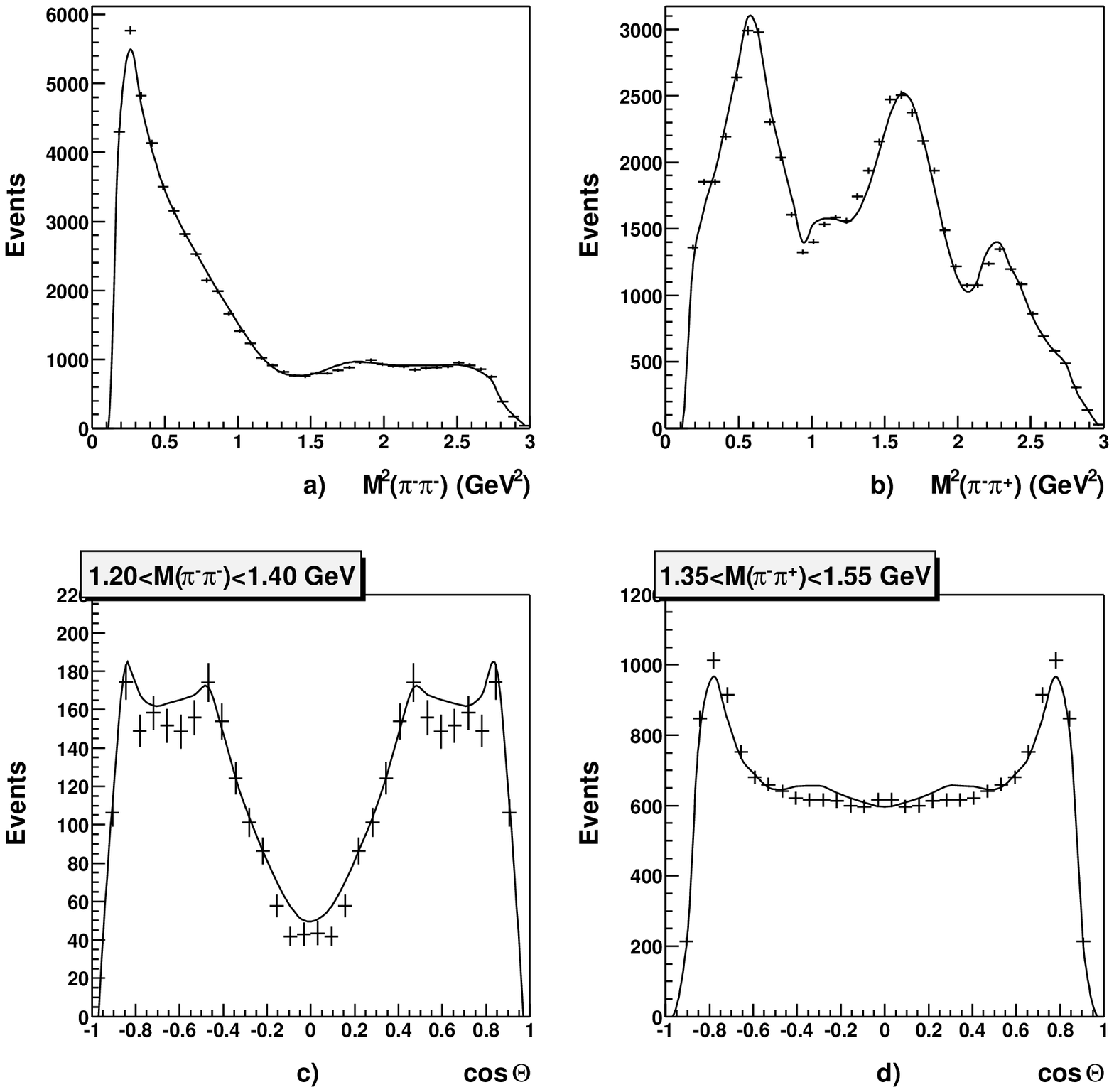,width=16.5cm}}
\caption{a,b) Mass projections of the acceptance-corrected
Dalitz plot for the $p\bar p$ annihilation into
$\pi^-\pi^-\pi^+$ in liquid $D_2$,
c) the angle distribution
between charged and neutral pions in c.m.s. of $\pi^-\pi^-$ system
taken at masses between 1.20 and  1.40 GeV,
d) the angle distribution
between charged pions in c.m.s. of $\pi^-\pi^0$ system
taken at masses between 1.35 and  1.55 GeV.}
\end{figure}

\begin{figure}
\centerline{\epsfig{file=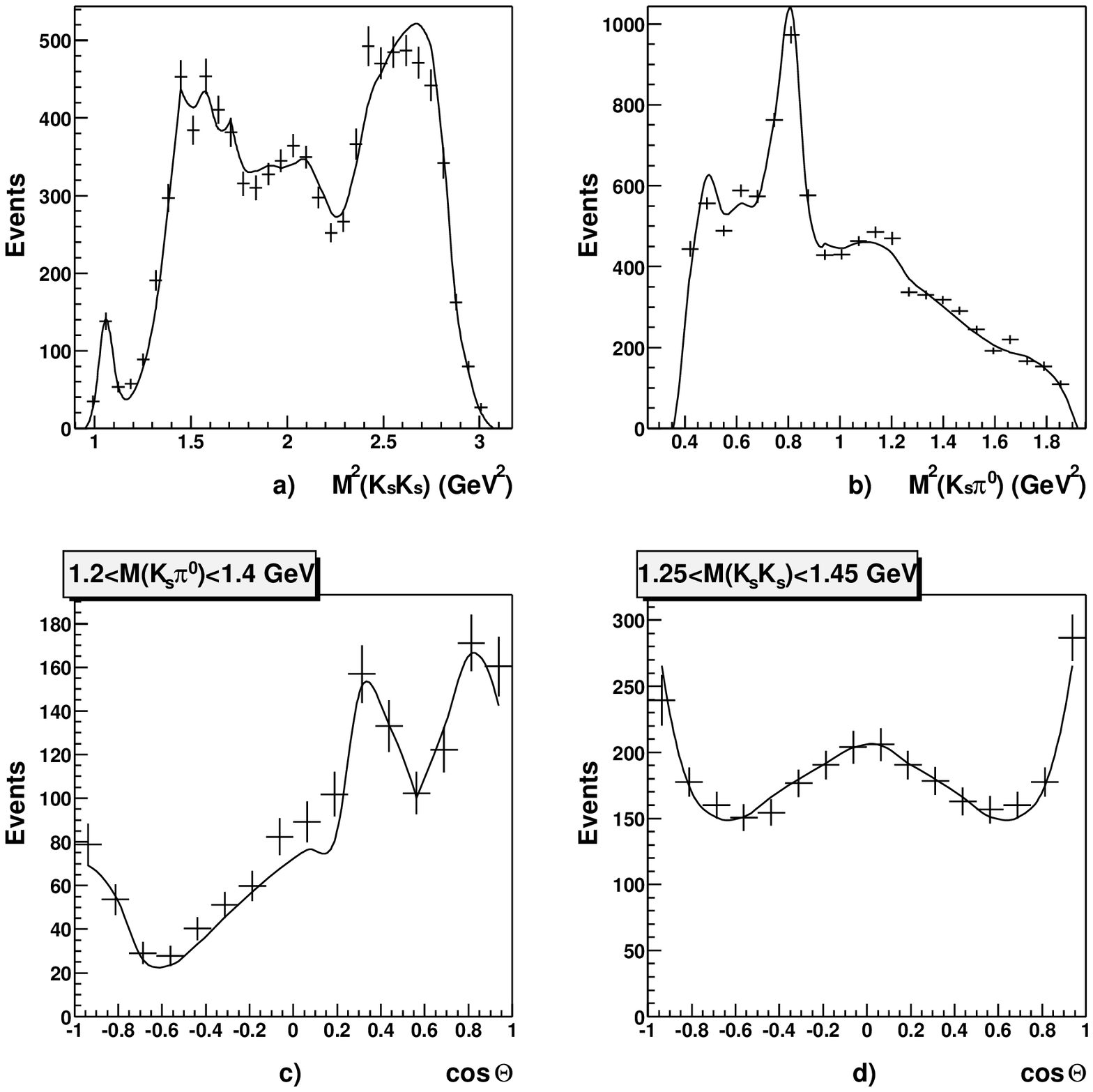,width=16.5cm}}
\caption{a,b) Mass projections of the acceptance-corrected
Dalitz plot for the $p\bar p$ annihilation into
$K_S K_S\pi^0$ in liquid $H_2$,
c) the angle distribution
between kaons in c.m.s. of $K_S\pi^0$ system
taken at masses between 1.20 and  1.40 GeV,
d) an angle distribution
between kaon pion in c.m.s. of $K_SK_S$ system
taken at masses between 1.25 and  1.45 GeV.}
\end{figure}

\begin{figure}
\centerline{\epsfig{file=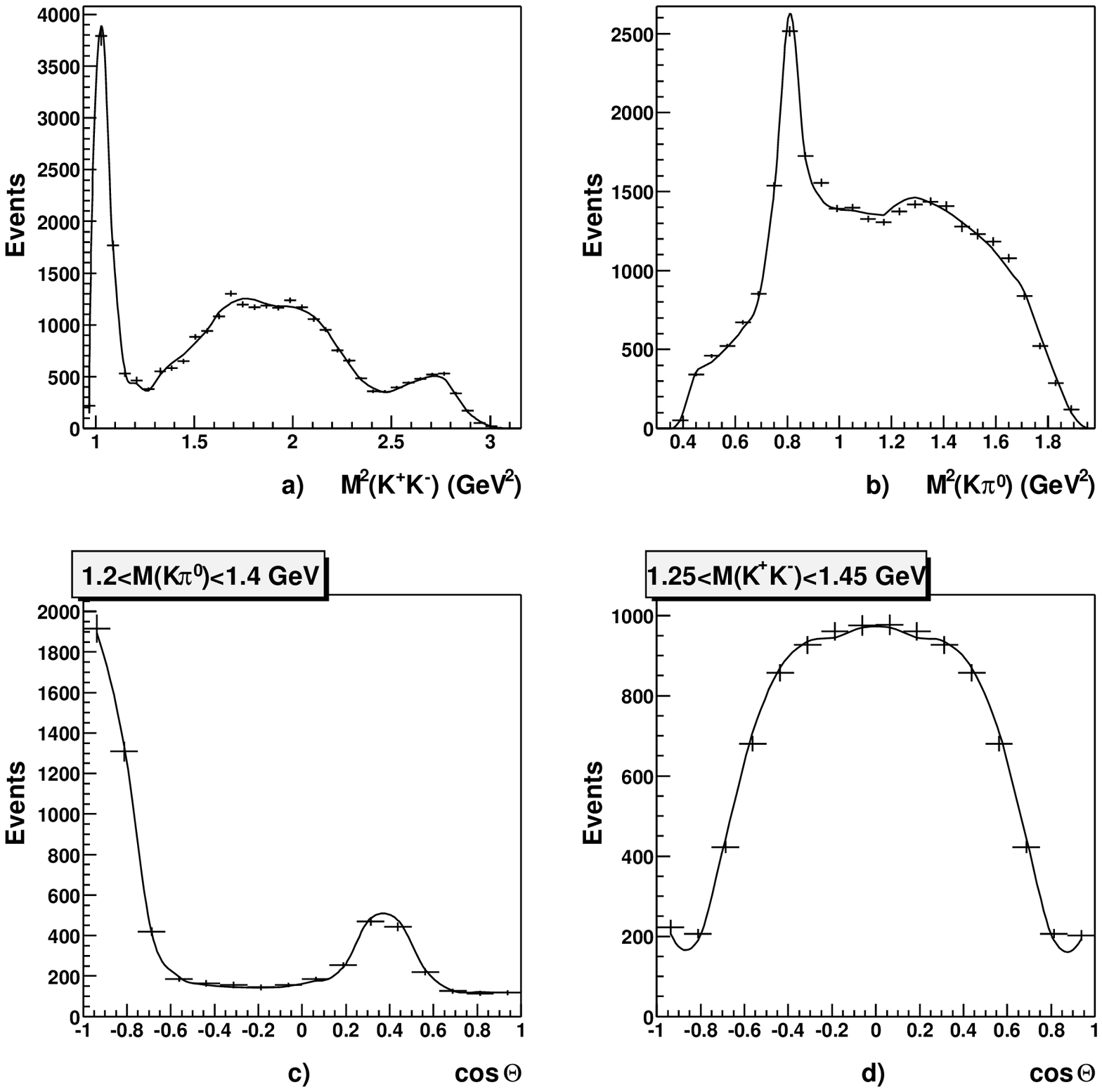,width=16.5cm}}
\caption{a,b) Mass projections of the acceptance-corrected
Dalitz plot for the $p\bar p$ annihilation into
$K^+ K^-\pi^0$ in liquid $H_2$,
c) the angle distribution
between kaons in c.m.s. of $K\pi^0$ system
taken at masses between 1.20 and  1.40 GeV,
d) the angle distribution
between kaon pion in c.m.s. of $K^+K^-$ system
taken at masses between 1.25 and  1.45 GeV.}
\end{figure}

\begin{figure}
\centerline{\epsfig{file=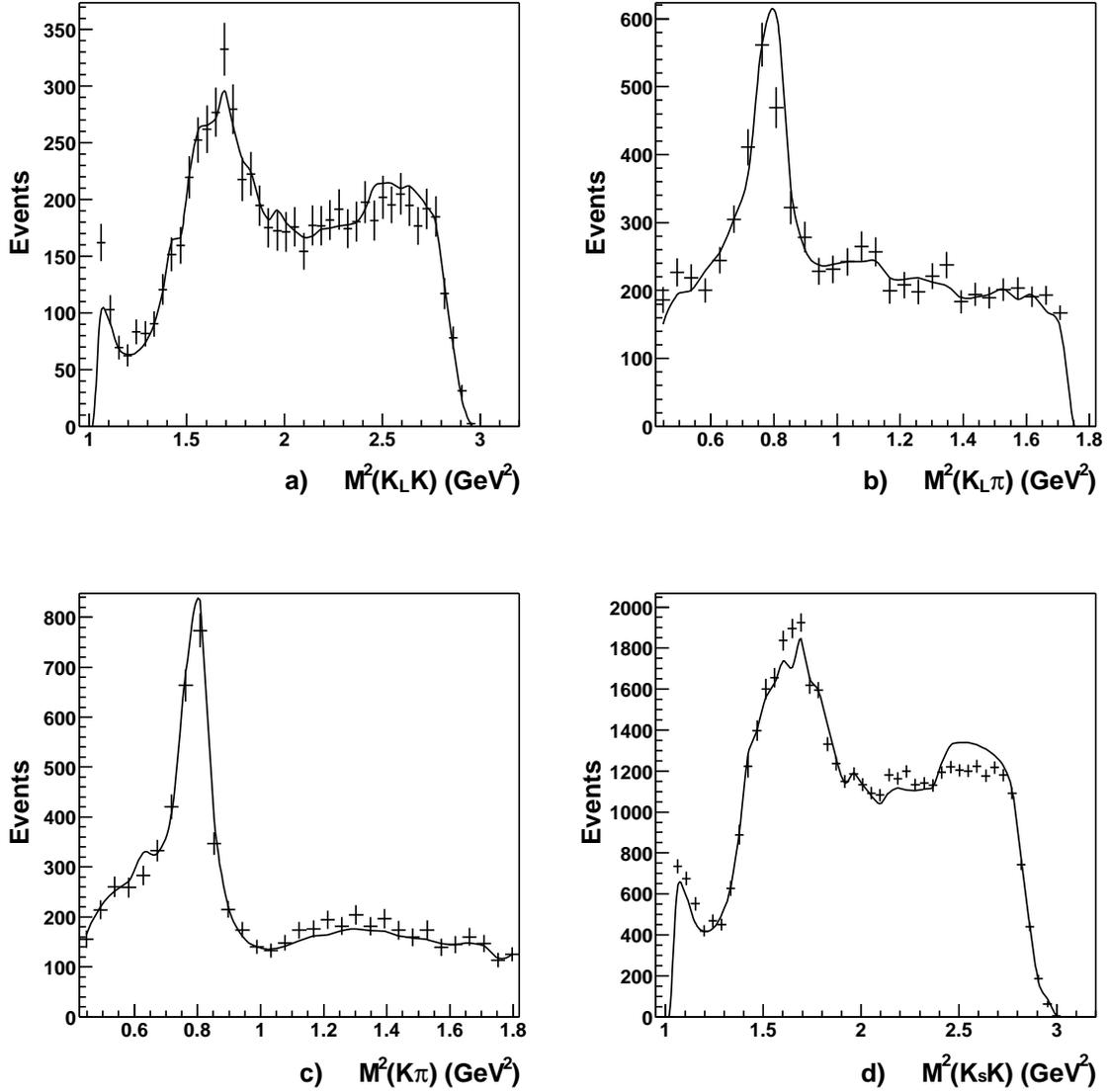,width=16.5cm}}
\caption{a,b,c) Mass projections of the acceptance-corrected
Dalitz plot for the $p\bar p$ annihilation into
$K_L K^-\pi^+$ ($K_L K^+\pi^-$) in liquid $H_2$,
d) $K_S K$ mass projection of the acceptance corrected
Dalitz plot for the $p\bar p$ annihilation into
$K_S K^-\pi^+$. This reaction has some problems with
acceptance correction and was not used in the analysis. The full
curve corresponds to the fit of $\bar p p\to K_L K^-\pi^+$ reaction
normalized to the number of $K_S K^-\pi^+$ events.}
\end{figure}

\begin{figure}
\centerline{\epsfig{file=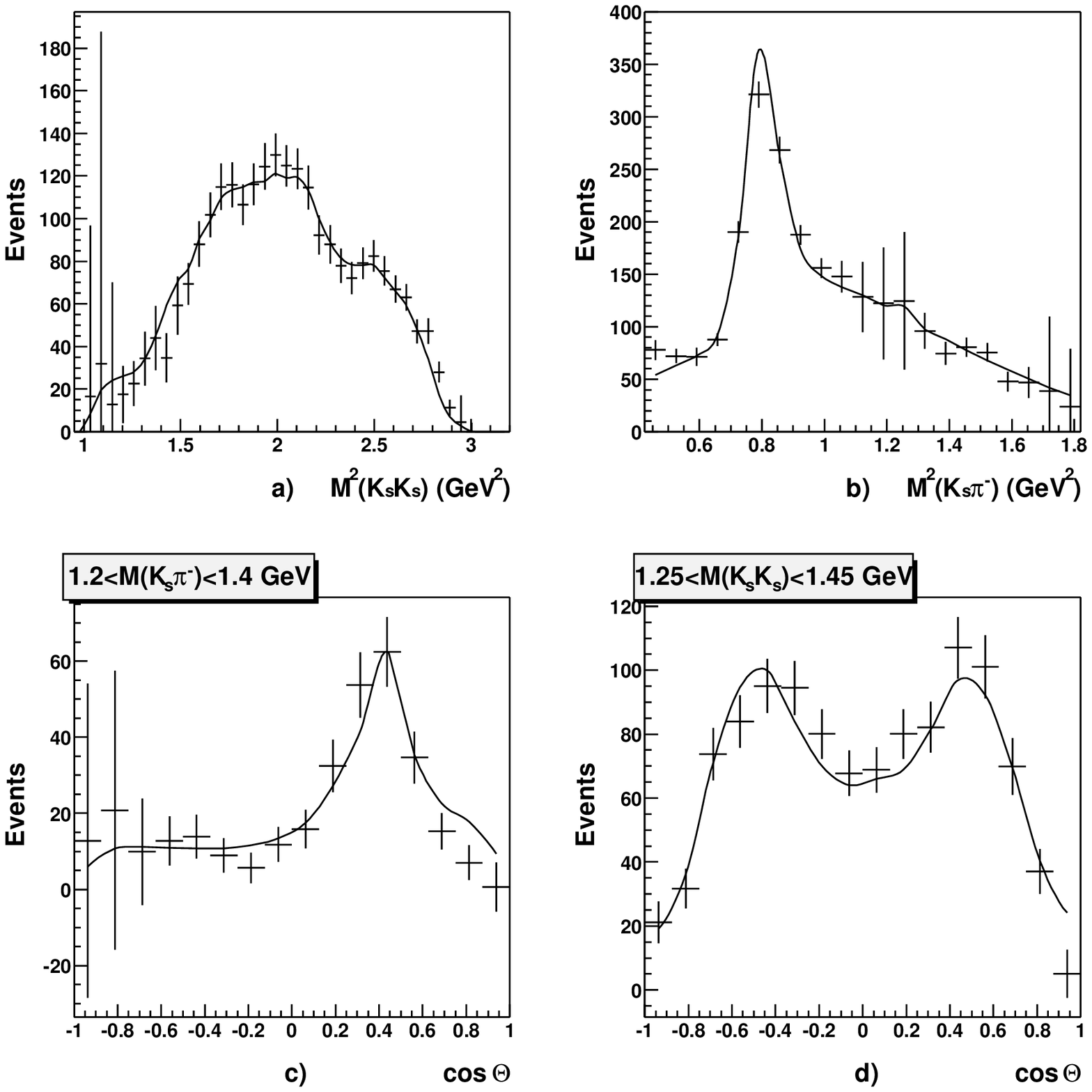,width=16.5cm}}
\caption{a,b) Mass projections of the acceptance-corrected
Dalitz plot for the $p\bar p$ annihilation into
$K_S K_S\pi^-$ in liquid $D_2$,
c) the angle distribution
between kaons in c.m.s. of $K_S\pi^-$ system
taken at masses between 1.20 and  1.40 GeV,
d) the angle distribution
between kaon and pion in c.m.s. of $K_S K_S$ system
taken at masses between 1.25 and  1.45 GeV.}
\end{figure}

\begin{figure}
\centerline{\epsfig{file=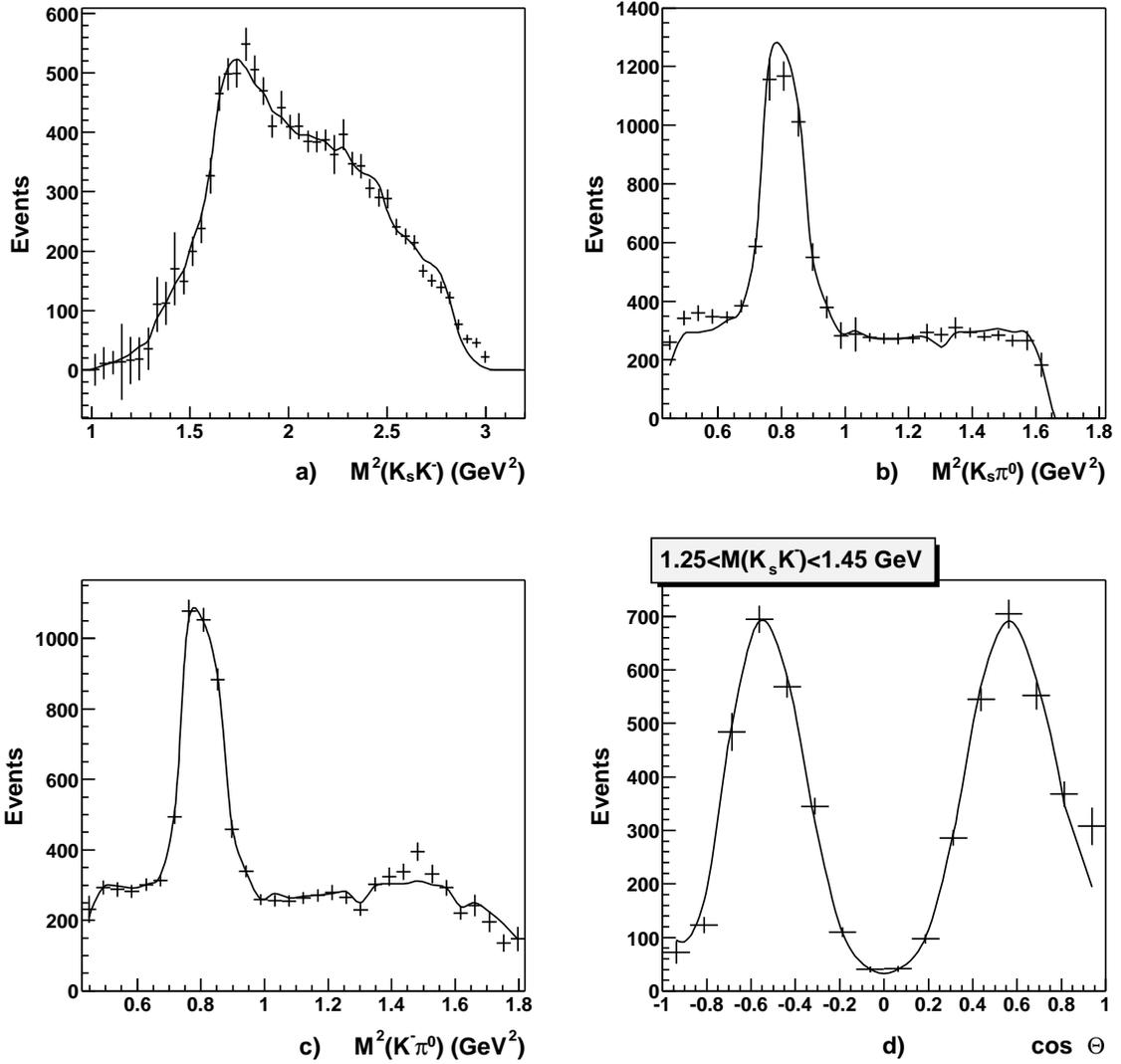,width=16.5cm}}
\caption{a,b,c) Mass projections of the acceptance-corrected
Dalitz plot for the $p\bar p$ annihilation into
$K_S K^-\pi^0$ in liquid $D_2$,
d) the angle distribution
between $K_S$ and $\pi^0$ in c.m.s. of $K_S K^-$ system
taken at masses between 1.25 and  1.45 GeV.}
\end{figure}

\begin{figure}
\begin{center}
\epsfig{file=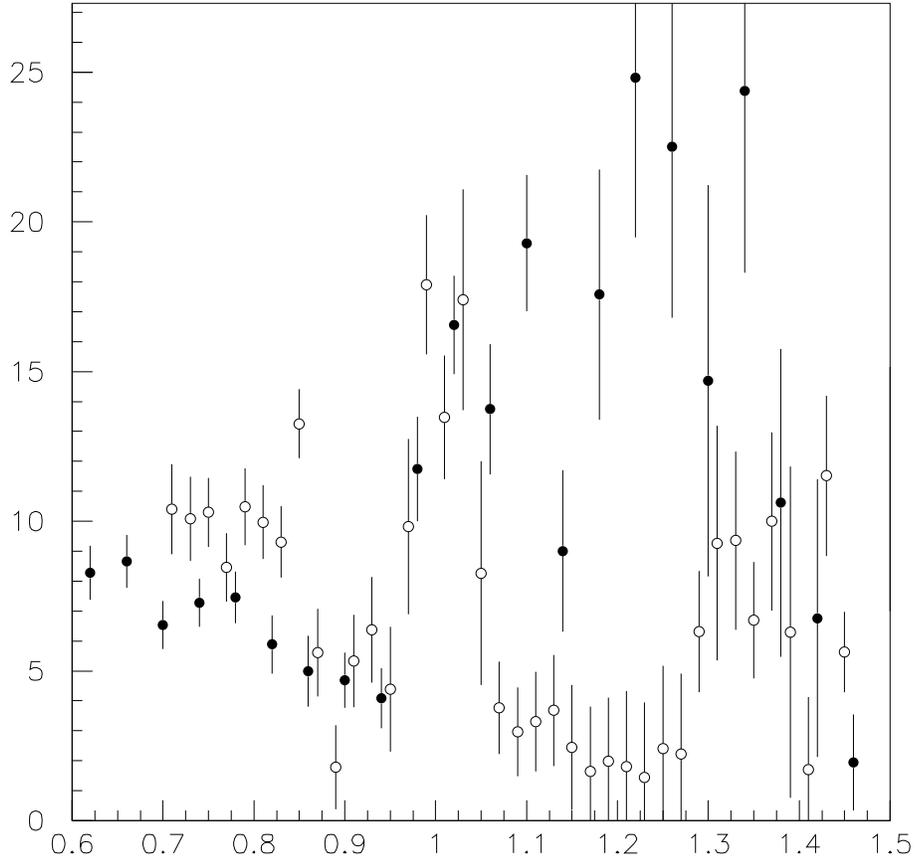,width=14.5cm}
\caption{Comparison of the GAMS and E852 data in the
$|t|$-interval $0.3\le |t|\le 0.4$ GeV$^2$. Full circles are
E852 data and open circles correspond to the substruction of the two
sets of GAMS data:
$N[0.3<|t|<1.0$ GeV$^2]$/(20 MeV) --
$N[0.4<|t|<1.0$ GeV$^2]$/(20 MeV).
}
\end{center}
\end{figure}

\begin{figure}
\begin{center}
\epsfig{file=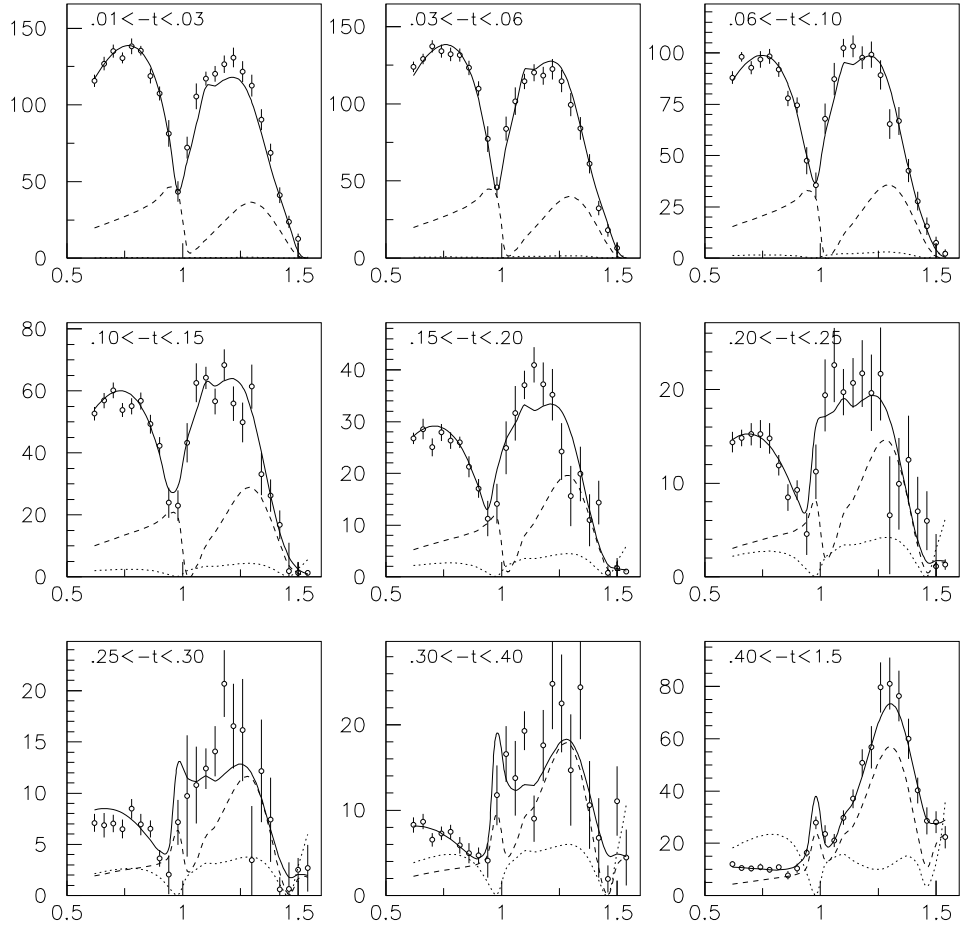,width=14.5cm}
\caption{ Description of the E852 data at different $t$-intervals.
Dashed curve shows the contribution from $a_1$-trajectory and dotted
curve from $\pi_{daughter}$-trajectory.}
\end{center}
\end{figure}

\begin{figure}[hp]
\begin{center}
\epsfig{file=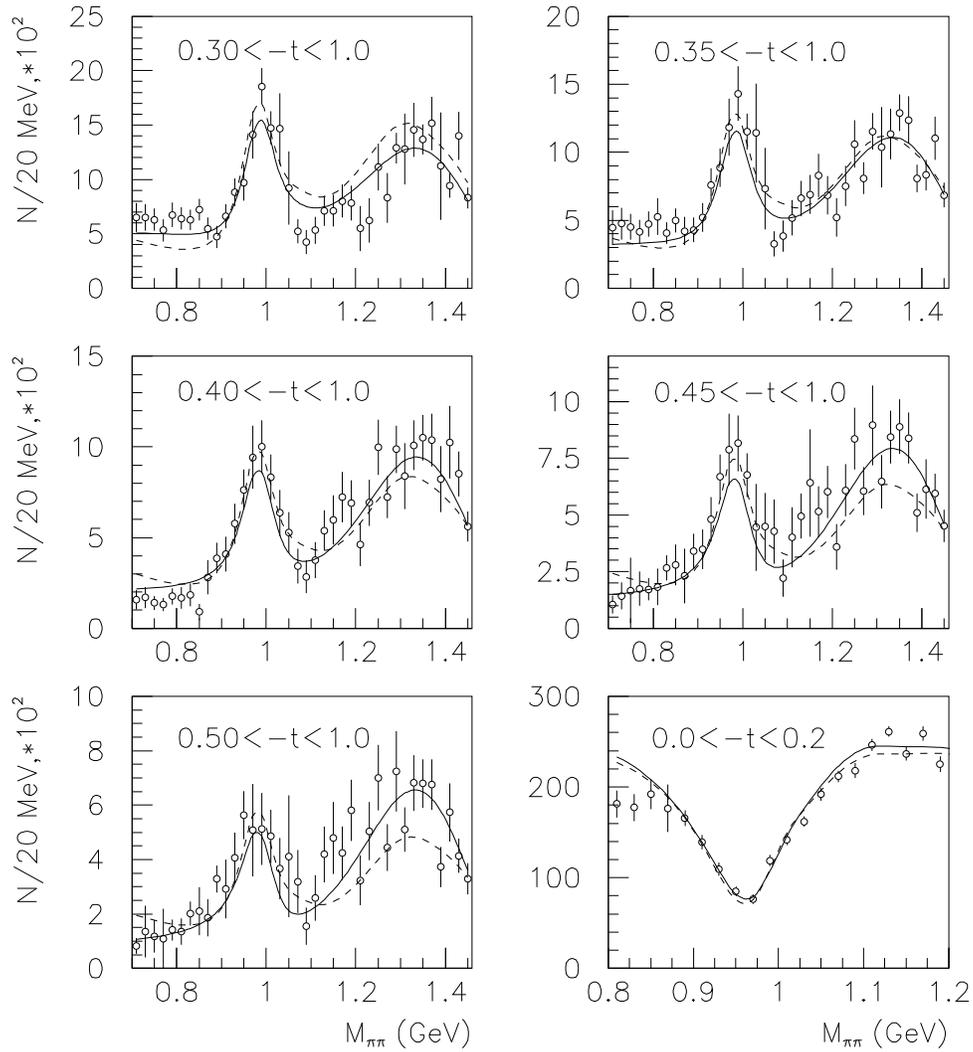,width=14.5cm}
\caption{Description of the GAMS data at different $t$-intervals.
The dashed curves refer to the solution in the previous analysis [11]. }
\end{center}
\end{figure}

\begin{figure}[hp]
\begin{center}
\epsfig{file=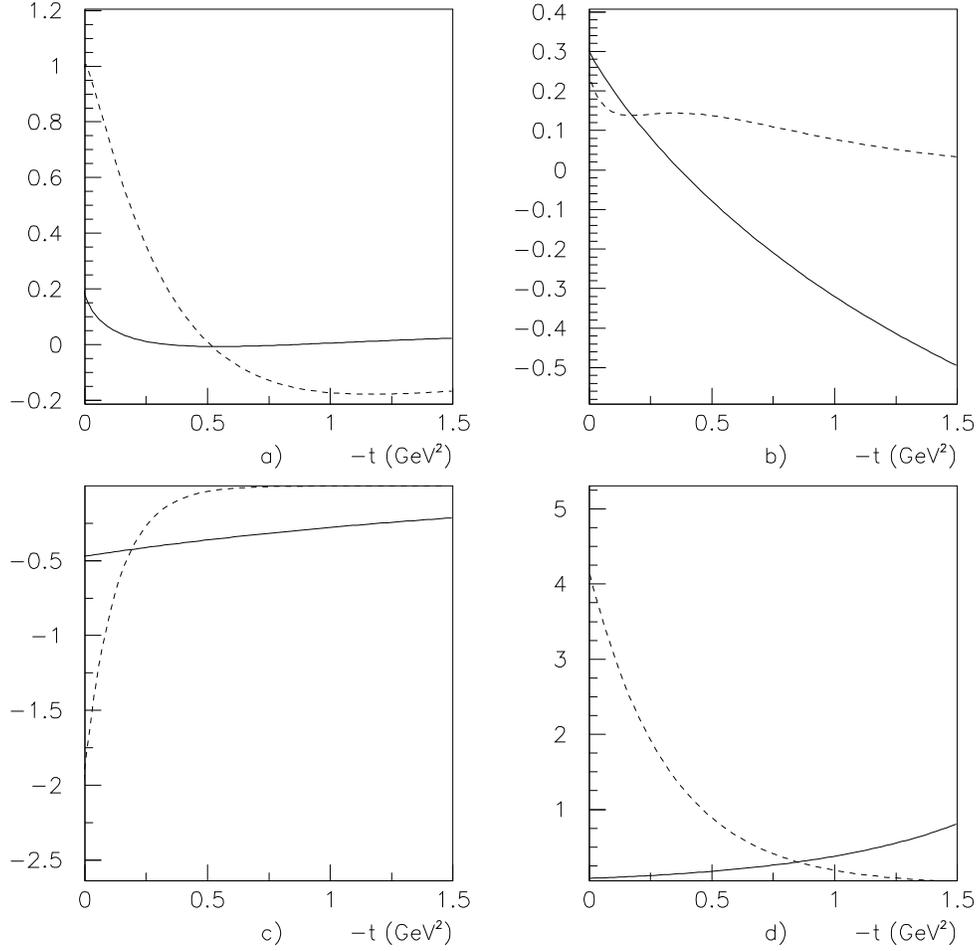,width=14.5cm}
\caption{The $t$-dependence of the K-matrix couplings.
a) $\pi$-exchanges:
full curve is for $f_0^{bare}(720)$ and dashed one for
$f_0^{bare}(1250)$,
b) $\pi$-exchanges: full curve is for $f_0^{bare}(1230)$ and dashed
curve for $f_0^{bare}(1600)$.
c,d) $a_1$-exchanges: $t$-dependence for the same states
as in figures $a$ and $b$.}
\end{center}
\end{figure}

\begin{figure}
\begin{center}
\epsfig{file=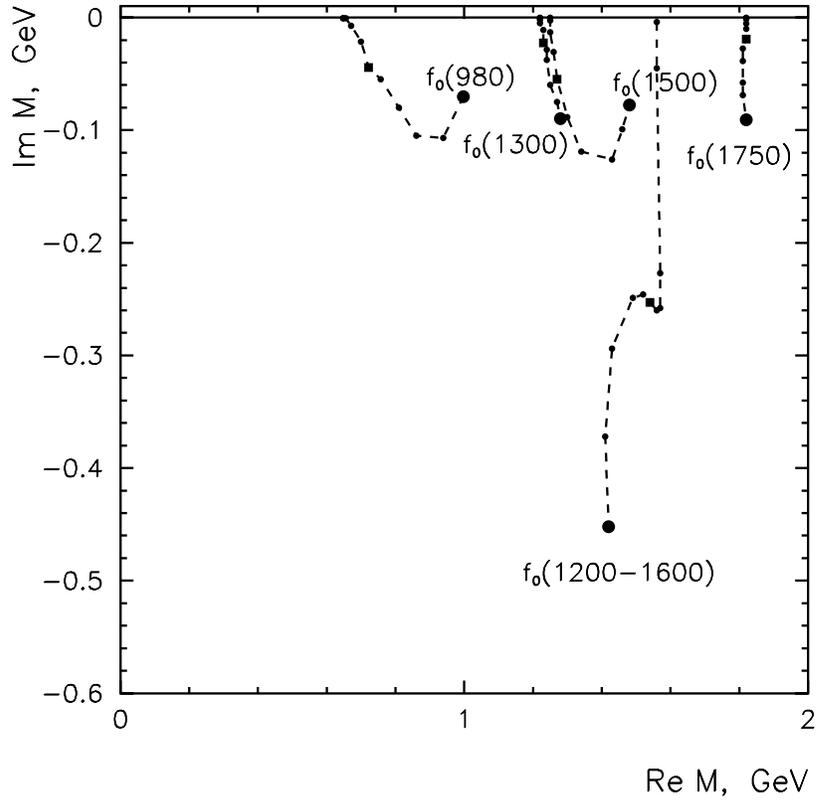,width=12cm}
\caption{Complex $M$-plane: trajectories of the poles
for $f_0(980)$, $f_0(1300)$, $f_0(1500)$,
$f_0(1750)$, $f_0(1200-1600)$
during gradual onset of the decay processes.}
\end{center}
\end{figure}

\begin{figure}
\centerline{\epsfig{file=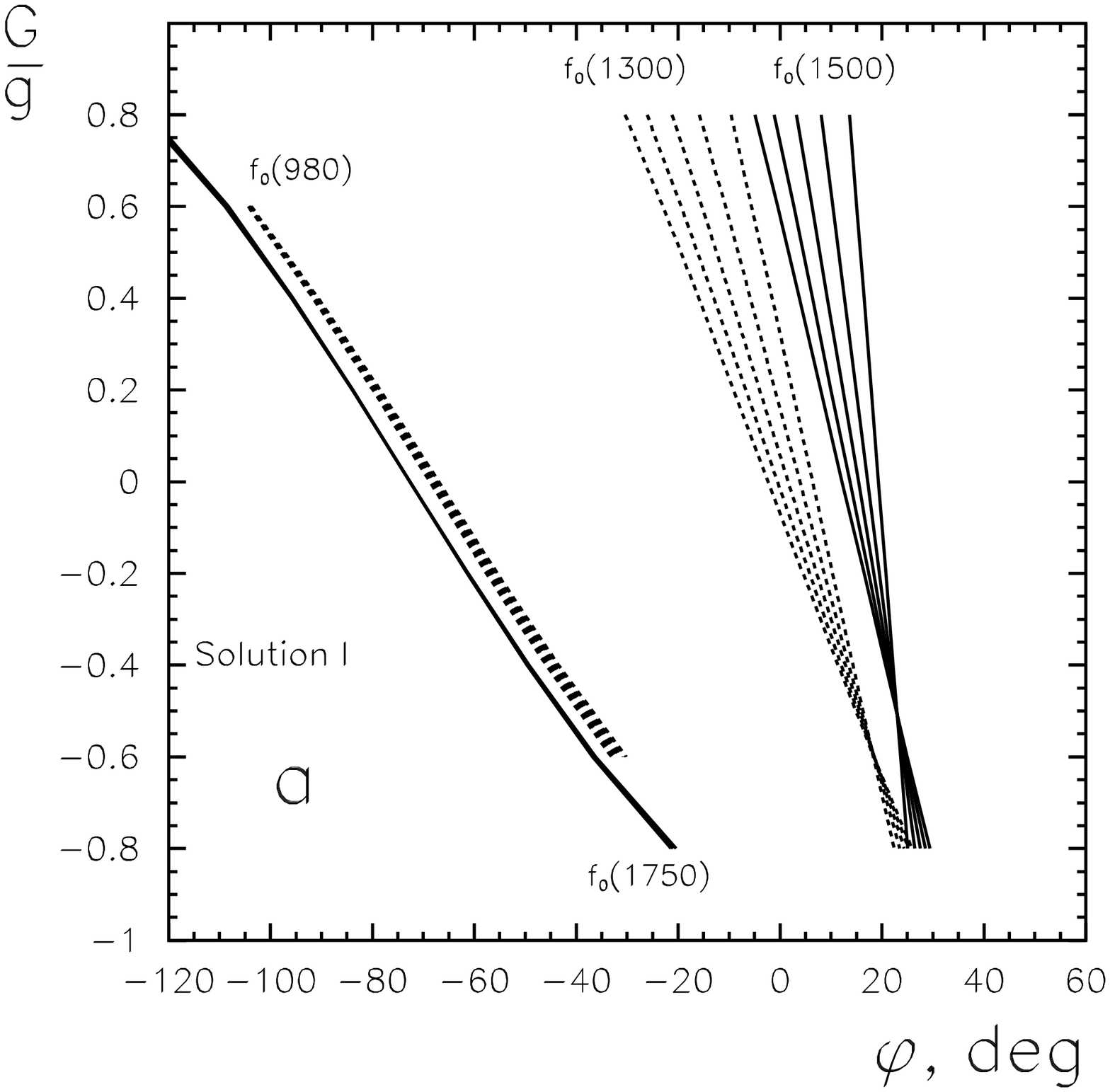,width=7cm}\hspace{0.5cm}
            \epsfig{file=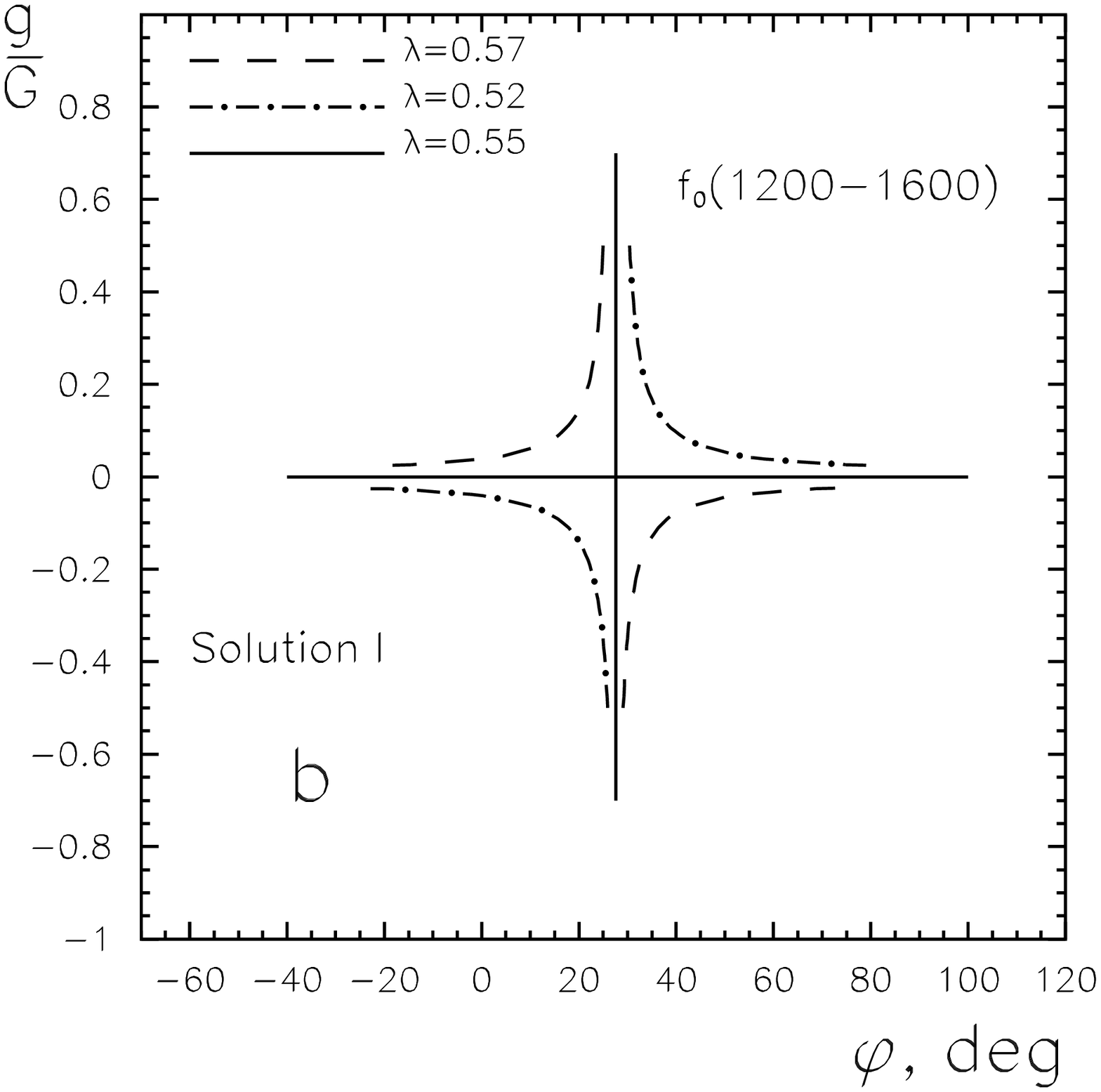,width=7cm}}
\centerline{\epsfig{file=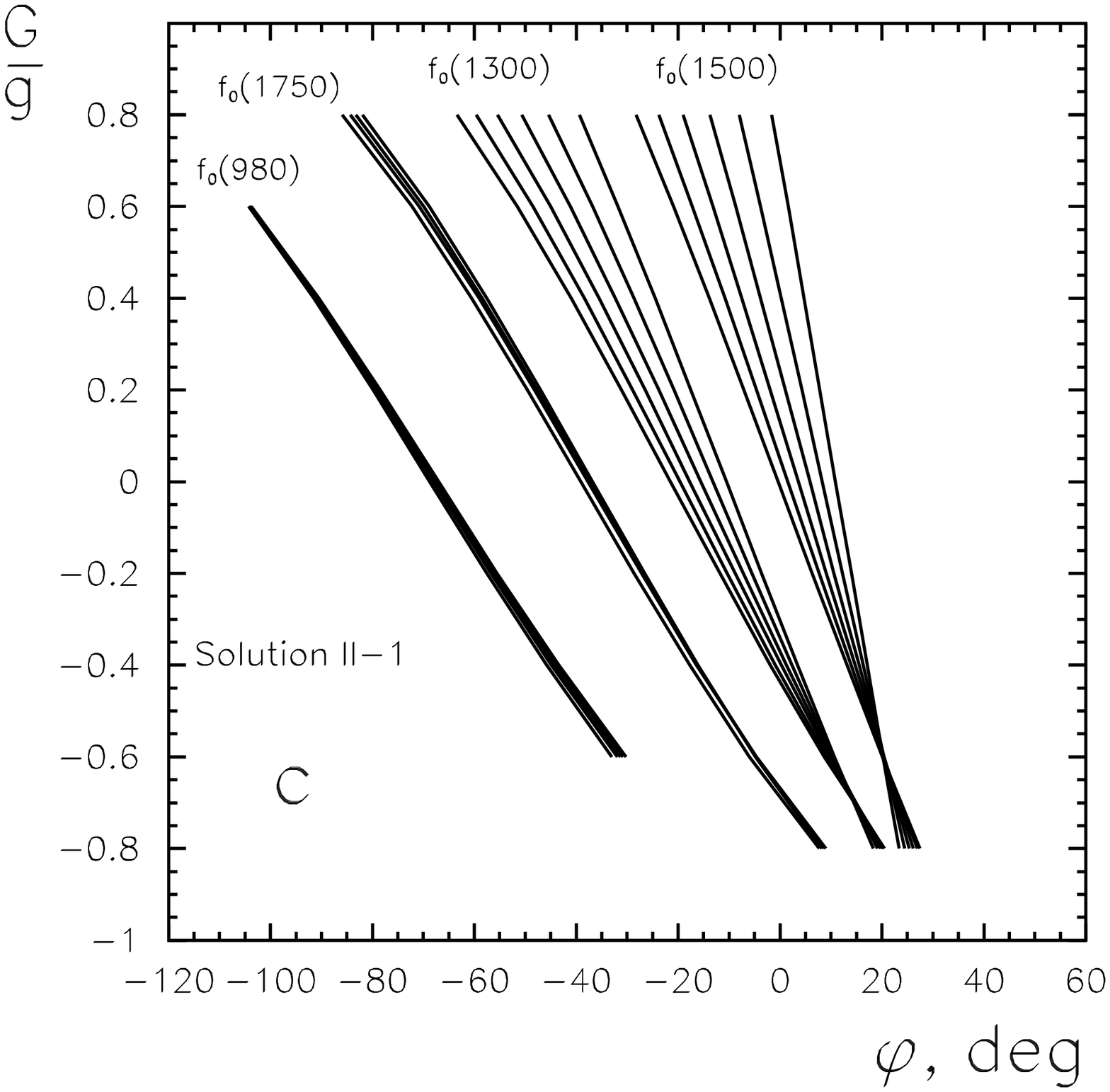,width=7cm}\hspace{0.5cm}
            \epsfig{file=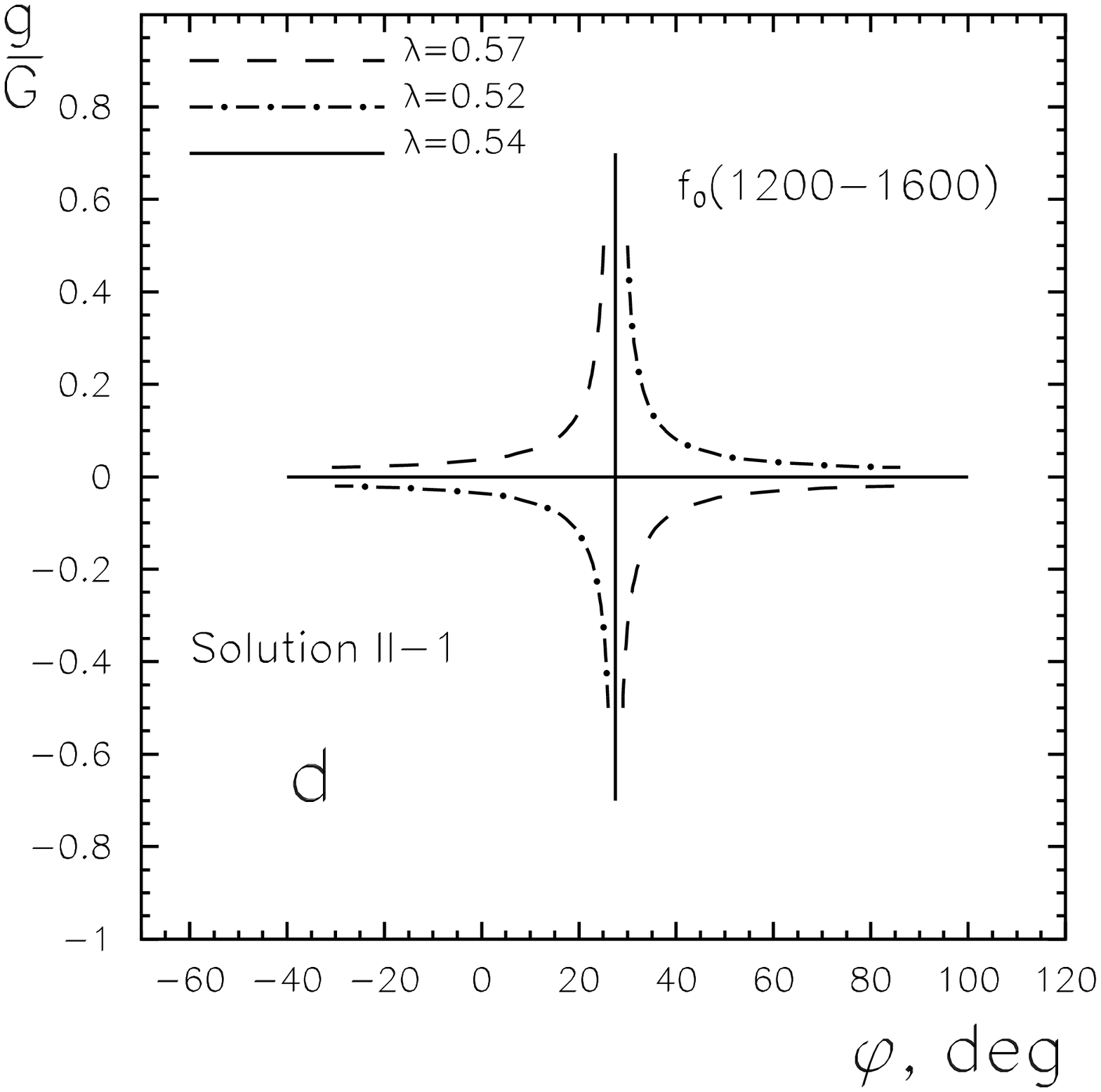,width=7cm}}
\centerline{\epsfig{file=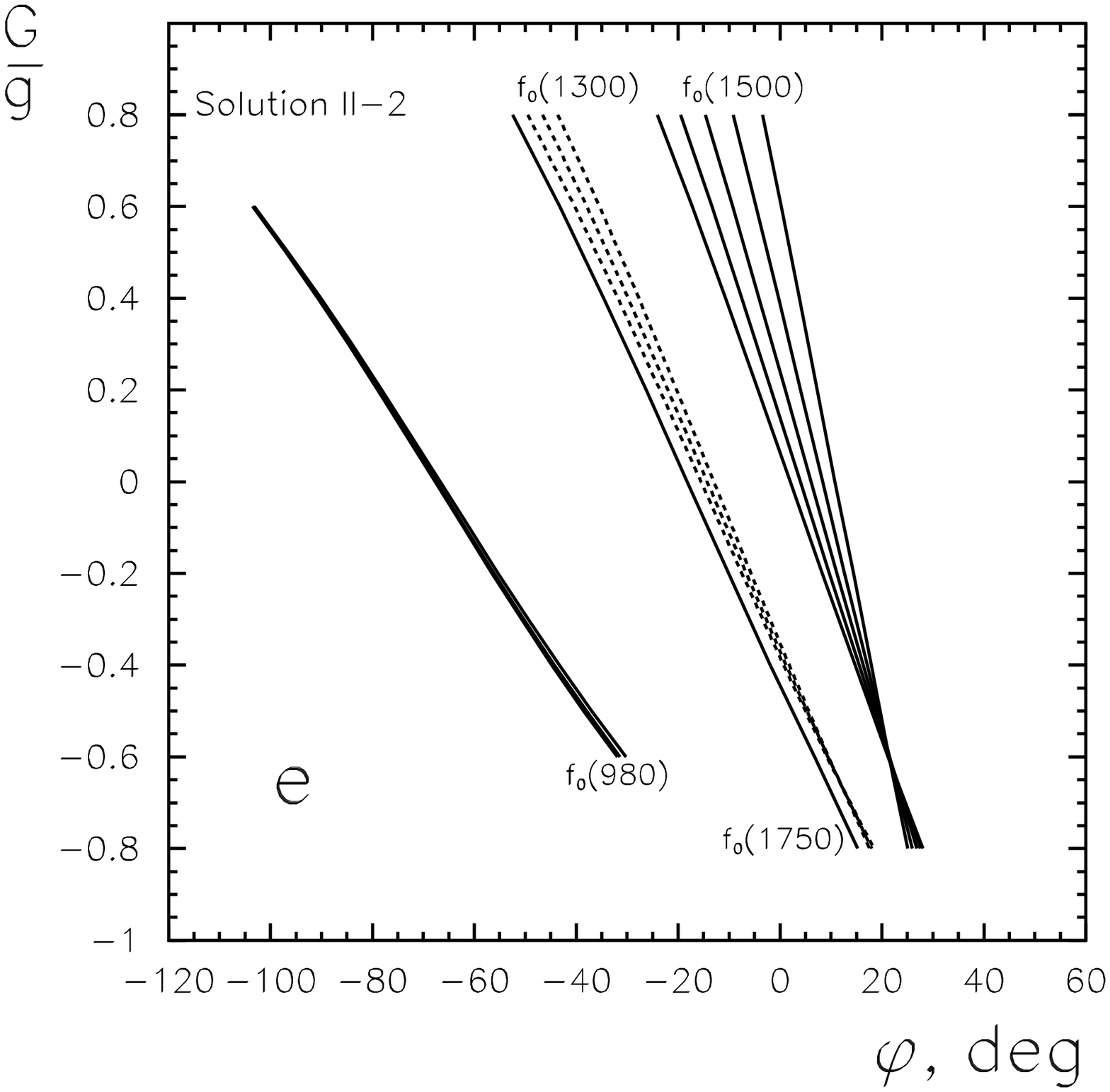,width=7cm}\hspace{0.5cm}
            \epsfig{file=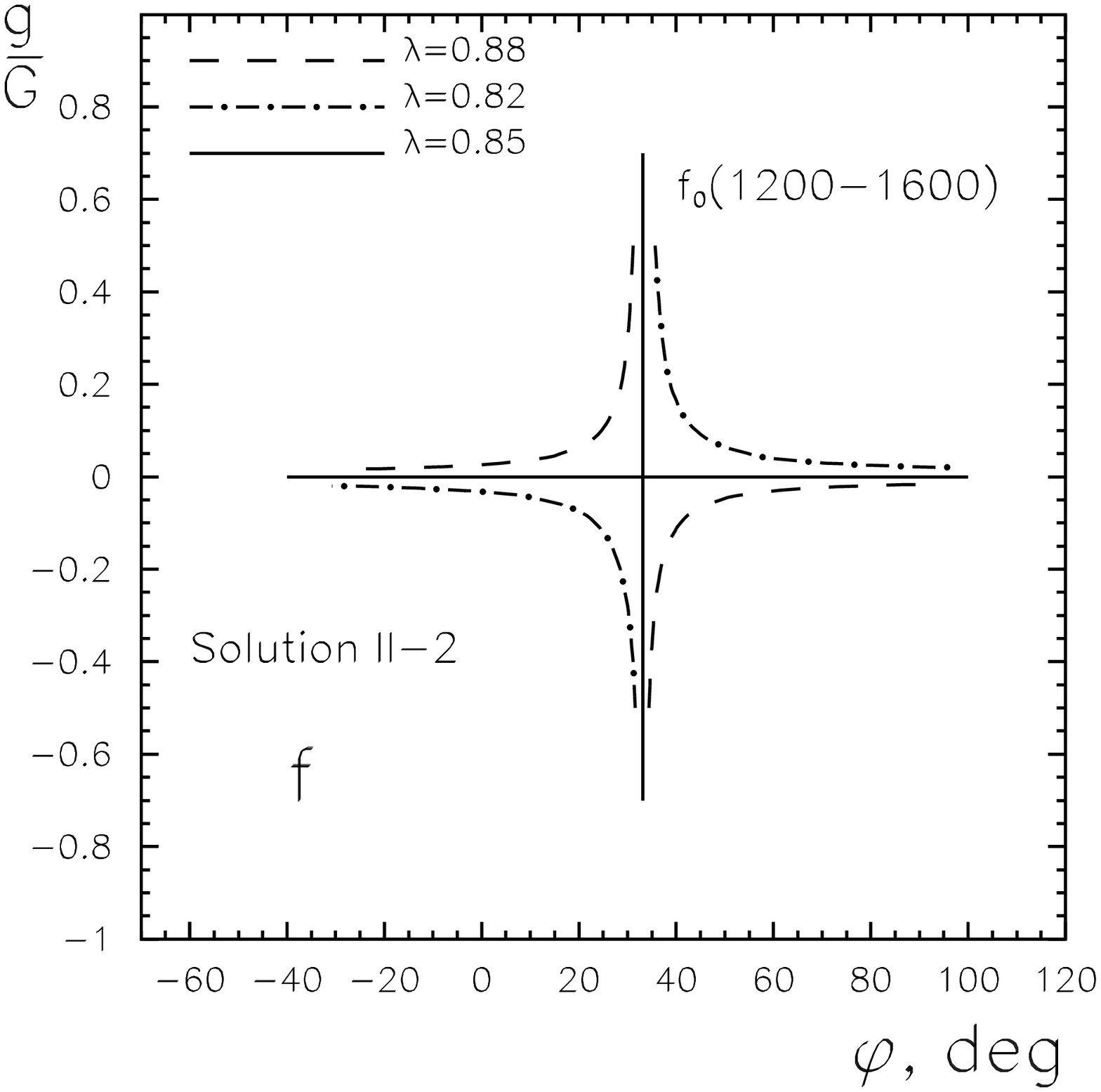,width=7cm}}
\caption{Correlation curves on the $(\varphi ,G/g)$ and $(\varphi ,g/G)$
plots for the description of the decay couplings of resonances (Table 5)
in terms of quark-combinatorics relations (22).
a,c,e) Correlation curves  for the
$q\bar q$-originated resonances: the curves with appropriate
$\lambda$'s cover strips on the $(\varphi ,G/g)$ plane.
b,d,f) Correlation curves  for the glueball descendant: the curves
at appropriate
$\lambda$'s form a cross on the $(\varphi ,g/G)$ plane
with the center near $\varphi \sim 30^\circ$,  $g/G\sim 0$.}
\end{figure}

\begin{figure}[h]
\centerline{\epsfig{file=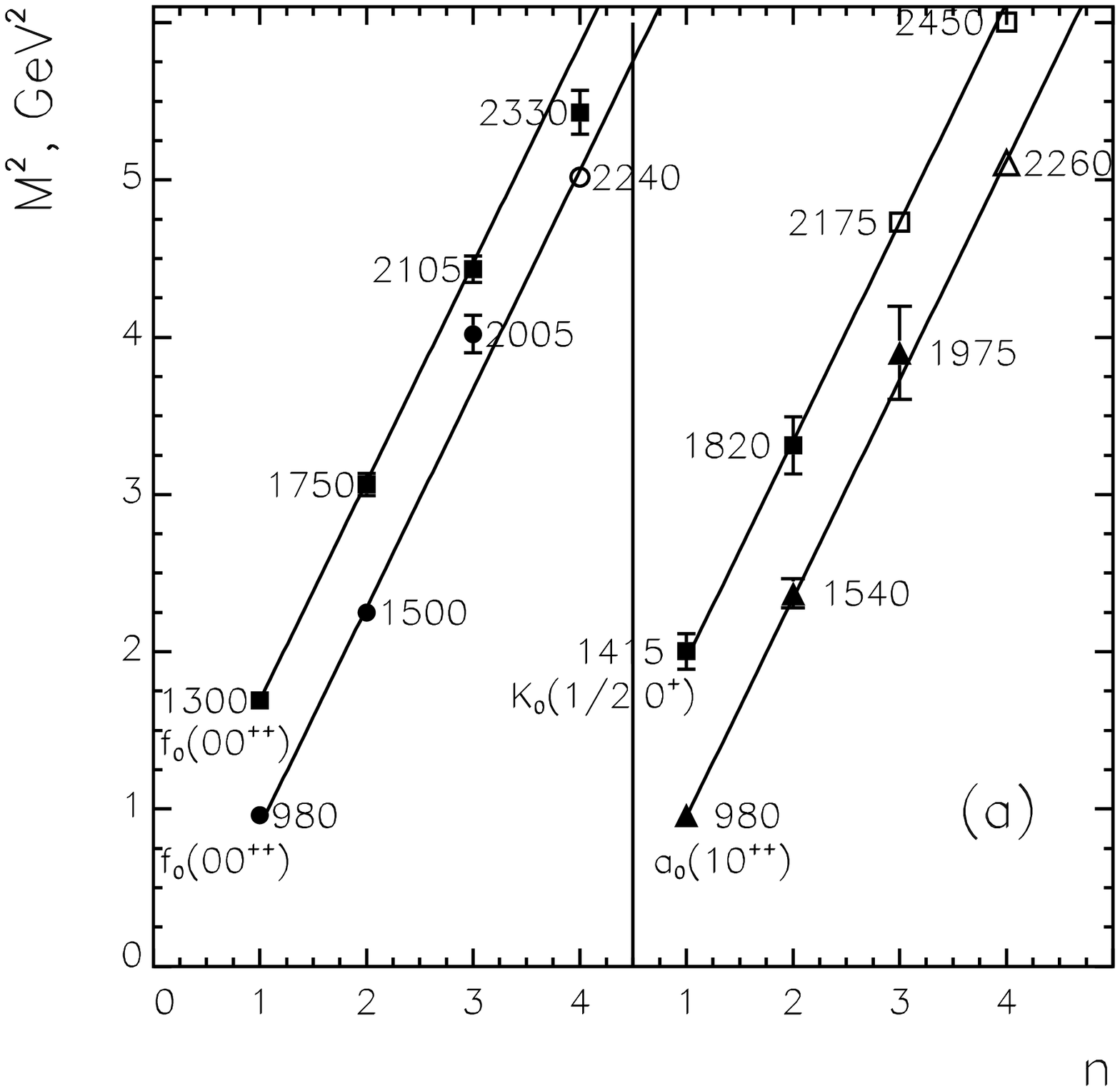,width=6.5cm}
            \epsfig{file=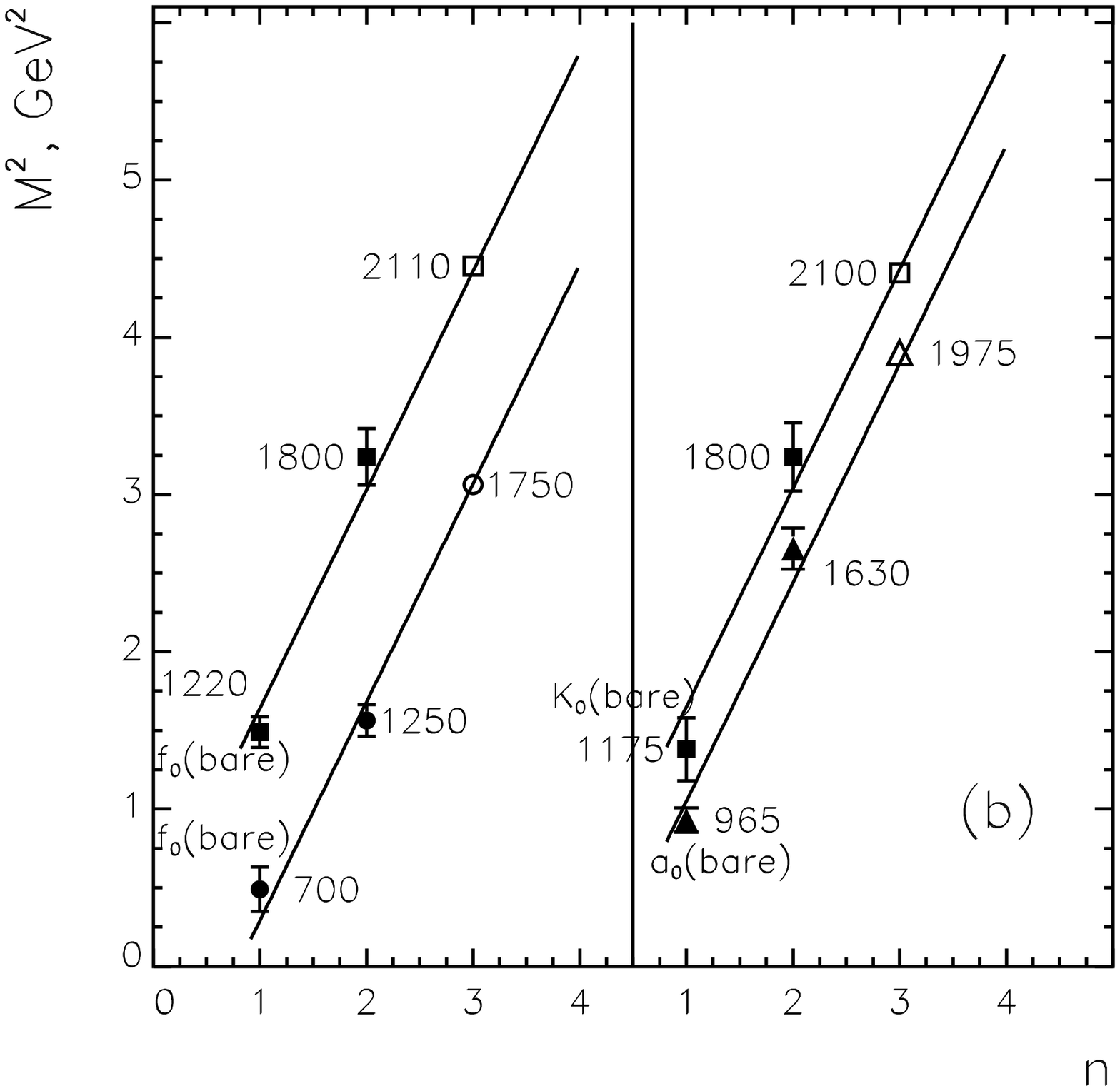,width=6.5cm}}
\caption{ Linear trajectoties on the $(n,M^2)$-plot
for scalar resonances (a) and bare scalar states (b). Open
points stand for the predicted states.}
\end{figure}

\end{document}